\documentclass[11pt,a4paper]{article}
\usepackage{jheppub}

\usepackage{dsfont}   
\newcommand{\dif}{\mathrm{d}} 

\usepackage{tikz}
\usetikzlibrary{arrows,decorations.pathmorphing,backgrounds,positioning,fit,shapes,chains,shapes.gates.logic.US,trees,tikzmark,decorations.text,calc,
mindmap,decorations.pathreplacing,decorations.markings}

\def\be{\begin{equation}}
\def\ee{\end{equation}}
\def\ba{\begin{eqnarray}}
\def\ea{\end{eqnarray}}
\def\nl{\nonumber\\}

\def\s{\sigma}

\def\a{\alpha}
\def\b{\beta}

\def\<{\langle}
\def\>{\rangle}

\usepackage{bm}

\usepackage{subfig}
\usepackage[colorinlistoftodos]{todonotes}
\usetikzlibrary{decorations.pathreplacing}

\usepackage{epstopdf}
\usepackage{verbdef}

\usepackage{tcolorbox}

\def\dotfill#1{\cleaders\hbox to #1{.}\hfill}
\newcommand\dotline[2][.5em]{\leavevmode\hbox to #2{\dotfill{#1}\hfil}}

\usepackage{amsthm}
\theoremstyle{plain}
\newtheorem{theorem}{Theorem}

\begin{document}

\title{Labelled tree graphs, Feynman diagrams
\\and disk integrals}

\author[a,b]{Xiangrui Gao}
\author[a,b]{,~Song He}
\author[a,b]{,~ Yong Zhang}
\affiliation[a]{CAS Key Laboratory of Theoretical Physics, Institute of Theoretical Physics, Chinese Academy of Sciences, Beijing 100190, China}
\affiliation[b]{School of Physical Sciences, University of Chinese Academy of Sciences, No.19A Yuquan Road, Beijing 100049, China}
\emailAdd{gaoxiangrui@itp.ac.cn,songhe@itp.ac.cn, yongzhang@itp.ac.cn}

\date{\today}

\abstract{In this note, we introduce and study a new class of ``half integrands" in Cachazo-He-Yuan (CHY) formula, which naturally generalize the so-called Parke-Taylor factors; these are dubbed Cayley functions as each of them corresponds to a labelled tree graph.  The CHY formula with a Cayley function squared gives a sum of Feynman diagrams, and we represent it
 by a combinatoric polytope whose vertices correspond to
 Feynman diagrams. We provide a simple graphic rule to derive the polytope from a labelled tree graph, and classify such polytopes ranging from the associahedron to the permutohedron. Furthermore, we study the linear space of such half integrands and find (1) a closed-form formula reducing any Cayley function to a sum of Parke-Taylor factors in the Kleiss-Kuijf basis (2) a set of Cayley functions as a new basis of the space;  each element has the remarkable property that its CHY formula with a given Parke-Taylor factor gives either a single Feynman diagram or zero. We also briefly discuss applications of Cayley functions and the new basis in certain disk integrals of superstring theory.
}

\maketitle

\section{Invitation: a new class of CHY half integrands}\label{invitation}

In 2013, F. Cachazo, E. Y. Yuan and one of the authors found a new formulation for tree-level S-matrices for a large variety of massless theories in arbitrary
dimensions~\cite{Cachazo:2013iaa,Cachazo:2013gna,Cachazo:2013hca,Cachazo:2013iea}
(for extension to more theories, see {\it i.e.} \cite{
Cachazo:2014nsa,Cachazo:2014xea,He:2016iqi,He:2016dol,Cachazo:2016njl}).  The key ingredient of the formulation is the so-called {\it scattering equations}, which link kinematics of $n$ massless particles to points in the moduli space of $n$-punctured Riemann spheres, ${\cal M}_{0,n}$~\cite{Cachazo:2013iaa,Cachazo:2013gna}:

\begin{equation}\label{se}
  \sum_{b\neq a}\frac{k_a\cdot k_b}{\sigma_a-\sigma_b}=0,\quad a\in\{1,2,...,n\},
\end{equation}
where $\sigma_a$ denotes the position of the $a^{th}$ puncture on the Riemann sphere. The tree-level S-matrix can be compactly formulated as an integral over ${\cal M}_{0,n}$ localized  on solutions of the scattering equations  \eqref{se}~\cite{Cachazo:2013hca}:
\be\label{sm}
{\cal M}^{\rm tree}_n (\{k, \epsilon\})=\int \dif {\bm \mu}_n\,{\cal \bf I}_n (\{k, \epsilon, \sigma\})\,, \quad {\rm with~} \dif {\bm \mu}_n:=\prod_{\substack{a=1\\a\neq i,j,k}}^n \left( \dif \sigma_a\,\delta(\sum_{b\neq a} \frac{k_a\cdot k_b}{\sigma_{a,b}})\right)\,\times (\sigma_{i,j}\sigma_{j,k}\sigma_{k,i})^2\,,
\ee
where $\sigma_{a,b}:=\sigma_a-\sigma_b$ and we have included delta functions imposing \eqref{se} in the measure $\dif {\bm \mu}$. Note that both the moduli space ${\cal M}_{0,n}$ and the scattering equations have an $\rm{SL}(2,\mathds{C})$ redundancy; our definition of $\dif {\bm \mu}$ means that we fix the $\rm{SL}(2,\mathds{C})$ redundancy by deleting three $\dif\sigma$'s and three delta-functions ({\it e.g.} both chosen to be $i,j,k$) with a compensation factor $(\sigma_{i,j}\sigma_{j,k}\sigma_{k,i})^2$. 

We will refer to \eqref{sm} as CHY formula for amplitudes in a given theory, where $\mathcal{\bf I}_n$ is the ``CHY integrand" of the theory that can generally also depends on momenta and polarizations.   Note that for \eqref{sm} to be well defined, the CHY integrand must transform covariantly,  with opposite weight as $\dif {\bm \mu}$, under a $\rm{SL}(2,\mathds{C})$ transformation 
(here $\alpha \delta-\beta\gamma=1$):
\be \label{trans} \sigma_a \to \frac{\alpha \sigma_a+\beta}{\gamma \sigma_a+ \delta}\,: \quad
\dif {\bm \mu}_n \to \prod_{a=1}^n\,(\gamma \sigma_a +\delta)^{-4} \dif {\bm \mu}_n \implies {\cal \bf I}_n \to \prod_{a=1}^n\,(\gamma \sigma_a +\delta)^4\,{\cal  \bf I}_n(\{\sigma_a\})\,,
\ee
and we will refer to this as the fact that ${\bf I}_n$ has weight 4. For most theories that admit CHY representations, the CHY integrand factorizes into two parts $\mathcal{\bf I}_n=\mathcal{\bf I}_n^{(L)}\,\mathcal{\bf I}_n^{(R)}$ where each of them transforms as in \eqref{trans} with weight 2 and we will refer to ${\bf I}_n^{(L)}$ and ${\bf I}_n^{(R)}$ as ``half integrands".

The simplest function with this transformation property is probably the so-called {\it Parke-Taylor} (PT) factor. Given an ordering of $n$ labels, $\alpha:=(\alpha(1), \alpha(2), \cdots, \alpha(n))$ we define
\begin{equation}\label{pt}
{\rm{\bf PT}}(\alpha):=\frac{1}{\sigma_{\alpha(1),\alpha(2)}\sigma_{\alpha(2),\alpha(3)}\cdot \cdot \cdot \sigma_{\alpha(n),\alpha(1)}}\,.
\end{equation}
It turns out that such Parke-Taylor factors play an important role in CHY formula for various theories (with ordering), and the simplest example is the so-called bi-adjoint $\phi^3$ theory~\cite{Cachazo:2013iea}. This is a theory with scalars in the adjoint of two flavor groups, {\it e.g.} U(N) $\times$ U(N'), with a cubic vertex $\sim f^{abc} f^{a'b'c'} \phi_{a, a'} \phi_{b, b'} \phi_{c, c'}$. By doing trace-decomposition in both groups, the so-called double-partial amplitude, $m(\alpha|\beta)$ for orderings $\alpha$ and $\beta$, is given by the sum of scalar Feynman diagrams (cubic tree graphs with $n$ external legs) that are compatible with both $\alpha$ and $\beta$:
\be\label{mab}
m(\alpha|\beta)=(-1)^{{\rm flip}(\alpha|\beta)}\,\sum_{g \in T(\alpha) \cap T(\beta) } \prod_{I \in P(g)} \frac 1 {s_I}\,,
\ee
where $T(\alpha)$ denotes the set of cubic tree graphs compatible with ordering $\alpha$ (similarly for $T(\beta)$), and for each graph $g$ we have the product of $n{-}3$ propagators labelled by $I$ (the collection of all poles/propagators of a Feynman diagram $g$ is denoted as $P(g)$) \footnote{Here thanks to cyclicity symmetry, without loss of general, we can let $\alpha$ and $\beta$ share the same end point and then ${\rm flip}(\alpha|\beta)$ denotes the number of flipped adjacent pairs, {\it i.e.} $\beta(i{+}1)$ precedes $\beta(i)$ in the ordering $\alpha$, for $i=1,\ldots, n{-}1$, see \cite{Cachazo:2013iea,Mizera:2016jhj}. The sign has also been discussed in \cite{Mafra:2016ltu}.}.
Although this $\phi^3$ theory is simple, it is remarkable that $m(\alpha|\beta)$ is given by the simplest CHY formula, with two PT factors:
\be
m(\alpha|\beta)=\int \dif {\bm \mu}_n~{\rm \bf PT}(\alpha)~{\rm \bf PT}(\beta)\,,
\ee
which is a rather non-trivial mathematical identity first proposed and shown in~\cite{Cachazo:2013iea}. In particular, if we choose $\alpha=\beta$ the CHY formula can be viewed as a map from a half integrand, {\bf PT}($\alpha$),  to the collection of Feynman diagrams that are compatible with ordering $\alpha$, $T(\alpha)$ (all planar cubic tree graphs with external legs in the ordering $\alpha$):
\be\label{map}
{\rm \bf PT}~\to~{\rm planar~cubic~tree~graphs}:\quad \int \dif {\bm \mu}_n~{\rm \bf PT}(\alpha)^2=\sum_{g~{\rm compatible~with}~\alpha}^{{\rm Cat}_{n{-}2}} \prod_I \frac 1 {s_I}\,,
\ee
where ${\rm Cat}_{n{-}2}$ denotes the
 \href{https://en.wikipedia.org/wiki/Catalan_number}
{Catalan number}
 $1, 2, 5, 14, 42, 132, \ldots$ for $n=3,4,5,6,7,8, \ldots$
 ~\cite{catalan}.
In this paper, we will study a new class of half integrands, which largely generalize Parke-Taylor factors and maps to collections of Feynman diagrams. In addition, they naturally appear in superstring disk integrals and we will discuss their applications in that direction as well.



\subsection{Cayley functions and the map to cubic Feynman graphs}
 The main character of our story is a new class of half-integrands that we call { Cayley functions}. Before proceeding, let us discuss a convenient way of fixing ${\rm SL}(2,{\mathbb C})$ in CHY formulas.   Recall that we need to fix three punctures: we can always choose $\sigma_n\to \infty$, and fix any two more punctures at finite positions, {\it e.g.} $\sigma_1=0$, $\sigma_{n{-}1}=1$ which won't be necessary to explicitly write down. With $\s_n\rightarrow \infty$, ${\rm SL}(2,{\mathbb C})$-fixed CHY formula reads
\be
{\cal M}_n=\int \dif {\mu}_n~{\cal I}_n\,, \quad \dif {\mu}_n:=\prod_{a=2}^{n{-}1} \dif\sigma_a\,\delta(\sum_{b\neq a} \frac{k_a\cdot k_b}{\sigma_{a,b}})\,,
\ee
where the four infinite factors containing $\sigma_n$ in $\dif {\bm \mu}_n$ cancel against those in ${\bf I}_n$ thus we can remove all $\sigma_n$-dependence in ${\rm SL}(2,{\mathbb C})$-fixed $\dif { \mu}_n$ and ${\cal I}_n$. For example, we define the ${\rm SL}(2,{\mathbb C})$-fixed PT factor as (since there are $(n{-}1)!$ $\alpha$'s we can always choose $n$ in the end)
\be \label{PT}
{\rm PT}(\alpha(1), \cdots , \alpha(n{-}1), n)=\frac{1}{\sigma_{\alpha(1),\alpha(2)}\sigma_{\alpha(2),\alpha(3)}\cdot \cdot \cdot \sigma_{\alpha(n{-}2),\alpha(n{-}1)}}\,.
\ee
From now on, we will mostly be using this ${\rm SL}(2,{\mathbb C})$-fixed form of CHY formulas and integrands, and only switch back to the covariant (boldface) form when necessary.

\begin{figure}[!htb]
    \centering
    \def \layersep {.9cm}
    \subfloat{
    \begin{tikzpicture}[shorten >=0pt,draw=black,
        node distance = \layersep,
        neuron/.style = {circle, minimum size=3pt, inner sep=0pt,  fill=black } ]
     \node[neuron] (1) {};
     \node[ neuron,right of = 1] (2)  {};
   \node[ neuron,right of = 2] (3)  {};
        \draw (1) node[below=4pt]{$1$} --(2)node[below=4pt]{$2$}
        --(3)node[below=4pt]{$3$};
    \end{tikzpicture}
     } \subfloat{
   \tikz{\node {~~~};}
     } \subfloat{
    \begin{tikzpicture}[shorten >=0pt,draw=black,
        node distance = \layersep,
        neuron/.style = {circle, minimum size=3pt, inner sep=0pt,  fill=black } ]
     \node[neuron] (1) {};
     \node[ neuron,right of = 1] (2)  {};
   \node[ neuron,right of = 2] (3)  {};
        \draw (1) node[below=4pt]{$1$} --(2)node[below=4pt]{$3$}
        --(3)node[below=4pt]{$2$};
    \end{tikzpicture}
     } \subfloat{
   \tikz{\node {~~~};}
     } \subfloat{
 \begin{tikzpicture}[shorten >=0pt,draw=black,
        node distance = \layersep,
        neuron/.style = {circle, minimum size=3pt, inner sep=0pt,  fill=black } ]
     \node[neuron] (1) {};
     \node[ neuron,right of = 1] (2)  {};
   \node[ neuron,right of = 2] (3)  {};
        \draw (1) node[below=4pt]{$3$} --(2)node[below=4pt]{$1$}
        --(3)node[below=4pt]{$2$};
    \end{tikzpicture}}
               \caption{\label{sijk123} Cayley functions for $n=4$}
          \end{figure}
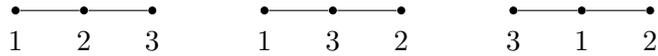

\begin{figure}[!htb]
    \centering
    \def \layersep {.9cm}
    \subfloat{
    \begin{tikzpicture}[shorten >=0pt,draw=black,
        node distance = \layersep,
        neuron/.style = {circle, minimum size=3pt, inner sep=0pt,  fill=black } ]
     \node[neuron] (1) {};
     \node[ neuron,right of = 1] (2)  {};
   \node[ neuron,right of = 2] (3)  {};
     \node[ neuron,right of = 3] (4)  {};
        \draw (1)--(4);
    \end{tikzpicture}
     } \subfloat{
   \tikz{\node {~~~};}
     } \subfloat{
    \begin{tikzpicture}[shorten >=0pt,draw=black,
        node distance = \layersep,
        neuron/.style = {circle, minimum size=3pt, inner sep=0pt,  fill=black } ]
     \node[neuron] (1) {};
     \node[ neuron,right of = 1] (2)  {};
   \node[ neuron,right of = 2] (3)  {};
     \node[ neuron,above of = 2] (4)  {};
        \draw (1)         --(3);
             \draw (2)         --(4);
    \end{tikzpicture}
}
          \caption{\label{sijkl123}  Two topologies for $n=5$ }
          \end{figure}
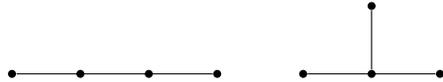

Now we can define our new half integrands in this $\sigma_n\to \infty$ form. Given any labelled tree graph with with points $1,2,\ldots, n{-}1$ (also called $(n{-}1)$-pt Cayley tree graph), we define Calyley function as the product of $n{-}2$ $\frac{1}{\s_i-\s_j}$, one for each edge $(i,j)$ of the tree $(1\leq i<j\leq n-1)$ \footnote{Note that a Cayley function is only defined for a {\it oriented} tree graph, since we assign $\frac 1 {\sigma_{i,j}}$ but not $\frac 1 {\sigma_{j,i}}$ for a directed edge (i,j). However, the difference is only an overall sign, and our convention is that if there is no arrow we simply choose $\frac 1 {\sigma_{i,j}}$ for $i<j$ . We will see that in certain cases it is convenient to rearrange orientations of edges, and there is a sign $(-1)^{r_{\rm flip}}$ where $r_{\rm flip}$ is the number of edges with flipped orientation.}.
\begin{equation}\label{C}
C_n(\{i,j\}):=\prod_{a=1}^{n{-}2} \frac 1 {\sigma_{i_a, j_a}}\,,
\end{equation}
where equivalently we can say that no cycle is formed with these $n{-}2$ pairs $\{i,j\}$. For example, for $n=4$ we can have the following three labelled trees , see figure \ref{sijk123} ,where {\it e.g.} Cayley function for the first one is $C_4(\{1,2\}, \{2,3\})=\frac 1 {\sigma_{12} \sigma_{23}}$.
 For $n=5$ there are 16 labelled trees, and here we give two examples of $C_5$ for the two topologies , see figure \ref{sijkl123}
\be
C_5 (\{1,2\}, \{2,3\}, \{3,4\})=\frac 1{\sigma_{12} \sigma_{23} \sigma_{34}}\,, \quad C_5(\{1,2\}, \{1,3\}, \{1,4\})=\frac 1 {\sigma_{12}\sigma_{13} \sigma_{14}}\,.
\ee

Since Cayley functions in this $\sigma_n \to \infty$ frame are in one-to-one correspondence with $(n{-}1)$-pt labelled trees, there are exactly $(n{-}1)^{n{-}3}$ of them. A basic question we are interested in is how many different classes of Cayley functions there are; two Cayley functions are said to be topologically equivalent if and only if their tree graphs can be brought to the same shape, which are just relabelling of each other, for example the there are two classes of Cayley functions for $n=5$.  Generally for any $n$, there are always these two extreme classes, given by the so-called Hamiltonian graph (a line) and star graph (a star-shaped tree, and we can choose the center to be {\it e.g.} 1) respectively (see figure \ref{S graph})
\be
C^{\rm H}_n
=\prod_{a=1}^{n-2} \frac 1 {\sigma_{\alpha(a), \alpha(a{+}1)}}\,, \quad C^{\rm S}_n
=\prod_{a=2}^{n-1} \frac 1 {\sigma_{1, a}}\,.
\ee
The former is nothing but the Parke-Taylor factor ${\rm PT}(\alpha(1), \alpha(2), \cdots, \alpha(n{-}1), n)$ in the ${\rm SL}(2, {\mathbb C})$-fixed form, \eqref{PT}, while the latter is totally symmetric in labels $2,3,\ldots, n{-}1$. \begin{figure}[!htb]
    \centering
    \subfloat[\label{Hamiltonian graph}Hamiltonian graph]{
   \begin{tikzpicture}[scale=.7]
    \draw (0,-1)--(1.5,-1)--(3,-1)--(4.5,-1);
  \draw (6.5,-1)--(8,-1);
    \fill (0,-1) circle (.1);
      \fill (1.5,-1) circle (.1);
        \fill  (4.5,-1) circle (.1);
          \fill (3,-1) circle (.1);
         \fill  (6.5,-1) circle (.1);
           \fill  (8,-1) circle (.1);
            \fill  (5,-1) circle (.04);
             \fill  (5.5,-1) circle (.04);
              \fill  (6,-1) circle (.04);
              \node at (3,-2) {~};
                 \node at (10,-2) {~};
                    \node at (-2,-2) {~};
    \end{tikzpicture}
           }             \subfloat[\label{Star graph}Star graph]{
   \begin{tikzpicture}[scale=.7]
   \draw (-1.5,0)--(0,0);
   \draw (0,1.5)--(0,-1.5);
      \draw (1,1)--(-1,-1);
       \draw (1,-1)--(-1,1);
      \fill (1.3,0.3)  circle (.04);
          \fill (1.4,0) circle (.04);
          \fill (1.3,-0.3) circle (.04);
           \fill (-1.5,0)  circle (.1);
          \fill (0,0) circle (.1);
          \fill (0,1.5) circle (.1);
           \fill (0,-1.5)  circle (.1);
          \fill (1,1) circle (.1);
          \fill (-1,-1) circle (.1);
           \fill (-1,1) circle (.1);
          \fill (1,-1) circle (.1);
       \end{tikzpicture}
}
\caption{\label{S graph}}
\end{figure}

For $n\geq 6$, we have new classes that are in between these two extremes, see examples in sec.\ref{sec2}. The number of distinct classes is nothing but the number of unlabelled trees, which equals $1,2,3,6,11,23,47,\cdots$ for $n{=}4,5,6,7,8,9,10,\cdots$ respectively (see
 \href{https://oeis.org/search?q=1%2C2%2C3%2C6%2C11%2C23%2C47%2C106
&sort=&language=english&go=Search}
{A000055}
~\cite{A000055}
)
. We will see that these classes play an important role in our following discussions.

Although we have defined Cayley functions in $\sigma_n\to \infty$ frame, it is straightforward to recover the ${\rm SL}(2,{\mathbb C})$ redundancy which results in the unique covariant form for each $C$:
\be
{\bf C}_n(\{i,j\}):=C_n(\{i,j\})~\sigma_{1,n}^{v_1-2}\,\sigma_{2,n}^{v_2-2}\,\cdots\,\sigma_{n{-}1,n}^{v_{n{-}1}-2}\,,
\ee
where $v_a$ denotes the valency of vertex $a=1,2,\cdots, n{-}1$ in the labelled tree, and by definition, $1\leq v_a \leq n{-}3$. It is easy to check that in ${\bf C}$ every $\sigma_a$ appears exactly twice in the denominator,
including $\sigma_n$; this is because $\sum_{a=1}^{n-1} (v_a-2)=2(n-2)-2(n-1)=-2$, thus ${\bf C}_n$ given above
 is indeed ${\rm SL}(2,{\mathbb C})$ covariant with weight 2. The covariant form of $C^H_n={\rm PT}_n$ is of course ${\bf PT}_n$ in \eqref{pt}, and that for $C^S_n$ is given by
\be\label{cccova}
{\bf C}^S_n=\prod_{a=2}^{n{-}1} \frac 1 {\sigma_{1,a}}~\sigma_{1,n}^{n{-}4}\,\prod_{a=2}^{n{-}1} \sigma_{a,n}^{-1} =\frac {\sigma_{1,n}^{n{-}4}}{\sigma_{1,2}\cdots \sigma_{1,n{-}1} \sigma_{2,n} \cdots \sigma_{n{-}1, n}}\,.
\ee

The Cayley functions generalize Parke-Taylor factors in an interesting way; it is well known that via partial-fraction identities,   they can be redued to PT's (see sec \ref{sec3p1}), but we find it intriguing and useful to study these functions directly, in the context of CHY formula and string integrals.
As we will see shortly, Cayley functions have the property that, similar to the case of PT factors \eqref{map}, via CHY formula it maps to a sum of certain cubic tree graphs with coefficient $+1$ only :
\be\label{cubicsum}
C_n(\{i,j\})~\to~{\rm cubic~tree~graphs}: \int \dif { \mu}_n~C_n^2(\{i,j\})=\sum_{g~{\rm ``compatible~with"}~\{i,j\}} \prod_I \frac 1 {s_I}\,.
\ee
As we will explain in sec 2, we say a Feynman diagram,$ g$, is ``compatible with'' $\{i,j\}$ if and only if the $n-3$ poles of $g$ correspond to $n-3$ mutually compatible connected subgraphs of the labelled tree $\{i,j\}$. We summarize this result as a theorem to be shown in sec \ref{sec2}:
 \begin{theorem}\label{identicalCc}
     \ba\label{identicalCc22}\boxed{
 \int \dif \mu_n C_n^2(\{i,j\})=\sum_{\substack{
  I_1,I_2,\cdots,I_{n-3} \text{~{\rm are}}\\\text{ ~{\rm compatible}~{\rm connected}~{\rm subgraphs} }}}\frac{1}{ s_{I_1}s_{I_2}\cdots s_{I_{n-3}}}\,.}
   \ea
    \end{theorem}

Our study of Cayley functions has been motivated by~\cite{Arkani-Hamedsonghe}, where the ``pushforward" of differential forms on ${\cal M}_{0,n}$ to Mandelstam space ({\it i.e.}  space spanned by some independent Mandelstam variables) has been considered. As explained in~\cite{Arkani-Hamedsonghe}, the pushforward of a half-integrand, which is a differential form in Mandelstam space , contains all the information of CHY formula of its square; in some sense, the discussions here are like the {\it combinatoric} version of the geometric story in~\cite{Arkani-Hamedsonghe} (the idea of studying the combinatorics of ``polytopes of Feynman diagrams" has been considered in \cite{JJCarrascoaafe} and also see \cite{GraphAssociahedra,ConciniCProcesi,AlexanderPostnikov,AlexandePostnikovVictorReiner} for some previous discussion  about graph associahedra , generalized permutohedra and so on).


\section{A map from Cayley functions to polytopes of Feynman diagrams}\label{sec2}

An important property of Cayley functions is that we can directly read off the pole structures and consequently the sum of Feynman diagrams of  their CHY formulas. We will see that the result provides an interesting map from any labelled tree to a polytope whose vertices correspond to Feynman diagrams. Note that these polytopes are only combinatoric, while in~\cite{Arkani-Hamedsonghe} one can actually construct polytopes, {\it e.g.} the associahedron in Mandelstam space, whose canonical form (also see \cite{Arkani-Hamed:2017tmz} for definition) turns out to be the pushforward of the canonical form of ${\cal M}_{0,n}$ (also an associahedron).

\subsection{CHY formulas for Cayley functions squared}\label{identicalCcc}

Here we explain  Theorem \ref{identicalCc} in two steps. First we show what poles are allowed on the RHS of Theorem  \ref{identicalCc}, and then
provide a way to obtain all Feynman diagrams  recursively .
Then we give a classification of Cayley functions and also present detailed examples, for the two extreme cases, $C_n^H$ and $C_n^S$,  as well as some other cases.

\subsubsection*{The pole structure}

The first result we present is the construction for the set of allowed poles, for CHY formula of $C_n^2$ , which we will denote as $P(C_n)$. Given the labelled tree, any of its non-trivial {\it connected subgraph} corresponds to a pole on the RHS of \eqref{identicalCc22}:

  \ba\label{pc}
P(C_n)=\Big\{s_{i_1,i_2,...,i_m}\Big|
\raisebox{-.5cm}{
\tikz[node distance =.5cm]{
\node (a) {there is a connected subgraph in the labelled tree of $C_n$   };
\node [below of=a]{whose vertices are $\{i_1,i_2,...,i_m\}$ for $m=2,3,\cdots,n-2$};
}}\,.
 \Big\}\,
 \ea
 This rule for the poles is  very intuitive, and it follows from the general analysis of pole structures of CHY formulas~\cite{Cachazo:2013gna,Baadsgaard:2015voa,Cachazo:2015nwa,Dolan:2013isa,Broedel:2013tta} (for $C_n^2$ we only have simple poles, see \cite{Huang:2016zzb,Cardona:2016gon} for discussions on higher-order poles). A connected subgraph with vertices $i_1, i_2, ...,i_m$ means that there are exactly $2(m-1)$ $\sigma_{i,j}$ with $i,j\in {i_1,...,i_m}$ in $C_n^2$, thus it will produce
 a pole $s_{i_1,i_2,...,i_m}$  according to the rule described in \cite{Baadsgaard:2015voa}. Note that we don't have $n$ contained in any subgraph as any pole containing $n$ can be expressed by its complement.

 Here we spell out some examples. For two-particle pole,
 $s_{i,j}\in P(C_n)$ iff
      \raisebox{-.6cm}{\begin{tikzpicture}[shorten >=0pt,draw=black,
        node distance = .55cm,
        neuron/.style = {circle, minimum size=3pt, inner sep=0pt,  fill=black } ]
     \node[neuron] (1) {};
     \node[ neuron,right of = 1] (2)  {};
        \draw (1) node[below=4pt]{$i$} --(2)node[below=4pt]{$j$}
       ;
    \end{tikzpicture}}
    is a edge in the labelled tree.
  For three particle pole,
  $s_{i,j,k}\in P(C_n)$ iff one of the graphs in figure \ref{sijk} exists in the labelled tree.

 \begin{figure}[!htb]
    \centering
    \def \layersep {.9cm}
    \subfloat{
    \begin{tikzpicture}[shorten >=0pt,draw=black,
        node distance = \layersep,
        neuron/.style = {circle, minimum size=3pt, inner sep=0pt,  fill=black } ]
     \node[neuron] (1) {};
     \node[ neuron,right of = 1] (2)  {};
   \node[ neuron,right of = 2] (3)  {};
        \draw (1) node[below=4pt]{$i$} --(2)node[below=4pt]{$j$}
        --(3)node[below=4pt]{$k$};
    \end{tikzpicture}
     } \subfloat{
   \tikz{\node {~~~};}
     } \subfloat{
    \begin{tikzpicture}[shorten >=0pt,draw=black,
        node distance = \layersep,
        neuron/.style = {circle, minimum size=3pt, inner sep=0pt,  fill=black } ]
     \node[neuron] (1) {};
     \node[ neuron,right of = 1] (2)  {};
   \node[ neuron,right of = 2] (3)  {};
        \draw (1) node[below=4pt]{$i$} --(2)node[below=4pt]{$k$}
        --(3)node[below=4pt]{$j$};
    \end{tikzpicture}
     } \subfloat{
   \tikz{\node {~~~};}
     } \subfloat{
 \begin{tikzpicture}[shorten >=0pt,draw=black,
        node distance = \layersep,
        neuron/.style = {circle, minimum size=3pt, inner sep=0pt,  fill=black } ]
     \node[neuron] (1) {};
     \node[ neuron,right of = 1] (2)  {};
   \node[ neuron,right of = 2] (3)  {};
        \draw (1) node[below=4pt]{$k$} --(2)node[below=4pt]{$i$}
        --(3)node[below=4pt]{$j$};
    \end{tikzpicture}}
          \caption{\label{sijk} $s_{i,j,k}$}
          \end{figure}

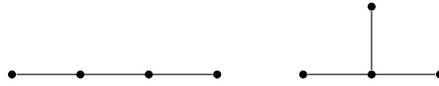
\begin{figure}[!htb]
    \centering
    \def \layersep {.9cm}
    \subfloat{
    \begin{tikzpicture}[shorten >=0pt,draw=black,
        node distance = \layersep,
        neuron/.style = {circle, minimum size=3pt, inner sep=0pt,  fill=black } ]
     \node[neuron] (1) {};
     \node[ neuron,right of = 1] (2)  {};
   \node[ neuron,right of = 2] (3)  {};
     \node[ neuron,right of = 3] (4)  {};
        \draw (1)--(4);
    \end{tikzpicture}
     } \subfloat{
   \tikz{\node {~~~};}
     } \subfloat{
    \begin{tikzpicture}[shorten >=0pt,draw=black,
        node distance = \layersep,
        neuron/.style = {circle, minimum size=3pt, inner sep=0pt,  fill=black } ]
     \node[neuron] (1) {};
     \node[ neuron,right of = 1] (2)  {};
   \node[ neuron,right of = 2] (3)  {};
     \node[ neuron,above of = 2] (4)  {};
        \draw (1)         --(3);
             \draw (2)         --(4);
    \end{tikzpicture}
}
     \caption{\label{sijkl}
        two   topologies of subgraphs
      for $s_{i,j,k,l}$}
          \end{figure}

   For $m>3$, there are more topologies of subgraphs. For example, for $s_{i,j,k,l}$ there are two different  topologies of subgraphs , see  figure \ref{sijkl}.

With the help of labelled tree, the relation of poles are also intuitive.
Two poles {\it i.e.} two connected subgraphs , are compatible iff the particle set of one is a subset of that of the other, see figure \ref{compatible subgraph} (1), or they have no intersection,see figure \ref{compatible subgraph} (2). Two incompatible poles, see figure \ref{incompatible subgraph} can't both appear in a Feynman diagram.

\def \layersep {.9cm}
         \begin{figure}[!htb]
    \centering
    \subfloat[{\label{compatible subgraph}compatible subgraphs}]{
    \tikz{\node {
    \begin{tikzpicture}[shorten >=0pt,draw=black,
        node distance = \layersep,
        neuron/.style = {circle, minimum size=3pt, inner sep=0pt,  fill=black } ]
     \node[neuron] (1) {};
     \node[ neuron,right of = 1] (2)  {};
   \node[ neuron,right of = 2] (3)  {};
     \node[ neuron,right of = 3] (4)  {};
     \node[ neuron,above of = 2] (5)  {};
        \node[ right of = 4] (6)  {,};
        \node at  ($(2)+(0.5,-.6)$) {(1)};
             \draw[line width=.25mm] (1)--(4);
        \draw[line width=.25mm] (2)--(5);
     \draw[blue,line width=.35mm] ($(1)+(0,-0.07)$)--($(3)+(0,-0.07)$);
        \draw[blue,line width=.35mm] ($(2)+(0.07,0)$)--($(5)+(0.07,0)$);
    \end{tikzpicture}
    }node[right =70pt,yshift=-9pt]{
       \begin{tikzpicture}[shorten >=0pt,draw=black,
        node distance = \layersep,
        neuron/.style = {circle, minimum size=3pt, inner sep=0pt,  fill=black } ]
     \node[neuron] (1) {};
     \node[ neuron,right of = 1] (2)  {};
   \node[ neuron,right of = 2] (3)  {};
      \node[ neuron,left of = 1] (4)  {};
          \node[ neuron,left of = 4] (5)  {};
            \node[ right of = 3] (6)  {~};
               \node at  ($(1)+(0,-.8)$) {(2)};
             \draw[line width=.25mm] (1)--(3);
        \draw[blue,line width=.35mm] (4)--(5);
         \end{tikzpicture}
};}}
 \subfloat[{\label{incompatible subgraph}incompatible subgraphs}]{
    \begin{tikzpicture}[shorten >=0pt,draw=black,
        node distance = \layersep,
        neuron/.style = {circle, minimum size=3pt, inner sep=0pt,  fill=black } ]
     \node[neuron] (1) {};
     \node[ neuron,right of = 1] (2)  {};
   \node[ neuron,right of = 2] (3)  {};
     \node[ neuron,above of = 2] (4)  {};
        \node[right of = 3]   {~};
         \node[left of = 1]   {~};
             \draw[line width=.25mm] (1)--(3);
     \draw[blue,line width=.35mm] ($(1)+(0,0.055)$)--($(2)+(0,0.055)$);
        \draw[blue,line width=.35mm] (2)--(4);
    \end{tikzpicture}
   }
   \caption{}
    \end{figure}
We say a set of poles are compatible iff any two of them are compatible.


\subsubsection*{Feynman diagrams from poles}

Any $n-3$ compatible poles from $P(C_n)$ should correspond to a Feynman diagram on RHS of \eqref{identicalCc22}. In this subsubsection, we provide a clever way to obtain all Feynman diagrams from those of lowers points recursively, which is  much more efficient than  to find all $n-3$ compatible pole sets by brute force in higher points.

The starting point is that any cubic Feynman diagram has two parts whose external particles are $I,{\bar I}$ respectively which join a vertex with $n$, in the way shown in figure \ref{factozation2}

According to \eqref{pc}, $s_I,s_{\bar I}$ corresponding to two connected subgraphs which together make up the whole labelled tree up to a link edge, see figure \ref{factozation}.
 \def\layersep{1cm}
 \begin{figure}[!htb]
 \centering
 \subfloat[\label{factozation}]{

}
\caption{}
 \end{figure}
Reversely, each edge of the labelled tree can be a linking edge, which corresponds to different Feynman diagram sets.
Using factorization, as shown in Appendix \ref{appa}  , they together make up all Feynman diagrams on the RHS of Theorem \ref{identicalCc} with coefficient $+1$,
\def\layersep{.8cm}
\ba\label{recursiongraph}
\raisebox{-.5cm}{
%
}
+\cdots\,.
\ea
While the two parts in figure \ref{factozation2} themselves are also Feynman diagrams , whose corresponding labelled trees are $C^1,C^2$ in  figure \ref{factozation}. That means we can obtain Feynman diagrams from those of lower points.
Let's give the recursion explicitly
\ba\label{recursion2}
T(C)= \bigsqcup_{ C^1\sqcup C^2 \sqcup - = C} \Big\{
  \def\layersep{.8cm}
  \raisebox{-.7cm}{
%
}
\Big|c^1\in T(C^1),c^2\in T(C^2)
\Big\}\,.
\ea
Here we use the symbol $\bigsqcup$ as disjoint union.

A direct  consequence is the recursion about the number of Feynman diagrams ,
 \ba\label{recursion22}
|T(C)|= \sum_{ C^1\sqcup C^2 \sqcup - = C} |T(C^1)||T(C^2)|\,.
\ea

Let's spell out some examples.

 \def\layersep{.4cm}
 \ba\label{example4pt}
 T(
 \raisebox{-.55cm}{
%
  }
     \Big\}\,.
 \ea
 5 planar Feynman diagrams with ordering $1,2,3,4,5$ as expected.

Similarly for $C_n^{S}(1)$, according to
 \def\layersep{.6cm}
 \ba
 \raisebox{-.55cm}{
 %
  }
    \Big\}
    \,,
\ea
which are 6 multi-peripheral Feynman diagrams with the permutations of $2,3,4$.

A more non-trivial example for $n=6$.
According to
\def\layersep{.6cm}
\ba
 \raisebox{-.55cm}{
%
}    \,,
\ea
we have the 18 Feynman diagrams
\ba\label{example6pt}
T(\raisebox{-.55cm}{
%
  }
    \Big\}
\,.
\ea
Here ${\color{blue}\big(}{\color{blue}\text{{ 1}}}\leftrightarrow {\color{blue}\text{{ 3}}}{\color{blue}\big)}$ means  five more Feynman diagrams owing to the symmetry of $1$ and $3$ in
the  labelled tree.

\subsubsection*{Hamiltonian graph , Star graph and beyond}

Above we have seen  examples \eqref{example4pt},\eqref{example5pt} for Hamiltonian graph,
which is the so-called Parke-Taylor graph.
All connected line segment except the labelled tree itself correspond to a pole
and all compatible $n-3$ connected line segment from these
correspond to a pole  Feynman diagrams.
  The exact pole sets and Feynman diagram sets are given by
\ba\label{p of H}
  P(\rm{PT}(1,2,...,n))&=&\{s_{i,i+1,...,j}| 1\leq i<j \leq n-1,\text{except} ~(i,j)=(1,n-1)\}\,\\
  T(\rm{PT}(1,2,...,n))&=&\{\text{all  planar Feynman diagrams with ordering $(1,2,...,n)$}\}\,.
 \ea
The number of poles are $\frac{n(n-3)}{2}$.
The number of Feynman diagrams  ${\rm Cat}_{n-2}$ can be seen from the recursion \eqref{recursion22} which gives the recursion of Catalan numbers directly.

 The
${\rm PT}(1,2,3,4)$
 in \eqref{example4pt} can  also  be seen as $C^S_4(2)$,
  \def\layersep{.4cm}
 \ba\label{example4pt22}
 T(
 \raisebox{-.55cm}{
\begin{tikzpicture}[shorten >=0pt,draw=black,
        node distance = \layersep,
        neuron/.style = {circle, minimum size=3pt, inner sep=0pt,  fill=black } ]

     \node[neuron] (1) {};
     \node[ neuron,right of = 1] (2)  {};
   \node[ neuron,right of = 2] (3)  {};
            \draw (1) node[below=4pt]{$1$} --(2)node[below=4pt]{$2$}
        --(3)node[below=4pt]{$3$};
    \end{tikzpicture}}
 )
 =
 \Big\{
   \raisebox{-1cm}{
 \begin{tikzpicture}[shorten >=0pt,draw=black,scale=.25,
        node distance = .2cm,
        neuron3/.style = {circle, minimum size=.1pt, inner sep=0pt,  fill=black } ]
     \node[neuron3] {}
     child {node[neuron3] {}node[below=0pt]{$1$}}
        child {node[neuron3] {}
        child {node[neuron3] {} node[below=0pt]{$3$}}
           child {node[neuron3] {}node[below=0pt]{$2$}}}
           ;
     \draw (0,0)--(.5,1) node[right=0pt]{$n$};
    \end{tikzpicture}  }
    ,
 \raisebox{-1cm}{
 \begin{tikzpicture}[shorten >=0pt,draw=black,scale=.25,
        node distance = .2cm,
        neuron3/.style = {circle, minimum size=.1pt, inner sep=0pt,  fill=black } ]
     \node[neuron3] {}
        child {node[neuron3] {}
        child {node[neuron3] {} node[below=0pt]{$2$}}
           child {node[neuron3] {}node[below=0pt]{$1$}}}
           child {node[neuron3] {}node[below=0pt]{$3$}};
     \draw (0,0)--(.5,1) node[right=0pt]{$n$};
    \end{tikzpicture}  }
    \Big\}\,,
 \ea
which are two multi-peripheral Feynman diagrams with the permutations of $1,3$. Another example  for star graph is given in  \eqref{example5ptstar} . For star graph,
 all particles   except the center one and $n$ are end points, which are symmetric.
Each of them corresponds to a line from center
  point to  it. Any nontrivial subset of these lines must
   make up a connected subgraph which corresponds to a pole. So there are $2^{n-2}-2$ poles
   in star graph.  What's more,  in star graph, any two connected
    subgraphs are compatible if and only if one is contained
    in the other. So start from two-particle pole,
     the next subgraph is compatible to the former if and only
      if it contains the former. So there are $(n-2)!$ sets of
       $n-3$ compatible poles, any of which corresponds to a
       multi-peripheral Feynman diagram .    The exact pole sets and Feynman diagram sets are given by
  \ba\label{p of S}
  P(C^{\rm{S}}_n(1))&=&\{s_{1,i_1,\cdots,i_m}|{i_1,\cdots,i_m}\in \{2,\cdots,n-1\},1\leq m\leq n-3  \}\,,\nl
  T(C^{\rm{S}}_n(1))&=&\{
\raisebox{-.5cm}{
 \begin{tikzpicture}
    \draw (0,0)--(4,0);
    \draw (1,0)--(1,1);
    \draw (3,0)--(3,1);
    \draw (1.5,0)--(1.5,1);
    \fill (2,.7) circle (.02);
    \fill (2.3,.7) circle (.02);
  \fill (2.6,.7) circle (.02);
    \node at (4.2,0) {$n$};
    \node at (-.2,0) {$1$};
        \node at (1,1.2) {$\rho_2$};
    \node at (1.5,1.2) {$\rho_3$};
   \node at (3,1.2) {$\rho_{n-1}$};
    \end{tikzpicture}
    }
    |\rho\in {\rm perms.~of ~} 2,3,\cdots,n-1
\}
\,.
 \ea

Starting from $n=6$,   Cayley functions of new kind comes out, see  \eqref{example6pt}. We can also extend it to general $n$,  next-to-Hamiltonian graph seen in figure \ref{nH graph} and next-to-Star graph seen in figure \ref{nS graph}.
  \begin{figure}[!htb]
    \centering
    \subfloat[\label{nH graph}Next-to-Hamiltonian graph]{
   \begin{tikzpicture}[scale=.7]
    \draw (0,-1)--(1.5,-1)--(3,-1)--(4.5,-1);
  \draw (6.5,-1)--(8,-1);
  \draw (1.5,-1)--(1.5,0.5);
    \fill (0,-1) circle (.1);
      \fill (1.5,-1) circle (.1);
        \fill  (4.5,-1) circle (.1);
          \fill (3,-1) circle (.1);
         \fill  (6.5,-1) circle (.1);
           \fill  (8,-1) circle (.1);
            \fill  (5,-1) circle (.04);
             \fill  (5.5,-1) circle (.04);
              \fill  (6,-1) circle (.04);
               \fill  (1.5,0.5) circle (.1);
               \node at (1,-1.5) {~};
               \node at (10,-1.5) {~};
               \node at (-2,-1.5) {~};
    \end{tikzpicture}
          }     \subfloat[\label{nS graph}Next-to-Star graph]{
   \begin{tikzpicture}[scale=.7]
   \draw(-3,0)--(-1.5,0)--(0,0);
   \draw (0,1.5)--(0,-1.5);
      \draw (1,1)--(-1,-1);
       \draw (1,-1)--(-1,1);
      \fill (1.3,0.3)  circle (.04);
          \fill (1.4,0) circle (.04);
          \fill (1.3,-0.3) circle (.04);
           \fill (-1.5,0)  circle (.1);
          \fill (0,0) circle (.1);
          \fill (0,1.5) circle (.1);
           \fill (0,-1.5)  circle (.1);
            \fill (-3,0)  circle (.1);
          \fill (1,1) circle (.1);
          \fill (-1,-1) circle (.1);
           \fill (-1,1) circle (.1);
          \fill (1,-1) circle (.1);

    \end{tikzpicture}
}
\caption{\label{nextnext}}
\end{figure}
With more patience, we can also list out their pole sets and Feynman diagram sets. Here we just show a nontrivial use of the recursion \eqref{recursion22} about $T(C_n)$ and
get a feeling about the value of
Parke-Taylors and star graph in the
analyzing a  more complicated Cayley function.
For next-to-Hamiltonian graph,
\ba
|T(C^{nH}_n)|= {\rm Cat}_{n-3}+{\rm Cat}_{n-3} +\sum_{r=0}^{n-5} |T(C^{nH}_{4+r})| {\rm Cat}_{n-5-r}\,,
\ea
where we have used $|T(C^{H}_n)|={\rm Cat}_{n-2}$.
So we obtain
\ba
|T(C^{nH}_n)|=\frac{6 (n-3) {\rm Cat}_{n-3}}{n-1}\,.
\ea
The counting of the number of poles is simple.  Using the  additional edge, we have $(n-3)\times 2^1-1$ more poles.
Then
\ba
|P(C^{nH}_n)|=
\frac{(n-1)(n-4)}{2}+1+(n-3)\times 2^1-1
=\frac{n(n-1)}{2}-4\,.
\ea
Similarly,
\ba
|T(C^{nS}_{n})|
=\frac{n (n-3)!}{2}
,\qquad
|P(C^{nS}_{n})|
=3\times 2^{n-4}-1\,.
\ea

More complicated Cayley functions can be analyzed with the help of simpler ones.
 Different kinds of Cayley functions have
  different pole structures $P(C_n)$ and Feynman diagram structures $T(C_n)$.
For $n\geq 6$, there are too many poles or Feynman diagrams for any Cayley function to put them here. We just make a list showing the number of poles and Feynman diagrams below.
\begin{table}[!htb]
\centering
\begin{tabular}{l|c|c}
  $n$& $|T(C_n)|$& $|P(C_n)|$  \\ \hline
4 & 2 & 2\\
5 & 5,6 & 5,6 \\
6  & 14,18,24 & 9,11,14\\
7        & 42,56,60,76,84,120  & 14,17,18,21,23,30 \\
8       &
132,180,200,222,248,280,288,324,408,480,720& 20,24,26,28,29,32,33,36,41,47,62\\
9 &
\raisebox{-.3cm}{
\tikz[node distance =.3cm]{
\node (a) {\scriptsize{429,\!\! 594,\!\! 675,\!\! 700,\!\! 794,\!\! 828,\!\! 950,\!\! 990,\!\! 1000,\!\! 1105,\!1144,\!\! 1188,\!\! ~~~~~~ }~~~~~~~};
\node [below of=a]{~~~~\scriptsize { 1374,\!1404,\!\! 1440,\!\! 1650,\!\! 1728,\!\! 1800,\!\! 2100,\!\! 2484,\!\! 2640,\!\!
3240,\!\! 5040}};
}}
&\raisebox{-.3cm}{
\tikz[node distance =.3cm]{
\node (a) {\scriptsize{27,\!\! 32,\!\! 35,\!\! 36,\!\! 39,\!\! 38,\!\! 42,\!\! {\color{red}44},\!\! {\color{red}44},\!\! 47,\!\! 48,\!\! 50 }~~~~~~~~~~~};
\node [below of=a]{~~~~~~~~~\scriptsize {55,\!\! 53,\!\! 54,\!\! 60,\!\! 62,\!\! 65,\!\! 72,\!\! 77,\!\! 81,\!\! 95,\!\! 126}};
}}
\end{tabular}
\caption{\label{tab:widgets}  $|T(C_n)|$ and $|P(C_n)|$ for $n\leq 9$ }
\end{table}
The lists are always like
 $|T(C_n^H)|,|T(C_n^{nH})|,\cdots,|T(C_{n}^{nS})|,|T(C_n^S)|$(  $|P(C_n^H)|,|P(C_n^{nH})|,\cdots,|P(C_{n}^{nS})|,|P(C_n^S)|$ ).
   Note that     $|T(C_n)|,|P(C_n)|$ are only rough descriptions of Cayley functions. Two Cayley functions from
 two kinds
 may share the same $|P(C_n)|$ or $|T(C_n)|$ , seen in figure \ref{samenumber}.
 \begin{figure}[!htb]
\centering
\def\layersep{.55cm}
\subfloat[$|T|\!\!=\!\!990,|P|\!\!=\!\!{\color{red}44}$ ]{
\begin{tikzpicture}[shorten >=0pt,draw=black,
        node distance = \layersep,
        neuron/.style = {circle, minimum size=3pt, inner sep=0pt,  fill=black } ]

     \node[neuron] (1) {};
       \node[ neuron,right of = 1] (2)  {};
       \node[ neuron,above of = 2] (3)  {};
   \node[ neuron,right of = 2] (4)  {};
      \node[ neuron,right of = 4] (5)  {};
       \node[ neuron,right of = 5] (6)  {};
         \node[ neuron,right of =6] (7)  {};
           \node[ neuron,below of =2] (8)  {};

        \draw (1) --(7);
        \draw (3)--(8);
    \end{tikzpicture}
}\subfloat[$|T|\!\!=\!\!1000,|P|\!\!=\!\!{\color{red}44}$ ]{
\begin{tikzpicture}[shorten >=0pt,draw=black,
        node distance = \layersep,
        neuron/.style = {circle, minimum size=3pt, inner sep=0pt,  fill=black } ]

     \node[neuron] (1) {};
       \node[ neuron,right of = 1] (2)  {};
       \node[ neuron,above of = 2] (3)  {};
   \node[ neuron,right of = 2] (4)  {};
      \node[ neuron,right of = 4] (5)  {};
       \node[ neuron,right of = 5] (6)  {};
         \node[ neuron,right of =6] (7)  {};
           \node[ neuron,below of =4] (8)  {};

        \draw (1) --(7);
        \draw (3)--(2);
         \draw (4)--(8);
    \end{tikzpicture}
}\subfloat{
\begin{tikzpicture}
\node {~~~~~};
\end{tikzpicture}
}\subfloat[$|T|={\color{red}17160},|P|=80$ ]{
\begin{tikzpicture}[shorten >=0pt,draw=black,
        node distance = \layersep,
        neuron/.style = {circle, minimum size=3pt, inner sep=0pt,  fill=black } ]

     \node[neuron] (1) {};
       \node[ neuron,right of = 1] (2)  {};
       \node[ neuron,above of = 2] (3)  {};
   \node[ neuron,right of = 2] (4)  {};
      \node[ neuron,right of = 4] (5)  {};
       \node[ neuron,right of = 5] (6)  {};
         \node[ neuron,right of =6] (7)  {};
           \node[ neuron,below of =2] (8)  {};
 \node[ neuron,right of =7] (10)  {};
  \node[ neuron,above of =7] (9)  {};
        \draw (1) --(10);
        \draw (3)--(8);
          \draw (7)--(9);
    \end{tikzpicture}
    }\subfloat[$|T|={\color{red}17160},|P|=87$ ]{
\begin{tikzpicture}[shorten >=0pt,draw=black,
        node distance = \layersep,
        neuron/.style = {circle, minimum size=3pt, inner sep=0pt,  fill=black } ]

     \node[neuron] (1) {};
       \node[ neuron,right of = 1] (2)  {};
       \node[ neuron,above of = 2] (3)  {};
   \node[ neuron,right of = 2] (4)  {};
      \node[ neuron,right of = 4] (5)  {};
       \node[ neuron,right of = 5] (6)  {};
         \node[ neuron,right of =6] (7)  {};
           \node[ neuron,below of =2] (8)  {};
 \node[ neuron,left of = 1] (9)  {};
  \node[ neuron,left of = 9] (10)  {};

        \draw (10) --(7);
        \draw (3)--(8);
    \end{tikzpicture}
}
\caption{  \label{samenumber}}
 \end{figure}

 A more  accurate way is to
  map a Cayley function  to a polytope composed by its
  Feynman diagrams and poles, making the map in \eqref{cubicsum} more intuitive as now we discuss.

  \subsection{Polytopes from Cayley function}

We have seen that CHY formula for $C_n^2$ produces a set of Feynman diagrams, $T(C_n)$, each with n-3 poles; two Feynman diagrams can share $n-4$ poles, and such objects with $n-4$ poles can start to share $n-5$ poles, and so on, until we reach the set of all poles $P(C_n)$. Combinatorically, they can be represented as a polytope in
$n-3$ dimensions. In this section, we describe the construction of such polytopes, and especially give a direct map from Cayely functions or labelled trees to these polytopes.

 \begin{tcolorbox}[colback=white,colframe=white!20!gray]
{\textbf {  Polytope of Feynman diagrams :}}
    Each vertex of this polytope corresponds to a Feynman diagram
     which is a set of $(n-3)$ compatible poles.
      Each edge  corresponds to a set of $(n-4)$ compatible poles.
      Two vertices are connected by an edge iff their Feynman diagrams share $n-4$ poles, which correspond to the intersection of the two diagrams ( so there are always $n-3$ edges stretching out from each vertex ).
         Similarly, a dimension-$r$ face corresponds to a set of $n-3-r$ compatible poles, which is the
     intersection of those of its boundaries, for $r=0,1,\cdots,n-4$.  In the end, each facet (dimension-$(n{-}4)$ face) corresponds to a pole.
 \end{tcolorbox}

 So far, this map is realized by CHY formula of Cayley functions squared. However, we can abstractly view it as constructing a polytope from subgraph structure of a labelled tree: every dimension-$r$ face of the polytope corresponds a collection of $n-3-r$ compatible connected subgraphs of the tree.

 \begin{figure}[!htb]
  \centering
  \subfloat{
  \begin{tikzpicture}[scale=.5]
        \draw[line width=.03cm] (0,0)  --+(0,2) --+(2,2)  --+(2,0) --+(0,0);
     \draw (0,0)--(2,2);
    \coordinate (a1) at (.5,1.5) ;
        \coordinate (a2) at (1.5,0.5) ;
    \draw[blue] (a1)--(a2);
    \draw[blue] (a1)--($(1,1)+(0,1.7)$) node [above=0pt]{2};
    \draw[blue] (a1)--($(1,1)+(-1.7,0)$) node [left=0pt]{1};
    \draw[blue] (a2)--($(1,1)+(1.7,0)$) node [right=0pt]{3};
    \draw[blue] (a2)--($(1,1)+(0,-1.7)$) node [below =0pt]{4};
    \end{tikzpicture}
     } \subfloat{
   \tikz{\node {~~~};}
     } \subfloat{
\begin{tikzpicture}[scale=.5]

 \draw[line width=.03cm]
 (0,0) coordinate (1)
 -- ++(108:2) coordinate (2)
  -- ++(36:2) coordinate (3)
   -- ++(-36:2) coordinate (4)
    --++(-108:2) coordinate (5)
    --+(-2,0);
    \draw  (2)-- (4);
    \draw (2)-- (5);
    \draw[blue] ($(3)+(0,-.6)$)--($(1)+(54:2)+(54:2.3)$)node [right=0pt]{3};
   \draw[blue] ($(3)+(0,-.6)$)--($(1)+(54:2)+(126:2.3)$)node [left=0pt]{2};
     \draw[blue] ($(2)+(-18:2.3)$)--($(3)+(0,-.6)$);
 \draw[blue] ($(1)+(54:.6)$)--($(1)+(54:2)+(-90:2.3)$)node [below=0pt]{5};
   \draw[blue] ($(1)+(54:.6)$)--($(1)+(54:2)+(198:2.3)$) node [left=0pt]{1};
    \draw[blue] ($(1)+(54:.6)$)--($(2)+(-18:2.3)$)--($(1)+(54:2)+(-18:2.3)$) node [right=0pt]{4};
    \end{tikzpicture}
        } \subfloat{
   \tikz{\node {~~~};}
     } \subfloat{
 \begin{tikzpicture}[scale=.5]
  \draw[line width=.03cm]  (0,0) coordinate (1)--++(120:2) coordinate (2)
  --++(60:2) coordinate (3)
  --++(0:2) coordinate (4)
  --++(-60:2) coordinate (5)
  --++(-120:2) coordinate (6)
  --+(180:2) ;
  \draw (1)--(3)--(5)--(1);

      \draw[blue] ($(2)+(0:.5)$)--($(1)+(60:2)+(150:2.3)$) node [left=0pt]{2};
       \draw[blue] ($(2)+(0:.5)$)--($(1)+(60:2)+(210:2.3)$) node [left=0pt]{1};
      \draw[blue] ($(4)+(-120:.5)$)--($(1)+(60:2)+(90:2.3)$) node [above=0pt]{3};
  \draw[blue] ($(4)+(-120:.5)$)--($(1)+(60:2)+(30:2.3)$) node [right=0pt]{4};

        \draw[blue] ($(6)+(120:.5)$)--($(1)+(60:2)+(-90:2.3)$) node [below=0pt]{6};
  \draw[blue] ($(6)+(120:.5)$)--($(1)+(60:2)+(-30:2.3)$) node [right=0pt]{5};

    \draw[blue]($(2)+(0:.5)$)-- ($(2)+(0:2)$)--($(6)+(120:.5)$);
      \draw[blue] ($(2)+(0:2)$)--($(4)+(-120:.5)$);
 \end{tikzpicture}
 }
 \caption{\label{triangulation}Dual graph}
    \end{figure}
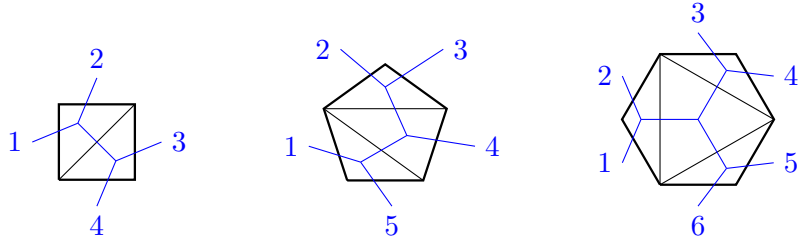

   For PT, the vertices of the corresponding polytope are all planar Feynman diagrams. The dual graph of each planar tree diagram is the  triangulation of a $n$-gon, see figure \ref{triangulation}.  Two vertices are connected by an edge iff their triangulations differ by a single flip. It is well known that a polytope with such vertices is the so-called \href{https://en.wikipedia.org/wiki/Associahedron} {\it associahedron} \cite{associahedron} living in $n-3$ dimensions, which we denote as ${\cal K}_{n-3}$ (in usual literature, it is called $K_{n-1}$). Therefore, we have mapped PT to an associahedron, and let's give some explicit examples.

     \begin{figure}[!htb]
    \centering

    \begin{tikzpicture}[shorten >=0pt,draw=black,scale=.8,
        node distance = 1cm,
        neuront/.style = {circle, minimum size=3pt, inner sep=-2pt } ]
   \node[neuront] (x1) {
   \tikz[scale=.1]{
 \draw[line width=.03cm]
 (0,0) coordinate (1)
 -- ++(108:2) coordinate (2)
  -- ++(36:2) coordinate (3)
   -- ++(-36:2) coordinate (4)
    --++(-108:2) coordinate (5)
    --+(-2,0);
    \draw  (2)-- (4);
    \draw (1)-- (4);}
   };
  \node[neuront] at ($(x1)+(108:2)$) (x2)
  {
     \tikz[scale=.1]{
 \draw[line width=.03cm]
 (0,0) coordinate (1)
 -- ++(108:2) coordinate (2)
  -- ++(36:2) coordinate (3)
   -- ++(-36:2) coordinate (4)
    --++(-108:2) coordinate (5)
    --+(-2,0);
    \draw  (1)-- (3);
    \draw (1)-- (4);}
  };
   \node[neuront] at ($(x2)+(36:2)$) (x3)
  {
     \tikz[scale=.1]{
 \draw[line width=.03cm]
 (0,0) coordinate (1)
 -- ++(108:2) coordinate (2)
  -- ++(36:2) coordinate (3)
   -- ++(-36:2) coordinate (4)
    --++(-108:2) coordinate (5)
    --+(-2,0);
    \draw  (1)-- (3);
    \draw (3)-- (5);}
  };
     \node[neuront] at ($(x3)+(-36:2)$) (x4)
  {
     \tikz[scale=.1]{
 \draw[line width=.03cm]
 (0,0) coordinate (1)
 -- ++(108:2) coordinate (2)
  -- ++(36:2) coordinate (3)
   -- ++(-36:2) coordinate (4)
    --++(-108:2) coordinate (5)
    --+(-2,0);
    \draw  (2)-- (5);
    \draw (3)-- (5);}
  };
       \node[neuront] at ($(x4)+(-108:2)$) (x5)
  {
     \tikz[scale=.1]{
 \draw[line width=.03cm]
 (0,0) coordinate (1)
 -- ++(108:2) coordinate (2)
  -- ++(36:2) coordinate (3)
   -- ++(-36:2) coordinate (4)
    --++(-108:2) coordinate (5)
    --+(-2,0);
    \draw  (2)-- (5);
    \draw (2)-- (4);}
  };

  \draw (x1)node[below left=0pt] {        \begin{tikzpicture}[node distance=.2cm]
 \coordinate (1);
 \coordinate [ right of = 1] (a1) ;
  \coordinate [ right of = a1] (a2) ;
   \coordinate [ right of = a2] (a3) ;
    \coordinate [ right of = a3] (5) ;
    \coordinate [ above of = a1] (2) ;
      \coordinate [ above of = a2] (3) ;
         \coordinate [ above of = a3] (4) ;
   \draw (1) node[below=-2.5pt]{{\tiny 4}} --(5)node[below=-2.5pt]{{\tiny 3}};
      \draw (a1)  --(2)node[above=-2.5pt]{{\tiny 5}};
         \draw (a2) --(3)node[above=-2.5pt]{{\tiny 1}};
          \draw (a3) --(4)node[above=-2.5pt]{{\tiny 2}};
\end{tikzpicture}}

  --(x2)
  node[left=0pt] {  \begin{tikzpicture}[node distance=.2cm]
 \coordinate (1);
 \coordinate [ right of = 1] (a1) ;
  \coordinate [ right of = a1] (a2) ;
   \coordinate [ right of = a2] (a3) ;
    \coordinate [ right of = a3] (5) ;
    \coordinate [ above of = a1] (2) ;
      \coordinate [ above of = a2] (3) ;
         \coordinate [ above of = a3] (4) ;
   \draw (1) node[below=-2.5pt]{{\tiny 1}} --(5)node[below=-2.5pt]{{\tiny 5}};
      \draw (a1)  --(2)node[above=-2.5pt]{{\tiny 2}};
         \draw (a2) --(3)node[above=-2.5pt]{{\tiny 3}};
          \draw (a3) --(4)node[above=-2.5pt]{{\tiny 4}};         \end{tikzpicture}}
          --(x3)
      node[above=0pt] {
          \begin{tikzpicture}[node distance=.2cm]
 \coordinate (1);
 \coordinate [ right of = 1] (a1) ;
  \coordinate [ right of = a1] (a2) ;
   \coordinate [ right of = a2] (a3) ;
    \coordinate [ right of = a3] (5) ;
    \coordinate [ above of = a1] (2) ;
      \coordinate [ above of = a2] (3) ;
         \coordinate [ above of = a3] (4) ;
   \draw (1) node[below=-2.5pt]{{\tiny 3}} --(5)node[below=-2.5pt]{{\tiny 2}};
      \draw (a1)  --(2)node[above=-2.5pt]{{\tiny 4}};
         \draw (a2) --(3)node[above=-2.5pt]{{\tiny 5}};
          \draw (a3) --(4)node[above=-2.5pt]{{\tiny 1}};
\end{tikzpicture}

  }
       --(x4)
         node[right=0pt] {
          \begin{tikzpicture}[node distance=.2cm]
 \coordinate (1);
 \coordinate [ right of = 1] (a1) ;
  \coordinate [ right of = a1] (a2) ;
   \coordinate [ right of = a2] (a3) ;
    \coordinate [ right of = a3] (5) ;
    \coordinate [ above of = a1] (2) ;
      \coordinate [ above of = a2] (3) ;
         \coordinate [ above of = a3] (4) ;
   \draw (1) node[below=-2.5pt]{{\tiny 5}} --(5)node[below=-2.5pt]{{\tiny 4}};
      \draw (a1)  --(2)node[above=-2.5pt]{{\tiny 1}};
         \draw (a2) --(3)node[above=-2.5pt]{{\tiny 2}};
          \draw (a3) --(4)node[above=-2.5pt]{{\tiny 3}};
\end{tikzpicture}

  }
       --(x5)
             node[below right=0pt] {
                 \begin{tikzpicture}[node distance=.2cm]
 \coordinate (1);
 \coordinate [ right of = 1] (a1) ;
  \coordinate [ right of = a1] (a2) ;
   \coordinate [ right of = a2] (a3) ;
    \coordinate [ right of = a3] (5) ;
    \coordinate [ above of = a1] (2) ;
      \coordinate [ above of = a2] (3) ;
         \coordinate [ above of = a3] (4) ;
   \draw (1) node[below=-2.5pt]{{\tiny 2}} --(5)node[below=-2.5pt]{{\tiny 1}};
      \draw (a1)  --(2)node[above=-2.5pt]{{\tiny 3}};
         \draw (a2) --(3)node[above=-2.5pt]{{\tiny 4}};
          \draw (a3) --(4)node[above=-2.5pt]{{\tiny 5}};
\end{tikzpicture}
  }
       --(x1);

 \draw[yellow,->] ($.5*(x3)+.5*(x4)+(.5,.5)$) .. controls  +(3,.3) ..+($-1.7*(x3)+2.5*(x4)$)
 node [below,black,-]
 {
 \begin{tikzpicture}[shorten >=0pt,draw=black,scale=1,
        node distance = 1cm,
        neuront/.style = {circle, minimum size=3pt, inner sep=-2pt } ]
 \node[neuront]  (xx3)
  {
     \tikz[scale=.1]{
 \draw[line width=.03cm]
 (0,0) coordinate (1)
 -- ++(108:2) coordinate (2)
  -- ++(36:2) coordinate (3)
   -- ++(-36:2) coordinate (4)
    --++(-108:2) coordinate (5)
    --+(-2,0);
    \draw[blue]  (1)-- (3);
    \draw (3)-- (5);}
  };
     \node[neuront] at ($(xx3)+(3:2.5)$) (xx4)
  {
     \tikz[scale=.1]{
 \draw[line width=.03cm]
 (0,0) coordinate (1)
 -- ++(108:2) coordinate (2)
  -- ++(36:2) coordinate (3)
   -- ++(-36:2) coordinate (4)
    --++(-108:2) coordinate (5)
    --+(-2,0);
    \draw[blue]  (2)-- (5);
    \draw (3)-- (5);}
  };
  \draw (xx3)
  node [below=0pt] {

            \begin{tikzpicture}[scale=.2]
          \draw (0,0) node[left=-2.5pt]{{\tiny 1}}
          -- ++(1,0) coordinate (a1)
          -- +(0,1) node[above=-2.5pt]{{\tiny 2}};
  \draw (a1)
          -- ++(1,0) coordinate (a2)
          -- +(0,-1) node[below=-2.5pt]{{\tiny 5}};
  \draw (a2)
          -- ++(1,0) coordinate (a3)
          -- +(-45:1) node[below right=-5pt]{{\tiny 4}};
    \draw (a3)
          -- +(45:1) node[above right=-5pt]{{\tiny 3}};
\end{tikzpicture}

  }
  --($.5*(xx3)+.5*(xx4)$)
  node [below]
  {
            \begin{tikzpicture}[scale=.2]
          \draw[blue] (0,0) node[left=-2.5pt]{{\tiny 1}}
          -- ++(1,0) coordinate (a1)
          -- +(0,-1) node[below=-2.5pt]{{\tiny 5}};
  \draw (a1)[blue]
          -- ++(0,0) coordinate (a2)
          -- +(0,1) node[above=-2.5pt]{{\tiny 2}};
  \draw (a2)
          -- ++(1,0) coordinate (a3)
          -- +(-45:1) node[below right=-5pt]{{\tiny 4}};
    \draw (a3)
          -- +(45:1) node[above right=-5pt]{{\tiny 3}};
\end{tikzpicture}

  }

 --(xx4)
 node [below]
 {

           \begin{tikzpicture}[scale=.2]
          \draw (0,0) node[left=-2.5pt]{{\tiny 1}}
          -- ++(1,0) coordinate (a1)
          -- +(0,-1) node[below=-2.5pt]{{\tiny 5}};
  \draw (a1)
          -- ++(1,0) coordinate (a2)
          -- +(0,1) node[above=-2.5pt]{{\tiny 2}};
  \draw (a2)
          -- ++(1,0) coordinate (a3)
          -- +(-45:1) node[below right=-5pt]{{\tiny 4}};
    \draw (a3)
          -- +(45:1) node[above right=-5pt]{{\tiny 3}};
\end{tikzpicture}

 }
 ;

   \end{tikzpicture}

 }
  ;

\end{tikzpicture}

          \caption{\label{H5a}  ${\mathcal{K}}_2$ from ${\rm PT}(1,2,3,4,5)$,}
          on the right we show one of its edges
          \end{figure}

    For ${\rm{PT}}(1,2,3,4)$,   it's mapped to  ${\mathcal{K}}_1$
    \raisebox{-1cm}{
\begin{tikzpicture}[shorten >=0pt,draw=black,scale=1.5,
        node distance = 1.5cm,
        neuront/.style = { minimum size=3pt, inner sep=0pt } ]
     \node[neuront] (01) {
     \tikz[scale=.1]{
        \draw[line width=.03cm] (0,0)--+(0,2)  --+(2,2)  --+(2,0) --+(0,0);
     \draw (0,0)--(2,2);}
     };
     \node[neuront,right of =01] (11)  {
          \tikz[scale=.1]{
        \draw[line width=.03cm] (0,0)--+(0,2)  --+(2,2)  --+(2,0) --+(0,0);
     \draw (0,2)--(2,0);}
     };
     \draw (01)--(11);

   \node at ($(01)+(0,-.35)$)  {
      \begin{tikzpicture}[node distance=.2cm]
 \coordinate (1);
 \coordinate [ right of = 1] (a1) ;
  \coordinate [ right of = a1] (a2) ;
   \coordinate [ right of = a2] (4) ;
    \coordinate [ above of = a1] (2) ;
      \coordinate [ above of = a2] (3) ;
   \draw (1) node[below=-2.5pt]{{\tiny 1}} --(4)node[below=-2.5pt]{{\tiny 4}};
      \draw (a1)  --(2)node[above=-2.5pt]{{\tiny 2}};
         \draw (a2) --(3)node[above=-2.5pt]{{\tiny 3}};
\end{tikzpicture}
                 };
           \node at ($(11)+(0,-.35)$) {
      \begin{tikzpicture}[node distance=.2cm]
 \coordinate (1);
 \coordinate [ right of = 1] (a1) ;
  \coordinate [ right of = a1] (a2) ;
   \coordinate [ right of = a2] (4) ;
    \coordinate [ above of = a1] (2) ;
      \coordinate [ above of = a2] (3) ;
   \draw (1) node[below=-2.5pt]{{\tiny 4}} --(4)node[below=-2.5pt]{{\tiny 3}};
      \draw (a1)  --(2)node[above=-2.5pt]{{\tiny 1}};
         \draw (a2) --(3)node[above=-2.5pt]{{\tiny 2}};
\end{tikzpicture}
};
\end{tikzpicture}
}.

   For ${\rm{PT}}(1,2,3,4,5)$, see \eqref{example5pt},
   it's mapped to ${\mathcal{K}}_2$, see figure \ref{H5a}.  Any two adjacent   vertices ,whose
   triangulations differs by a flip ,
   share a common pole represented as an edge.

   For ${\rm{PT}}(1,2,3,4,5,6)$,  it's mapped to
   ${\mathcal{K}}_3$,see figure \ref{H6a}.  Any two adjacent  vertices ,whose
   triangulations differs by a flip,
   share  two common poles represented as edge. Any adjacent  two edges share a common pole represented as a face. Those Feynman diagrams sharing a common pole sit on the same face. Note that a pentagon  corresponds to a two-particle pole and  a square corresponds to a three-particle pole.

\begin{figure}[!htb]
\centering


};

     \node[neuront] (f9) at (2.46,-6.) {

      \tikz[scale=.1]{
  \draw[line width=.03cm]  (0,0) coordinate (w1) --++(120:2) coordinate (w2)
  --++(60:2) coordinate (w3)
  --++(0:2) coordinate (w4)
  --++(-60:2) coordinate (w5)
  --++(-120:2) coordinate (w6)
  --+(180:2) ;
  \draw (w1)--(w5)--(w2)-- (w4);
    }

    };

      \node[neuront] (f7) at (4.80,-6.) {

            \tikz[scale=.1]{
  \draw[line width=.03cm]  (0,0) coordinate (w1) --++(120:2) coordinate (w2)
  --++(60:2) coordinate (w3)
  --++(0:2) coordinate (w4)
  --++(-60:2) coordinate (w5)
  --++(-120:2) coordinate (w6)
  --+(180:2) ;
  \draw (w4)--(w2)--(w6);
    \draw (w2)--(w5);
    }

   };
 \draw (f8)--(f7)--(f9)--(f10);
 \node at ($.5*(f10)+.5*(f7) $) {

     \begin{tikzpicture}[scale=.2]
          \draw (0,0)[red] node[right=-2.5pt]{{\tiny 3}}
          -- ++(0,-1) coordinate (a1)
          -- +(1,0) node[right=-2.5pt]{{\tiny 4}};
  \draw (a1)[red]
          -- ++(0,0) coordinate (a2)
          -- +(-1,0) node[left=-2.5pt]{{\tiny 2}};
  \draw (a2)
          -- ++(0,-1) coordinate (a3);
  \draw[blue] (a3)
          -- +(1,0) node[right=-2.5pt]{{\tiny 5}};
\draw[blue] (a3)
          -- ++(0,-0) coordinate (a4)
          -- +(-1,0) node[left=-2.5pt]{{\tiny 1}};
    \draw[blue] (a4)
          --+ (0,-1) node[right=-2.5pt]{{\tiny 6}};
\end{tikzpicture}

};
  \end{tikzpicture}

 }
 ;

 \end{tikzpicture}
\caption{\label{H6a} ${\mathcal{K}}_3$ from ${\rm PT}(1,2,3,4,5,6)$,
}
on the right we show its  two faces, one pentagon and one square
\end{figure}


As we have discussed, a star graph
 corresponds to $(n-2)!$ multi-peripheral Feynman diagrams, each  characterized by a permutation of $n-2$ particles.
For star graph, the vertices of the corresponding polytope are $(n-2)!$ multi-peripheral Feynman diagrams, each  characterized by a permutation of $n-2$ particles. Two vertices are connected by an edge iff their permutations differ by a relabeling of two adjacent particles. It is known that a polytope with such vertices is the so-called
 \href{https://en.wikipedia.org/wiki/Permutohedron}{\it permutohedron} \cite{permutohedron}
 living in $n-3$ dimensions, which we denoted as
 ${\mathcal{P}}_{n-2}$
 (in usual literature, it is called $P_{n-1}$).
 Therefore. we have mapped a star graph to
a star graph to a  permutohedron, and let's give some explicit examples.

  For
  $n=4$,
   ${\rm PT}(1,2,3,4)=C^S_4(2)$,
   so the associahedron is also a permutohedron, but in a different view,
     ${\mathcal{P}}_1$
       \raisebox{-1cm}{ \begin{tikzpicture}
     \draw (0,0)--(1,0);
      \fill  (0,0) circle (.1);
       \fill  (1,0) circle (.1);
         \node at (0,-.5) {
      \begin{tikzpicture}[node distance=.2cm]
 \coordinate (1);
 \coordinate [ right of = 1] (a1) ;
  \coordinate [ right of = a1] (a2) ;
   \coordinate [ right of = a2] (4) ;
    \coordinate [ above of = a1] (2) ;
      \coordinate [ above of = a2] (3) ;
   \draw (1) node[below=-2.5pt]{{\tiny 2}} --(4)node[below=-2.5pt]{{\tiny 4}};
      \draw (a1)  --(2)node[above=-2.5pt]{{\tiny 3}};
         \draw (a2) --(3)node[above=-2.5pt]{{\tiny 1}};
\end{tikzpicture}
                 };
           \node at (1,-.5) {
      \begin{tikzpicture}[node distance=.2cm]
 \coordinate (1);
 \coordinate [ right of = 1] (a1) ;
  \coordinate [ right of = a1] (a2) ;
   \coordinate [ right of = a2] (4) ;
    \coordinate [ above of = a1] (2) ;
      \coordinate [ above of = a2] (3) ;
   \draw (1) node[below=-2.5pt]{{\tiny 2}} --(4)node[below=-2.5pt]{{\tiny 4}};
      \draw (a1)  --(2)node[above=-2.5pt]{{\tiny 1}};
         \draw (a2) --(3)node[above=-2.5pt]{{\tiny 3}};
\end{tikzpicture}
};
      \end{tikzpicture}}
      , corresponding to the permutation of $1,3$.

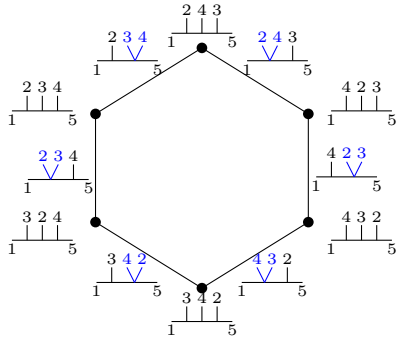
\begin{figure}[!htb]
    \centering
   \begin{tikzpicture}[scale=.7]
       \node at (5,0) {~};
    \draw (-1,-1.25)--(1,0)--(3,-1.25)--(3,-3.3)--(1,-4.55)--(-1,-3.3)--(-1,-1.25);
    \node at (-2,-1.05) {

     \begin{tikzpicture}[node distance=.2cm]
 \coordinate (1);
 \coordinate [ right of = 1] (a1) ;
  \coordinate [ right of = a1] (a2) ;
   \coordinate [ right of = a2] (a3) ;
    \coordinate [ right of = a3] (5) ;
    \coordinate [ above of = a1] (2) ;
      \coordinate [ above of = a2] (3) ;
         \coordinate [ above of = a3] (4) ;
   \draw (1) node[below=-2.5pt]{{\tiny 1}} --(5)node[below=-2.5pt]{{\tiny 5}};
      \draw (a1)  --(2)node[above=-2.5pt]{{\tiny 2}};
         \draw (a2) --(3)node[above=-2.5pt]{{\tiny 3}};
          \draw (a3) --(4)node[above=-2.5pt]{{\tiny 4}};
\end{tikzpicture}

};
    \node at (1,0.42) {

     \begin{tikzpicture}[node distance=.2cm]
 \coordinate (1);
 \coordinate [ right of = 1] (a1) ;
  \coordinate [ right of = a1] (a2) ;
   \coordinate [ right of = a2] (a3) ;
    \coordinate [ right of = a3] (5) ;
    \coordinate [ above of = a1] (2) ;
      \coordinate [ above of = a2] (3) ;
         \coordinate [ above of = a3] (4) ;
   \draw (1) node[below=-2.5pt]{{\tiny 1}} --(5)node[below=-2.5pt]{{\tiny 5}};
      \draw (a1)  --(2)node[above=-2.5pt]{{\tiny 2}};
         \draw (a2) --(3)node[above=-2.5pt]{{\tiny 4}};
          \draw (a3) --(4)node[above=-2.5pt]{{\tiny 3}};
\end{tikzpicture}

};
    \node at (4,-1.05) {

         \begin{tikzpicture}[node distance=.2cm]
 \coordinate (1);
 \coordinate [ right of = 1] (a1) ;
  \coordinate [ right of = a1] (a2) ;
   \coordinate [ right of = a2] (a3) ;
    \coordinate [ right of = a3] (5) ;
    \coordinate [ above of = a1] (2) ;
      \coordinate [ above of = a2] (3) ;
         \coordinate [ above of = a3] (4) ;
   \draw (1) node[below=-2.5pt]{{\tiny 1}} --(5)node[below=-2.5pt]{{\tiny 5}};
      \draw (a1)  --(2)node[above=-2.5pt]{{\tiny 4}};
         \draw (a2) --(3)node[above=-2.5pt]{{\tiny 2}};
          \draw (a3) --(4)node[above=-2.5pt]{{\tiny 3}};
\end{tikzpicture}

};
    \node at (4,-3.5) {

             \begin{tikzpicture}[node distance=.2cm]
 \coordinate (1);
 \coordinate [ right of = 1] (a1) ;
  \coordinate [ right of = a1] (a2) ;
   \coordinate [ right of = a2] (a3) ;
    \coordinate [ right of = a3] (5) ;
    \coordinate [ above of = a1] (2) ;
      \coordinate [ above of = a2] (3) ;
         \coordinate [ above of = a3] (4) ;
   \draw (1) node[below=-2.5pt]{{\tiny 1}} --(5)node[below=-2.5pt]{{\tiny 5}};
      \draw (a1)  --(2)node[above=-2.5pt]{{\tiny 4}};
         \draw (a2) --(3)node[above=-2.5pt]{{\tiny 3}};
          \draw (a3) --(4)node[above=-2.5pt]{{\tiny 2}};
\end{tikzpicture}

};
    \node at (1,-5.04) {

                 \begin{tikzpicture}[node distance=.2cm]
 \coordinate (1);
 \coordinate [ right of = 1] (a1) ;
  \coordinate [ right of = a1] (a2) ;
   \coordinate [ right of = a2] (a3) ;
    \coordinate [ right of = a3] (5) ;
    \coordinate [ above of = a1] (2) ;
      \coordinate [ above of = a2] (3) ;
         \coordinate [ above of = a3] (4) ;
   \draw (1) node[below=-2.5pt]{{\tiny 1}} --(5)node[below=-2.5pt]{{\tiny 5}};
      \draw (a1)  --(2)node[above=-2.5pt]{{\tiny 3}};
         \draw (a2) --(3)node[above=-2.5pt]{{\tiny 4}};
          \draw (a3) --(4)node[above=-2.5pt]{{\tiny 2}};
\end{tikzpicture}

};
     \node at (-2,-3.5) {

                     \begin{tikzpicture}[node distance=.2cm]
 \coordinate (1);
 \coordinate [ right of = 1] (a1) ;
  \coordinate [ right of = a1] (a2) ;
   \coordinate [ right of = a2] (a3) ;
    \coordinate [ right of = a3] (5) ;
    \coordinate [ above of = a1] (2) ;
      \coordinate [ above of = a2] (3) ;
         \coordinate [ above of = a3] (4) ;
   \draw (1) node[below=-2.5pt]{{\tiny 1}} --(5)node[below=-2.5pt]{{\tiny 5}};
      \draw (a1)  --(2)node[above=-2.5pt]{{\tiny 3}};
         \draw (a2) --(3)node[above=-2.5pt]{{\tiny 2}};
          \draw (a3) --(4)node[above=-2.5pt]{{\tiny 4}};
\end{tikzpicture}

};
    \node at (-0.4,-0.1)  {

     \begin{tikzpicture}[node distance=.2cm]
 \coordinate (1);
 \coordinate [ right of = 1] (a1) ;
  \coordinate [ right of = a1] (a2) ;
   \coordinate [ right of = a2] (a3) ;
    \coordinate [ right of = a3] (5) ;
    \coordinate [ above of = a1] (2) ;
      \coordinate [ above of = a2] (3) ;
         \coordinate [ above of = a3] (4) ;
   \draw (1) node[below=-2.5pt]{{\tiny 1}} --(5)node[below=-2.5pt]{{\tiny 5}};
      \draw (a1)  --(2)node[above=-2.5pt]{{\tiny 2}};
         \draw[blue] ($.5*(a2)+.5*(a3)$) --(3)node[above=-2.5pt]{{\tiny 3}};
          \draw[blue] ($.5*(a2)+.5*(a3)$) --(4)node[above=-2.5pt]{{\tiny 4}};
\end{tikzpicture}

};
    \node at (2.42,-0.1)  {

             \begin{tikzpicture}[node distance=.2cm]
 \coordinate (1);
 \coordinate [ right of = 1] (a1) ;
  \coordinate [ right of = a1] (a2) ;
   \coordinate [ right of = a2] (a3) ;
    \coordinate [ right of = a3] (5) ;
    \coordinate [ above of = a1] (2) ;
      \coordinate [ above of = a2] (3) ;
         \coordinate [ above of = a3] (4) ;
   \draw (1) node[below=-2.5pt]{{\tiny 1}} --(5)node[below=-2.5pt]{{\tiny 5}};
      \draw[blue] ($.5*(a1)+.5*(a2)$)  --(2)node[above=-2.5pt]{{\tiny 2}};
         \draw[blue] ($.5*(a1)+.5*(a2)$) --(3)node[above=-2.5pt]{{\tiny 4}};
          \draw (a3) --(4)node[above=-2.5pt]{{\tiny 3}};
\end{tikzpicture}

};
    \node at (3.72,-2.3) {

      \begin{tikzpicture}[node distance=.2cm]
 \coordinate (1);
 \coordinate [ right of = 1] (a1) ;
  \coordinate [ right of = a1] (a2) ;
   \coordinate [ right of = a2] (a3) ;
    \coordinate [ right of = a3] (5) ;
    \coordinate [ above of = a1] (2) ;
      \coordinate [ above of = a2] (3) ;
         \coordinate [ above of = a3] (4) ;
   \draw (1) node[below=-2.5pt]{{\tiny 1}} --(5)node[below=-2.5pt]{{\tiny 5}};
      \draw (a1)  --(2)node[above=-2.5pt]{{\tiny 4}};
         \draw[blue] ($.5*(a2)+.5*(a3)$) --(3)node[above=-2.5pt]{{\tiny 2}};
          \draw[blue] ($.5*(a2)+.5*(a3)$) --(4)node[above=-2.5pt]{{\tiny 3}};
\end{tikzpicture}

};
    \node at(2.32,-4.3)  {

             \begin{tikzpicture}[node distance=.2cm]
 \coordinate (1);
 \coordinate [ right of = 1] (a1) ;
  \coordinate [ right of = a1] (a2) ;
   \coordinate [ right of = a2] (a3) ;
    \coordinate [ right of = a3] (5) ;
    \coordinate [ above of = a1] (2) ;
      \coordinate [ above of = a2] (3) ;
         \coordinate [ above of = a3] (4) ;
   \draw (1) node[below=-2.5pt]{{\tiny 1}} --(5)node[below=-2.5pt]{{\tiny 5}};
      \draw[blue] ($.5*(a1)+.5*(a2)$)  --(2)node[above=-2.5pt]{{\tiny 4}};
         \draw[blue] ($.5*(a1)+.5*(a2)$) --(3)node[above=-2.5pt]{{\tiny 3}};
          \draw (a3) --(4)node[above=-2.5pt]{{\tiny 2}};
\end{tikzpicture}

};
      \node at(-0.41,-4.3) {

        \begin{tikzpicture}[node distance=.2cm]
 \coordinate (1);
 \coordinate [ right of = 1] (a1) ;
  \coordinate [ right of = a1] (a2) ;
   \coordinate [ right of = a2] (a3) ;
    \coordinate [ right of = a3] (5) ;
    \coordinate [ above of = a1] (2) ;
      \coordinate [ above of = a2] (3) ;
         \coordinate [ above of = a3] (4) ;
   \draw (1) node[below=-2.5pt]{{\tiny 1}} --(5)node[below=-2.5pt]{{\tiny 5}};
      \draw (a1)  --(2)node[above=-2.5pt]{{\tiny 3}};
         \draw[blue] ($.5*(a2)+.5*(a3)$) --(3)node[above=-2.5pt]{{\tiny 4}};
          \draw[blue] ($.5*(a2)+.5*(a3)$) --(4)node[above=-2.5pt]{{\tiny 2}};
\end{tikzpicture}

};
      \node at (-1.7,-2.35) {

               \begin{tikzpicture}[node distance=.2cm]
 \coordinate (1);
 \coordinate [ right of = 1] (a1) ;
  \coordinate [ right of = a1] (a2) ;
   \coordinate [ right of = a2] (a3) ;
    \coordinate [ right of = a3] (5) ;
    \coordinate [ above of = a1] (2) ;
      \coordinate [ above of = a2] (3) ;
         \coordinate [ above of = a3] (4) ;
   \draw (1) node[below=-2.5pt]{{\tiny 1}} --(5)node[below=-2.5pt]{{\tiny 5}};
      \draw[blue] ($.5*(a1)+.5*(a2)$)  --(2)node[above=-2.5pt]{{\tiny 2}};
         \draw[blue] ($.5*(a1)+.5*(a2)$) --(3)node[above=-2.5pt]{{\tiny 3}};
          \draw (a3) --(4)node[above=-2.5pt]{{\tiny 4}};
\end{tikzpicture}

};

    \fill  (-1,-1.25) circle (.1);
      \fill (1,0) circle (.1);
       \fill (3,-1.25) circle (.1);
        \fill (3,-3.3)circle (.1);
         \fill  (1,-4.55) circle (.1);
     \fill  (-1,-3.3) circle (.1);
\end{tikzpicture}
              \caption{\label{per of S5} ${\mathcal{P}}_2$ from $C^S_5(1)$}
          \end{figure}

 For
 $C^S_5(1)$,  see \eqref{example5ptstar},
 it's mapped to ${\mathcal{P}}_2$ ,see figure \ref{per of S5} .
 The leg 1 and 5 in each Feynman diagram are particular and all other legs take part in the permutations. Any two adjacent vertices,
  which differs by a relabeling of two adjacent particles,
  ,share a common pole represented by an edge.

As for the $C^S_6(1)$,
it's mapped to ${\mathcal{P}}_3$
  ,see figure \ref{S6p},
 \begin{figure}[!htb]
\centering

\begin{tikzpicture}[shorten >=0pt,draw=black,scale=1.2,
        node distance = \layersep,
        neuront/.style = { minimum size=3pt, inner sep=0pt } ]
     \node[neuront] (2345) at (3.23,-6.01) {\tiny {2354}};
     \node[neuront] (2354) at (4.20,-5.44) {\tiny{2345}};
      \node[neuront] (3254) at (2.73,-4.89) {\tiny{3245}};
   \node[neuront] (3245) at (1.71,-5.44) {\tiny{3254}};

      \node[neuront] (2453) at (5.43,-4.19) {\tiny{2435}};
      \node[neuront] (2543) at (5.76,-3.47) {\tiny{2453}};
      \node[neuront] (2534) at (4.85,-3.95) {\tiny{2543}};

      \node[neuront] (3425) at (2.12,-3.98) {\tiny{5234}};
      \node[neuront] (3524) at (3.46,-2.69) {\tiny{5243}};
      \node[neuront] (2435) at (3.48,-5.35) {\tiny{2534}};
       \node[neuront] (4325) at (0.62,-3.44) {\tiny{5324}};
       \node[neuront] (4235) at (.48,-4.18) {\tiny{3524}};

   \node[neuront] (3542) at (5.39,-1.75) {\tiny{4253}};
\node[neuront] (4532) at (4.15,-.53) {\tiny{4523}};
\node[neuront] (4523) at (3.21,-.88) {\tiny{5423}};

   \node[neuront] (5423) at (1.74,-.52) {\tiny{5432}};
   \node[neuront] (5324) at (.49,-1.76) {\tiny{5342}};

     \node[neuront] (5432) at (2.72,-.18) {\tiny{4532}};
    \node[neuront] (5234) at (.36,-2.49) {\tiny{3542}};

     \node[neuront] (4253) at (2.51,-3.23) {\tiny{3425}};
      \node[neuront] (4352) at (3.65,-2.17) {\tiny{4325}};
       \node[neuront] (3452) at (5.08,-2.51) {\tiny{4325}};

   \node[neuront] (5342) at (2.53,-1.01) {\tiny{4352}};
       \node[neuront] (5243) at (1.46,-2.13) {\tiny{3452}};

      \draw (2345)--(2354)--(2453)--(2543)--(2534)--(2435)--(2345);
      \draw (2345)--(2435)--(3425)--(4325)--(4235)--(3245)--(2345);
  \draw (2534)--(3524)--(3425);
   \draw (2543)--(3542)--(4532)-- (4523)--(3524);
     \draw (4523)--(5423)--(5324)--(4325);
     \draw (4532)--(5432)--(5423);
          \draw (5324)--  (5234)--(4235);

          \draw [dashed,blue] (2354)--(3254)--(3245);

     \draw [dashed,blue] (3254)--(4253)--(4352)-- (3452)--(2453);

   \draw [dashed,blue] (4352)--(5342)--(5243)--(4253);

    \draw [dashed,blue] (3452)--(3542);

    \draw [dashed,blue] (5432)--(5342);

       \draw [dashed,blue] (5234)--(5243);

    \draw[yellow,->] ($(4532)-.5*(5432)+.5*(4532)$) .. controls  +(2,.3) ..+($2*(4352)-2*(5243)$)
    coordinate (hahah);
   \draw (hahah)--(hahah) node [black,-,below]{

  \begin{tikzpicture}[shorten >=0pt,draw=black,scale=.95,
        node distance = \layersep,
        neuront/.style = { minimum size=3pt, inner sep=0pt } ]
 \node    [neuront] (f5423)  {\tiny {5423}};
 \node  at ($(f5423)+(5:2)$)    (f4523)  {\tiny {4523}};
  \node  at ($ (f4523)+(-60:2)$)  (f4253)  {\tiny {4253}};
\node  at ($ (f4253)+(-120:2) $)   (f2453)  {\tiny {2453}};
\node  at ($  (f2453)+(-186:2) $)   (f2543)  {\tiny {2543}};

\node  at ($   (f2543) +(120:2.4) $)  (f5243)  {\tiny {5243}}  ;

\node  at ($   (f5423)  +(90:2) $)  (f5432)  {\tiny {5432}}  ;
\node  at ($   (f5432)  +(5:2) $)  (f4532)  {\tiny {4532}}  ;

   \draw (f5423)
   --($.5*(f5423)+.5*(f4523)$)
   node [above=-2pt]{

  \begin{tikzpicture}[scale=.2]
\draw (0,0)--++(1,0) coordinate (a1);
\draw  (a1)[blue]--++(120:1);
\draw  (a1)[blue]--+(60:1);
\draw  (a1)--++(1,0) coordinate (a2)--++(90:1);
\draw  (a2)--++(1,0) coordinate (a3)--++(90:1);
\draw  (a3)--+(1,0) ;
\end{tikzpicture}
   }

   --(f4523)

     --($.5*(f4253)+.5*(f4523)$)
   node [right=-1pt]{
   \begin{tikzpicture}[scale=.2]
\draw (0,0)--++(1,0) coordinate (a1)--++(90:1);
\draw  (a1)--++(1,0) coordinate (a2);

\draw  (a2)[blue]--++(120:1);
\draw  (a2)[blue]--+(60:1);

\draw  (a2)--++(1,0) coordinate (a3)--++(90:1);
\draw  (a3)--+(1,0) ;
\end{tikzpicture}
}

   -- (f4253)

        --($.5*(f4253)+.5*(f2453)$)
   node [right=-2pt]{
     \begin{tikzpicture}[scale=.2]
\draw (0,0)--++(1,0) coordinate (a1);
\draw  (a1)[blue]--++(120:1);
\draw  (a1)[blue]--+(60:1);
\draw  (a1)--++(1,0) coordinate (a2)--++(90:1);
\draw  (a2)--++(1,0) coordinate (a3)--++(90:1);
\draw  (a3)--+(1,0) ;
\end{tikzpicture}
}

   --(f2453)

           --($.5*(f2543)+.5*(f2453)$)
   node [below=-2pt]{
   \begin{tikzpicture}[scale=.2]
\draw (0,0)--++(1,0) coordinate (a1)--++(90:1);
\draw  (a1)--++(1,0) coordinate (a2);

\draw  (a2)[blue]--++(120:1);
\draw  (a2)[blue]--+(60:1);

\draw  (a2)--++(1,0) coordinate (a3)--++(90:1);
\draw  (a3)--+(1,0) ;
\end{tikzpicture}
}

   -- (f2543)

           --($.5*(f2543)+.5*(f5243)$)
   node [left=-2pt]{
     \begin{tikzpicture}[scale=.2]
\draw (0,0)--++(1,0) coordinate (a1);
\draw  (a1)[blue]--++(120:1);
\draw  (a1)[blue]--+(60:1);
\draw  (a1)--++(1,0) coordinate (a2)--++(90:1);
\draw  (a2)--++(1,0) coordinate (a3)--++(90:1);
\draw  (a3)--+(1,0) ;
\end{tikzpicture}
}

   --  (f5243)

              --($.5*(f5243)+.5*(f5423)$)
   node [left=-1pt]{
   \begin{tikzpicture}[scale=.2]
\draw (0,0)--++(1,0) coordinate (a1)--++(90:1);
\draw  (a1)--++(1,0) coordinate (a2);

\draw  (a2)[blue]--++(120:1);
\draw  (a2)[blue]--+(60:1);

\draw  (a2)--++(1,0) coordinate (a3)--++(90:1);
\draw  (a3)--+(1,0) ;
\end{tikzpicture}
}

   --(f5423)

        --($.5*(f5432)+.5*(f5423)$)
   node [left=-2pt]{

   \begin{tikzpicture}[scale=.2]
\draw (0,0)--++(1,0) coordinate (a1)--++(90:1);
\draw  (a1)--++(1,0) coordinate (a2)--++(90:1);
\draw  (a2)--++(1,0) coordinate (a3);
\draw  (a3)[blue]--++(120:1);
\draw  (a3)[blue]--+(60:1);
\draw  (a3)--+(1,0) ;
\end{tikzpicture}

}

   --(f5432)

       --($.5*(f5432)+.5*(f4532)$)
   node [above=-2pt]{

     \begin{tikzpicture}[scale=.2]
\draw (0,0)--++(1,0) coordinate (a1);
\draw  (a1)[blue]--++(120:1);
\draw  (a1)[blue]--+(60:1);
\draw  (a1)--++(1,0) coordinate (a2)--++(90:1);
\draw  (a2)--++(1,0) coordinate (a3)--++(90:1);
\draw  (a3)--+(1,0) ;
\end{tikzpicture}

}

   --(f4532)

    --($.5*(f4523)+.5*(f4532)$)
   node [right=-2pt]{
   \begin{tikzpicture}[scale=.2]
\draw (0,0)--++(1,0) coordinate (a1)--++(90:1);
\draw  (a1)--++(1,0) coordinate (a2)--++(90:1);
\draw  (a2)--++(1,0) coordinate (a3);
\draw  (a3)[blue]--++(120:1);
\draw  (a3)[blue]--+(60:1);
\draw  (a3)--+(1,0) ;
\end{tikzpicture}
}

   --  (f4523);
   \node at ($.5*(f5423)+.5*(f2453)$) {
   \begin{tikzpicture}[scale=.2]
\draw[blue] (0,0)--++(1,0) coordinate (a1);
\draw  (a1)[blue]--++(135:1);
\draw  (a1)[blue]--+(90:1);
\draw  (a1)[blue]--+(45:1);
\draw  (a1)--++(1.3,0) coordinate (a2)--++(90:1);
\draw  (a2)--+(1,0) ;
\end{tikzpicture}
};
      \node at ($.5*(f5423)+.5*(f4532)$) {
      \begin{tikzpicture}[scale=.2]
\draw (0,0)[blue]--++(1,0) coordinate (a1);
\draw  (a1)[blue]--++(120:1);
\draw  (a1)[blue]--+(60:1);
\draw  (a1)--++(1.4,0) coordinate (a2);
\draw  (a2)[pink]--++(120:1);
\draw  (a2)[pink]--+(60:1);
\draw  (a2)[pink]--++(1,0)  ;
\end{tikzpicture}

};
 \end{tikzpicture}

    }
    ;

    \end{tikzpicture}
\caption{\label{S6p} ${\mathcal{P}}_3$ from $C^S_6(1)$,}
on the right we show its  two faces, one square and one hexagon
\end{figure}
 which corresponds to the permutations of $2,3,4,5$.
   Any two adjacent vertices,
  which differ by a relabeling of two adjoint particles
  ,share two common poles represented by an edge.
  Any adjacent two edges share a common pole represented as a face. Note that  a three-particle pole corresponds to a square while different from the case of associahedron ${\cal K}_3$, a two-particle pole  corresponds to a  hexagon.

Similarly, for arbitrary Cayley function,
 we can always draw its polytope
by the map.
One more  example about  the polytope , see figure \ref{ahfewiofawei},
from
 \raisebox{-.55cm}{
\begin{tikzpicture}[shorten >=0pt,draw=black,
        node distance = .4cm,
        neuron/.style = {circle, minimum size=3pt, inner sep=0pt,  fill=black } ]

     \node[neuron] (1) {};
       \node[ neuron,right of = 1] (2)  {};
       \node[ neuron,above of = 2] (3)  {};
   \node[ neuron,right of = 2] (4)  {};
      \node[ neuron,right of = 4] (5)  {};
        \draw (1) node[below=4pt]{$1$} --(2)node[below=4pt]{$2$}
        --(3)node[above=4pt]{$3$};
        \draw (2)--(4) node[below=4pt]{$4$}  --(5)node[below=4pt]{$5$};
    \end{tikzpicture}}
 , see eq\eqref{H6a}.
\begin{figure}[!htb]
\centering
\begin{tikzpicture}[shorten >=0pt,draw=black,
        node distance = \layersep,
        neuron/.style = {circle, minimum size=3pt, inner sep=0pt,  fill=black } ]
     \node[neuron] (2) at  (2.44,-5.8){};
\node[neuron] (18) at (6.21,-6.41)  {};
\node[neuron] (16) at (7.01,-5.93) {};
\node[neuron] (8) at   (5.29,-5.29){};
\node[neuron] (7) at   (3.73,-5.16){};
\node[neuron] (17) at   (7.39,-4.45){};
\node[neuron] (15) at   (8.21,-3.92){};
\node[neuron] (1) at   (6.14,-3.26){};
\node[neuron] (4) at   (2.79,-2.86){};
\node[neuron] (3) at   (0.57,-3.78){};
\node[neuron] (11) at   (6.30,-1.17){};
\node[neuron] (12) at   (7.95,-1.76){};
\node[neuron] (14) at   (6.46,-.70){};
\node[neuron] (13) at   (4.94,-.23){};
\node[neuron] (6) at   (2.75,-.64){};
\node[neuron] (5) at   (0.36,-1.52){};
\node[neuron] (4) at   (2.79,-2.86){};
\node[neuron] (9) at   (1.20,-1.77){};
\node[neuron] (10) at   (5.35,-1.52){};
\draw (2)

node  [below=-5pt]{

\begin{tikzpicture}[node distance=.2cm,color=red]
 \coordinate (b1);
 \coordinate [ right of = b1] (a1) ;
  \coordinate [ right of = a1] (a2) ;
   \coordinate [ right of = a2] (a3) ;
    \coordinate [ right of = a3] (b5) ;
    \coordinate [ above of = a1] (b2) ;
      \coordinate [ above of = a2] (a4) ;
       \coordinate [  above right of = a4] (b6) ;
           \coordinate [  above left of = a4] (b3) ;
         \coordinate [ above of = a3] (b4) ;
   \draw (b1) node[below=-2.5pt]{{{\tiny 1}}} --(b5)node[below=-2.5pt]{{\tiny 5}};
      \draw (a1)  --(b2)node[left=-2.5pt]{{\tiny 2}};
         \draw (a2) --(a4)--(b3)node[above=-2.5pt]{{\tiny 3}};
           \draw (a4)--(b6)node[above=-2.5pt]{{\tiny 6}};
          \draw (a3) --(b4)node[right=-2.5pt]{{{\tiny  4}}};
\end{tikzpicture}

}

--(18)
node  [below=-5pt]{

\begin{tikzpicture}[node distance=.2cm,color=red]
 \coordinate (b1);
 \coordinate [ right of = b1] (a1) ;
  \coordinate [ right of = a1] (a2) ;
   \coordinate [ right of = a2] (a3) ;
    \coordinate [ right of = a3] (a4) ;
     \coordinate [ right of = a4] (b6) ;
    \coordinate [ above of = a1] (b2) ;
      \coordinate [ above of = a2] (b3) ;
         \coordinate [ above of = a3] (b4) ;
    \coordinate [ above of = a4] (b5) ;

     \draw (b1) node[below=-2.5pt]{{\tiny 4}}  --(b6)node[below=-2.5pt]{{\tiny 6}};
      \draw (a1)  --(b2)node[above=-2.5pt]{{\tiny 5}};
         \draw (a2) --(b3)node[above=-2.5pt]{{\tiny 2}};
          \draw (a3) --(b4)node[above=-2.5pt]{{\tiny 1}};
          \draw (a4) --(b5)node[above=-2.5pt]{{\tiny 3}};
\end{tikzpicture}

}

--(16)

node  [below right=-5pt]{
\begin{tikzpicture}[node distance=.2cm,color=red]
 \coordinate (b1);
 \coordinate [ right of = b1] (a1) ;
  \coordinate [ right of = a1] (a2) ;
   \coordinate [ right of = a2] (a3) ;
    \coordinate [ right of = a3] (a4) ;
     \coordinate [ right of = a4] (b6) ;
    \coordinate [ above of = a1] (b2) ;
      \coordinate [ above of = a2] (b3) ;
         \coordinate [ above of = a3] (b4) ;
    \coordinate [ above of = a4] (b5) ;

     \draw (b1) node[below=-2.5pt]{{\tiny 2}}  --(b6)node[below=-2.5pt]{{\tiny 6}};
      \draw (a1)  --(b2)node[above=-2.5pt]{{\tiny 4}};
         \draw (a2) --(b3)node[above=-2.5pt]{{\tiny 5}};
          \draw (a3) --(b4)node[above=-2.5pt]{{\tiny 1}};
          \draw (a4) --(b5)node[above=-2.5pt]{{\tiny 3}};
\end{tikzpicture}
}

--(15)

node  [right=-5pt]{
\begin{tikzpicture}[node distance=.2cm,color=red]
 \coordinate (b1);
 \coordinate [ right of = b1] (a1) ;
  \coordinate [ right of = a1] (a2) ;
   \coordinate [ right of = a2] (a3) ;
    \coordinate [ right of = a3] (a4) ;
     \coordinate [ right of = a4] (b6) ;
    \coordinate [ above of = a1] (b2) ;
      \coordinate [ above of = a2] (b3) ;
         \coordinate [ above of = a3] (b4) ;
    \coordinate [ above of = a4] (b5) ;

     \draw (b1) node[below=-2.5pt]{{\tiny 2}}  --(b6)node[below=-2.5pt]{{\tiny 6}};
      \draw (a1)  --(b2)node[above=-2.5pt]{{\tiny 4}};
         \draw (a2) --(b3)node[above=-2.5pt]{{\tiny 5}};
          \draw (a3) --(b4)node[above=-2.5pt]{{\tiny 3}};
          \draw (a4) --(b5)node[above=-2.5pt]{{\tiny 1}};
\end{tikzpicture}
}

--(17)

node  [left=-5pt]{
\begin{tikzpicture}[node distance=.2cm,color=red]
 \coordinate (b1);
 \coordinate [ right of = b1] (a1) ;
  \coordinate [ right of = a1] (a2) ;
   \coordinate [ right of = a2] (a3) ;
    \coordinate [ right of = a3] (a4) ;
     \coordinate [ right of = a4] (b6) ;
    \coordinate [ above of = a1] (b2) ;
      \coordinate [ above of = a2] (b3) ;
         \coordinate [ above of = a3] (b4) ;
    \coordinate [ above of = a4] (b5) ;

     \draw (b1) node[below=-2.5pt]{{\tiny 4}}  --(b6)node[below=-2.5pt]{{\tiny 6}};
      \draw (a1)  --(b2)node[above=-2.5pt]{{\tiny 5}};
         \draw (a2) --(b3)node[above=-2.5pt]{{\tiny 2}};
          \draw (a3) --(b4)node[above=-2.5pt]{{\tiny 3}};
          \draw (a4) --(b5)node[above=-2.5pt]{{\tiny 1}};
\end{tikzpicture}
}

--(18);
\draw (2)--(3)

node  [left=-5pt]{
\begin{tikzpicture}[node distance=.2cm,color=red]
 \coordinate (b1);
 \coordinate [ right of = b1] (a1) ;
  \coordinate [ right of = a1] (a2) ;
   \coordinate [ right of = a2] (a3) ;
    \coordinate [ right of = a3] (a4) ;
     \coordinate [ right of = a4] (b6) ;
    \coordinate [ above of = a1] (b2) ;
      \coordinate [ above of = a2] (b3) ;
         \coordinate [ above of = a3] (b4) ;
    \coordinate [ above of = a4] (b5) ;

     \draw (b1) node[below=-2.5pt]{{\tiny 1}}  --(b6)node[below=-2.5pt]{{\tiny 5}};
      \draw (a1)  --(b2)node[above=-2.5pt]{{\tiny 2}};
         \draw (a2) --(b3)node[above=-2.5pt]{{\tiny 3}};
          \draw (a3) --(b4)node[above=-2.5pt]{{\tiny 6}};
          \draw (a4) --(b5)node[above=-2.5pt]{{\tiny 4}};
\end{tikzpicture}
}

--(4)

node  [below=-3pt]{
\begin{tikzpicture}[node distance=.2cm,color=red]
 \coordinate (b1);
 \coordinate [ right of = b1] (a1) ;
  \coordinate [ right of = a1] (a2) ;
   \coordinate [ right of = a2] (a3) ;
    \coordinate [ right of = a3] (a4) ;
     \coordinate [ right of = a4] (b6) ;
    \coordinate [ above of = a1] (b2) ;
      \coordinate [ above of = a2] (b3) ;
         \coordinate [ above of = a3] (b4) ;
    \coordinate [ above of = a4] (b5) ;

     \draw (b1) node[below=-2.5pt]{{\tiny 2}}  --(b6)node[below=-2.5pt]{{\tiny 5}};
      \draw (a1)  --(b2)node[above=-2.5pt]{{\tiny 3}};
         \draw (a2) --(b3)node[above=-2.5pt]{{\tiny 1}};
          \draw (a3) --(b4)node[above=-2.5pt]{{\tiny 6}};
          \draw (a4) --(b5)node[above=-2.5pt]{{\tiny 4}};
\end{tikzpicture}
}

--(1)--(17);
\draw (1)
node  [right=-5pt]{
\begin{tikzpicture}[node distance=.2cm,color=red]
 \coordinate (b1);
 \coordinate [ right of = b1] (a1) ;
  \coordinate [ right of = a1] (a2) ;
   \coordinate [ right of = a2] (a3) ;
    \coordinate [ right of = a3] (b5) ;
    \coordinate [ above of = a1] (b2) ;
      \coordinate [ above of = a2] (a4) ;
       \coordinate [  above right of = a4] (b6) ;
           \coordinate [  above left of = a4] (b3) ;
         \coordinate [ above of = a3] (b4) ;
   \draw (b1) node[below=-2.5pt]{{{\tiny  2}}} --(b5)node[below=-2.5pt]{{\tiny 5}};
      \draw (a1)  --(b2)node[left=-2.5pt]{{\tiny 3}};
         \draw (a2) --(a4)--(b3)node[above=-2.5pt]{{\tiny 1}};
           \draw (a4)--(b6)node[above=-2.5pt]{{\tiny 6}};
          \draw (a3) --(b4)node[right=-2.5pt]{{{\tiny  4}}};
\end{tikzpicture}
}
--(11)

node  [below right=-5pt]{
\begin{tikzpicture}[node distance=.2cm,color=red]
 \coordinate (b1);
 \coordinate [ right of = b1] (a1) ;
  \coordinate [ right of = a1] (a2) ;
   \coordinate [ right of = a2] (a3) ;
    \coordinate [ right of = a3] (a4) ;
     \coordinate [ right of = a4] (b6) ;
    \coordinate [ above of = a1] (b2) ;
      \coordinate [ above of = a2] (b3) ;
         \coordinate [ above of = a3] (b4) ;
    \coordinate [ above of = a4] (b5) ;

     \draw (b1) node[below=-2.5pt]{{\tiny 2}}  --(b6)node[below=-2.5pt]{{\tiny 6}};
      \draw (a1)  --(b2)node[above=-2.5pt]{{\tiny 3}};
         \draw (a2) --(b3)node[above=-2.5pt]{{\tiny 4}};
          \draw (a3) --(b4)node[above=-2.5pt]{{\tiny 5}};
          \draw (a4) --(b5)node[above=-2.5pt]{{\tiny 1}};
\end{tikzpicture}
}

--(12)

node  [right=-5pt]{
\begin{tikzpicture}[node distance=.2cm,color=red]
 \coordinate (b1);
 \coordinate [ right of = b1] (a1) ;
  \coordinate [ right of = a1] (a2) ;
   \coordinate [ right of = a2] (a3) ;
    \coordinate [ right of = a3] (a4) ;
     \coordinate [ right of = a4] (b6) ;
    \coordinate [ above of = a1] (b2) ;
      \coordinate [ above of = a2] (b3) ;
         \coordinate [ above of = a3] (b4) ;
    \coordinate [ above of = a4] (b5) ;

     \draw (b1) node[below=-2.5pt]{{\tiny 2}}  --(b6)node[below=-2.5pt]{{\tiny 6}};
      \draw (a1)  --(b2)node[above=-2.5pt]{{\tiny 4}};
         \draw (a2) --(b3)node[above=-2.5pt]{{\tiny 3}};
          \draw (a3) --(b4)node[above=-2.5pt]{{\tiny 5}};
          \draw (a4) --(b5)node[above=-2.5pt]{{\tiny 1}};
\end{tikzpicture}
}

--(14)

node  [above right=-5pt]{
\begin{tikzpicture}[node distance=.2cm,color=red]
 \coordinate (b1);
 \coordinate [ right of = b1] (a1) ;
  \coordinate [ right of = a1] (a2) ;
   \coordinate [ right of = a2] (a3) ;
    \coordinate [ right of = a3] (a4) ;
     \coordinate [ right of = a4] (b6) ;
    \coordinate [ above of = a1] (b2) ;
      \coordinate [ above of = a2] (b3) ;
         \coordinate [ above of = a3] (b4) ;
    \coordinate [ above of = a4] (b5) ;

     \draw (b1) node[below=-2.5pt]{{\tiny 2}}  --(b6)node[below=-2.5pt]{{\tiny 6}};
      \draw (a1)  --(b2)node[above=-2.5pt]{{\tiny 4}};
         \draw (a2) --(b3)node[above=-2.5pt]{{\tiny 3}};
          \draw (a3) --(b4)node[above=-2.5pt]{{\tiny 1}};
          \draw (a4) --(b5)node[above=-2.5pt]{{\tiny 5}};
\end{tikzpicture}
}

--(13)

node  [above=-5pt]{
\begin{tikzpicture}[node distance=.2cm,color=red]
 \coordinate (b1);
 \coordinate [ right of = b1] (a1) ;
  \coordinate [ right of = a1] (a2) ;
   \coordinate [ right of = a2] (a3) ;
    \coordinate [ right of = a3] (a4) ;
     \coordinate [ right of = a4] (b6) ;
    \coordinate [ above of = a1] (b2) ;
      \coordinate [ above of = a2] (b3) ;
         \coordinate [ above of = a3] (b4) ;
    \coordinate [ above of = a4] (b5) ;

     \draw (b1) node[below=-2.5pt]{{\tiny 2}}  --(b6)node[below=-2.5pt]{{\tiny 6}};
      \draw (a1)  --(b2)node[above=-2.5pt]{{\tiny 3}};
         \draw (a2) --(b3)node[above=-2.5pt]{{\tiny 4}};
          \draw (a3) --(b4)node[above=-2.5pt]{{\tiny 1}};
          \draw (a4) --(b5)node[above=-2.5pt]{{\tiny 5}};
\end{tikzpicture}
}

--(11);
\draw (13)--(6)

node  [above=-5pt]{
\begin{tikzpicture}[node distance=.2cm,color=red]
 \coordinate (b1);
 \coordinate [ right of = b1] (a1) ;
  \coordinate [ right of = a1] (a2) ;
   \coordinate [ right of = a2] (a3) ;
    \coordinate [ right of = a3] (a4) ;
     \coordinate [ right of = a4] (b6) ;
    \coordinate [ above of = a1] (b2) ;
      \coordinate [ above of = a2] (b3) ;
         \coordinate [ above of = a3] (b4) ;
    \coordinate [ above of = a4] (b5) ;

     \draw (b1) node[below=-2.5pt]{{\tiny 2}}  --(b6)node[below=-2.5pt]{{\tiny 6}};
      \draw (a1)  --(b2)node[above=-2.5pt]{{\tiny 3}};
         \draw (a2) --(b3)node[above=-2.5pt]{{\tiny 1}};
          \draw (a3) --(b4)node[above=-2.5pt]{{\tiny 4}};
          \draw (a4) --(b5)node[above=-2.5pt]{{\tiny 5}};
\end{tikzpicture}
}

--(5)

node  [above left=-5pt]{
\begin{tikzpicture}[node distance=.2cm,color=red]
 \coordinate (b1);
 \coordinate [ right of = b1] (a1) ;
  \coordinate [ right of = a1] (a2) ;
   \coordinate [ right of = a2] (a3) ;
    \coordinate [ right of = a3] (a4) ;
     \coordinate [ right of = a4] (b6) ;
    \coordinate [ above of = a1] (b2) ;
      \coordinate [ above of = a2] (b3) ;
         \coordinate [ above of = a3] (b4) ;
    \coordinate [ above of = a4] (b5) ;

     \draw (b1) node[below=-2.5pt]{{\tiny 1}}  --(b6)node[below=-2.5pt]{{\tiny 6}};
      \draw (a1)  --(b2)node[above=-2.5pt]{{\tiny 2}};
         \draw (a2) --(b3)node[above=-2.5pt]{{\tiny 3}};
          \draw (a3) --(b4)node[above=-2.5pt]{{\tiny 4}};
          \draw (a4) --(b5)node[above=-2.5pt]{{\tiny 5}};
\end{tikzpicture}
}

--(3)--(4)--(6);
\draw (12)--(15);
\draw (4)--(1);
\draw[dashed,blue](9)--(5);
\draw(9)--(9)
node  [below=-5pt]{
\begin{tikzpicture}[node distance=.2cm,color=red!50!green]
 \coordinate (b1);
 \coordinate [ right of = b1] (a1) ;
  \coordinate [ right of = a1] (a2) ;
   \coordinate [ right of = a2] (a3) ;
    \coordinate [ right of = a3] (a4) ;
     \coordinate [ right of = a4] (b6) ;
    \coordinate [ above of = a1] (b2) ;
      \coordinate [ above of = a2] (b3) ;
         \coordinate [ above of = a3] (b4) ;
    \coordinate [ above of = a4] (b5) ;

     \draw (b1) node[below=-2.5pt]{{\tiny 1}}  --(b6)node[below=-2.5pt]{{\tiny 6}};
      \draw (a1)  --(b2)node[above=-2.5pt]{{\tiny 2}};
         \draw (a2) --(b3)node[above=-2.5pt]{{\tiny 4}};
          \draw (a3) --(b4)node[above=-2.5pt]{{\tiny 3}};
          \draw (a4) --(b5)node[above=-2.5pt]{{\tiny 5}};
\end{tikzpicture}
};

\draw[dashed,blue] (10)--(14);
\draw (10)--(10)
node  [below left=-5pt]{
\begin{tikzpicture}[node distance=.2cm,color=red!50!green]
 \coordinate (b1);
 \coordinate [ right of = b1] (a1) ;
  \coordinate [ right of = a1] (a2) ;
   \coordinate [ right of = a2] (a3) ;
    \coordinate [ right of = a3] (a4) ;
     \coordinate [ right of = a4] (b6) ;
    \coordinate [ above of = a1] (b2) ;
      \coordinate [ above of = a2] (b3) ;
         \coordinate [ above of = a3] (b4) ;
    \coordinate [ above of = a4] (b5) ;

     \draw (b1) node[below=-2.5pt]{{\tiny 2}}  --(b6)node[below=-2.5pt]{{\tiny 6}};
      \draw (a1)  --(b2)node[above=-2.5pt]{{\tiny 4}};
         \draw (a2) --(b3)node[above=-2.5pt]{{\tiny 1}};
          \draw (a3) --(b4)node[above=-2.5pt]{{\tiny 3}};
          \draw (a4) --(b5)node[above=-2.5pt]{{\tiny 5}};
\end{tikzpicture}
};

\draw (7)--(7)

node  [above=-5pt]{
\begin{tikzpicture}[node distance=.2cm,color=red!50!green]
 \coordinate (b1);
 \coordinate [ right of = b1] (a1) ;
  \coordinate [ right of = a1] (a2) ;
   \coordinate [ right of = a2] (a3) ;
    \coordinate [ right of = a3] (a4) ;
     \coordinate [ right of = a4] (b6) ;
    \coordinate [ above of = a1] (b2) ;
      \coordinate [ above of = a2] (b3) ;
         \coordinate [ above of = a3] (b4) ;
    \coordinate [ above of = a4] (b5) ;

     \draw (b1) node[below=-2.5pt]{{\tiny 1}}  --(b6)node[below=-2.5pt]{{\tiny 6}};
      \draw (a1)  --(b2)node[above=-2.5pt]{{\tiny 2}};
         \draw (a2) --(b3)node[above=-2.5pt]{{\tiny 4}};
          \draw (a3) --(b4)node[above=-2.5pt]{{\tiny 5}};
          \draw (a4) --(b5)node[above=-2.5pt]{{\tiny 3}};
\end{tikzpicture}
};

\draw[dashed,blue] (7)--(9)--(10)--(8);

\draw (8)--(8)

node  [below=-5pt]{
\begin{tikzpicture}[node distance=.2cm,color=red!50!green]
 \coordinate (b1);
 \coordinate [ right of = b1] (a1) ;
  \coordinate [ right of = a1] (a2) ;
   \coordinate [ right of = a2] (a3) ;
    \coordinate [ right of = a3] (a4) ;
     \coordinate [ right of = a4] (b6) ;
    \coordinate [ above of = a1] (b2) ;
      \coordinate [ above of = a2] (b3) ;
         \coordinate [ above of = a3] (b4) ;
    \coordinate [ above of = a4] (b5) ;

     \draw (b1) node[below=-2.5pt]{{\tiny 2}}  --(b6)node[below=-2.5pt]{{\tiny 6}};
      \draw (a1)  --(b2)node[above=-2.5pt]{{\tiny 4}};
         \draw (a2) --(b3)node[above=-2.5pt]{{\tiny 1}};
          \draw (a3) --(b4)node[above=-2.5pt]{{\tiny 5}};
          \draw (a4) --(b5)node[above=-2.5pt]{{\tiny 3}};
\end{tikzpicture}
}

;
\draw[dashed,blue] (16)--(8)--(7)--(2);
    \end{tikzpicture}
 \caption{\label{ahfewiofawei} Polytope from
  \protect\raisebox{-.5cm}{
 \protect\tikz[shorten >=0pt,draw=black,
        node distance = .4cm,
        neuron/.style = {circle, minimum size=3pt, inner sep=0pt,  fill=black } ]{
        \protect  \node[neuron] (1) {};
         \protect \node[ neuron,right of = 1] (2)  {};
       \protect   \node[ neuron,above of = 2] (3)  {};
  \protect \node[ neuron,right of = 2] (4)  {};
     \protect \node[ neuron,right of = 4] (5)  {};
       \protect \draw (1) node[below=0pt]{$1$} --(2)node[below=0pt]{$2$}
        --(3)node[above=0pt]{$3$};
       \protect \draw (2)--(4) node[below=0pt]{$4$}  --(5)node[below=0pt]{$5$};
}}}
\end{figure}
Any two adjacent vertices share two common poles represented by an edge.
  Any adjacent two edges share a common pole represented as a face. Note that A three-particle pole corresponds to a square , so there are 4 squares.  While
   the case of two-particle pole
 are between associahedron ${\cal K}_{3}$ and permutohedron ${\cal P}_3$: 4 two-particle poles
 correspond to  pentagons  and 3  correspond to  hexagons .

We can see symmetries of polytope reflects that of the covariant form of Cayley functions, see some examples in figure \ref{cova}.
\def \layersep {.7cm}
\begin{figure}[!htb]
  \centering
  \subfloat{
  \begin{tikzpicture}[shorten >=0pt,draw=black,scale=.8,
        node distance = \layersep,
        neuron/.style = {circle, minimum size=3pt, inner sep=0pt,  fill=black },
        blueneuron/.style = {circle, minimum size=3pt, inner sep=0pt,  fill=blue } ]
\draw (0,0) node[neuron] (1) {}  node[left=4pt]{$1$}
     -- ++(-90:1) node[neuron] (2) {}  node[left=4pt]{$2$}
     -- ++(-30:1) node[neuron] (3) {}  node[below=4pt]{$3$}
     -- ++(30:1) node[neuron] (4) {}  node[right=4pt]{$4$}
     -- +(90:1) node[neuron] (5) {}  node[right=4pt]{$5$}
     ;
\draw[blue] (5)
 -- +(150:1) node[blueneuron] (6) {}  node[above=4pt]{$6$}
--(1);
\node at ($(4)+(1,0)$) {~};
\node at ($(3)+(0,-1.2)$) {${\rm PT}(1,2,\cdots,6)$};
\node at ($(3)+(0,-1.7)$) {cyclic symmetry};
\end{tikzpicture}

  }\subfloat{
  \begin{tikzpicture}[shorten >=0pt,draw=black,scale=.56,
        node distance = \layersep,
        neuron/.style = {circle, minimum size=3pt, inner sep=0pt,  fill=black },
        blueneuron/.style = {circle, minimum size=3pt, inner sep=0pt,  fill=blue } ]
    \draw[white] (0,0) coordinate (1)   node[below=4pt]{$1$}
     -- +(15:2) node[neuron] (4) {}  ;
       \draw[white] (1)
     -- +(-15:2) node[neuron] (5) {} ;
       \draw [white](1)
     -- +(165:2) node[neuron] (3) {} ;
        \draw[white] (1)
     -- +(195:2) node[neuron] (2) {}  ;

     \node[ blueneuron] at ($(1)+(0,1.5)$) (6)  {};
     \node[ neuron] at ($(1)+(0,-1.5)$) (11)  {};
     \draw [blue] (2)--(6) node[above=4pt]{$6$} ;
     \draw [blue] (3) --(6);
     \draw [blue] (4)--(6);
     \draw [blue] (5) --(6);
     \draw (11)node[below=4pt]{$1$}--(2)node[left=4pt]{$2$};
     \draw (11)--(3)node[left=4pt]{$3$};
     \draw (11)--(4)node[right=4pt]{$4$};
     \draw (11)--(5)node[right=4pt]{$5$};
     \node at ($(4)+(1,0)$) {~};
     \node at ($(11)+(0,-1.4)$) {$C_6^S(1)$,~1,6 are symmetric,};
      \node at ($(11)+(0,-2.1)$) { others  are symmetric};
    \end{tikzpicture}
  }\subfloat{
\begin{tikzpicture}[shorten >=0pt,draw=black,scale=.8,
        node distance = \layersep,
        neuron/.style = {circle, minimum size=3pt, inner sep=0pt,  fill=black },
        blueneuron/.style = {circle, minimum size=3pt, inner sep=0pt,  fill=blue } ]
\draw (0,0) node[neuron] (2) {} node[below=4pt]{$2$}
--++(1,0) node[neuron] (4) {} node[below=4pt]{$4$}
--+(1,0)node[neuron] (5) {} node[below=4pt]{$5$}
 ;
 \draw (2)--+(160:1) node[neuron] (1) {} node[left=4pt]{$1$};
\draw (2)--+(200:1) node[neuron] (3) {} node[left=4pt]{$3$};
\node[neuron] at ($.5*(2)+.5*(4)+(0,1.5)$) (6) {};

\draw[blue] (1)--(6) node[blueneuron]  {} node[above=4pt]{$6$}
--(5);
\draw [blue] (3)--(6);
 \node at ($.5*(2)+.5*(4)+(0,-1.4)$) {1,3 are symmetric};
\end{tikzpicture}
  }
  \caption{\label{cova} symmetries of covariant form of some Cayley functions
}
\end{figure}

\section{Linear space of Cayley functions }
In this section, we study the linear space spanned by all Cayley functions, which is of dimension $(n{-}2)!$. We first show that any Cayley function can be written as a linear combination of $(n{-}2)!$ Parke-Taylor factors, known as the ``Kleiss-Kuijf (KK) basis". More importantly, we find a new basis of the space which consists of elements we call $C^{\rm single}$ and $C^{\rm kernel}$. The remarkable property of the new basis is that, given a PT, the CHY formula of any $C^{\rm single}$ and PT gives a single Feynman diagram, while that of $C^{\rm kernel}$ and PT gives zero.

\subsection{Reduction to PT factors and KK basis}\label{sec3p1}
The main result here is a remarkable formula expressing C as sum of certain PT's:
\ba\label{id}
C(\{i_1,j_1\},\cdots,\{i_{n-2},j_{n-2}\})=\sum_{\substack{\rho\in S_{n-1}\\
\rho^{-1}(i_1)<\rho^{-1}(j_1)\\ \cdots\\ \rho^{-1}(i_{n-2})<\rho^{-1}(j_{n-2})
}} {\rm PT}(\rho(1),\rho(2),\cdots,\rho(n-1),n)\,.
\ea
Here $\rho^{-1}(i)<\rho^{-1}(j)$ means $i$ is in the left of $j$ in $\rho$.
It is not surprising that by partial fraction, a Cayley function can be reduced to those of Hamilton graphs (see \cite{Stieberger:2013hza}), but here we see that the result takes such a simple form, with coefficient only +1!

This identity can be easily proved by recursion. We remove any of the $n{-}2$ pairs denoted as $\{i_r,j_r\}$ and make $i_r,j_r$  identical, then the remaining $n{-}3$ pairs still compose a C. More intuitively, we shrink any line  $\{i_r,j_r\}$ in the labelled tree of C, and it is still a Cayley graph. If any of the C of $n{-}1$ points satisfy \eqref{id}, that is to say any residue of C in LHS of \eqref{id} equals to that of RHS. Obviously, C function doesn't have pole at infinity, and \eqref{id} is correct for $n=4$, so we finish the proof.

For example,
\ba\label{cflip}
C(\{1,3\},\{2,3\},\{3,4\})&=&{\rm PT}(1,2,3,4,5)+
{\rm PT}(2,1,3,4,5) \,,\nl
C(\{3,1\},\{3,2\},\{3,4\})&=&\sum_{\rho\in S_3}{\rm PT}(3,\rho(1,2,4),5) \,,\nl
C(\{1,2\},\{2,3\},\{2,4\},\{3,5\})&=&{\rm PT}(1, 2, 3, 4, 5, 6)+{\rm PT}(1, 2, 3, 5, 4, 6)\nl
&&+{\rm PT}(1, 2, 4, 3, 5, 6)\,,\nl
C(\{2,4\},\{4,3\},\{3,5\},\{5,1\},\{1,6\})&=&{\rm PT}(2,4,3,5,1,6,7)\,.
\ea

Note that we can flip some pairs in $C$, which at most changes its overall sign, since $C(\cdots,\{j,i\},\cdots)=-C(\cdots,\{i,j\},\cdots)$; while the summation on the RHS of \eqref{id} changes completely.
For example we can see that on the first  line of \eqref{cflip}, by flipping two pairs, the RHS differs from that of the second line of \eqref{cflip}.
This is not surprising since the PT's are not linearly independent but satisfy relations known as ``Kleiss-Kuijf(KK) relations". However, it is remarkable that we have a canonical way to land on a basis; we give each edge   an orientation
 such that the whole flow are  from 1 to end points, see figure \ref{stretch1},
 \tikzset{
particle/.style={draw=black, postaction={decorate},
    decoration={markings,mark=at position .5 with {\arrow[draw=blue,scale=1.5]{>}}}}
 }
\begin{figure}[!htb]
\centering
\begin{tikzpicture}[shorten >=0pt,draw=black,
        node distance = .7cm,
        neuron/.style = {circle, minimum size=3pt, inner sep=0pt,  fill=black } ]

\draw[particle] (0,0) node [neuron] (1) {} node[below=0pt]{$1$}
--+(-1,0)node [neuron] (a) {};

\draw[particle] (a)--+(0,-1)node [neuron]  {};
\draw[particle]  (a)
--+(-1,0)node [neuron] (a1) {};
\draw[particle]  (a1)
--+(0,1)node [neuron] (a2) {};
\draw[particle]  (a2)
--+(45:1)node [neuron]  {};
\draw[particle] (a2)
--+(135:1)node [neuron] {};
\draw[particle] (a1)
--+(-1,0)node [neuron] (a3) {};
\draw[particle] (a3)
--+(-1,0)node [neuron] {};
\draw[particle] (1)
--+(0,1)node [neuron] (a1) {};
\draw[particle] (a1)
--+(-1,0)node [neuron] {};
\draw[particle] (a1)
--+(1,0)node [neuron] {};
\draw[particle] (a1)
--+(0,1)node [neuron] {};
\draw[particle] (1)
--+(1,0)node [neuron] (a1){};
\draw[particle] (a1)
--+(1,0)node [neuron] (a2){};
\draw[particle] (a2)
--+(1,0)node [neuron] {};
\draw[particle] (a2)
--+(0,-1)node [neuron] {};
    \end{tikzpicture}
\caption{\label{stretch1}Oriented from 1}
\end{figure}
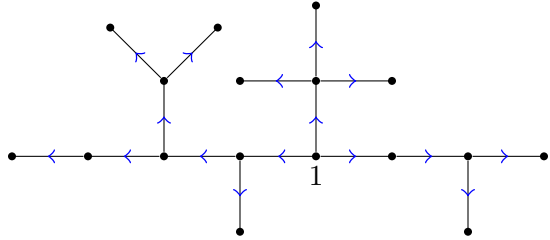
which makes sure that 1 is always the left-most particle in each contributing PT factor in the summation on the RHS of  \eqref{id}. We denote the deformed  Cayley function as
  $C'(\{i'_1,j'_1\},\cdots,\{i'_{n-2},j'_{n-2}\})$ with $\{i',j'\}$ equal to either $\{i,j\}$ or $\{j,i\}$ and count the number of flip pairs as $r_{\rm flip}$. Then these two Cayley functions differ by a overall sign $(-1)^{r_{\rm flip}}$ .
  Thus we expand any Cayley function into  $(n-2)!$  PT factors with an overall sign,
\ba\label{idkk}
&&C(\{i_1,j_1\},\cdots,\{i_{n-2},j_{n-2}\})
=(-1)^{r_{\rm flip}}
C(\{i'_1,j'_1\},\cdots,\{i'_{n-2},j'_{n-2}\})\nl
&=&(-1)^{r_{\rm flip}}\sum_{\substack{\rho\in S_{n-2}\\
\rho^{-1}(i')<\rho^{-1}(j')
}} {\rm PT}(1,\rho(2),\cdots,\rho(n-1),n)\,.
\ea
As we know, these $(n-2)!$ PT factors of KK basis are linearly independent algebraically,
so the rank of all C is $(n{-}2)!$ .

For example, $C(\{1,3\},\{2,3\},\{3,4\})$ on the first two lines of \eqref{cflip} are expanded to KK basis this way,
\ba\label{cflip}
C(\{1,3\},\{2,3\},\{3,4\})&=&-C(\{1,3\},\{3,2\},\{3,4\})\nl
&=&
-{\rm PT}(1,3,2,4,5)-
{\rm PT}(1,3,4,2,5) \,.
\ea
Here is a more example,
\ba
C_6(\{2,3\},\{3,5\},\{3,1\},\{1,4\})&=&
C_6(\{3,2\},\{3,5\},\{1,3\},\{1,4\})\nl
&=&\;\text{PT}(1,3,2,4,5,6)+\text{PT}(1,3,2,5,4,6)\nl
&&+\text{PT}(1,3,4,2,5,6)+\text{PT}(1,3,4,5,2,6)\nl
&&+\text{PT}(1,3,5,2,4,6)+\text{PT}(1,3,5,4,2,6)\nl
&&+\text{PT}(1,4,3,2,5,6)+\text{PT}(1,4,3,5,2,6)\,.
 \ea

\subsection{Interlude: CHY formulas with two distinct Cayley functions}

Here we present another theorem which states that
the Feynman diagrams obtained by the CHY integral of two distinct Cayley functions is just the intersection of those obtained by the CHY integral of
 of Cayley function squared, up to a overall sign which we know how to determine now. The diagrams can be directly obtained by finding all $n-3$ compatible poles of the intersection of their pole sets,
         \begin{theorem}\label{identicalCcp}
      \ba\boxed{
  \int \dif \mu_n C_n C'_n=
  (-1)^f\sum_{\substack{
  s_{I_1},s_{I_2},\cdots,s_{I_{n-3}} \in P(C_n)\cap P(C'_n)\\
~\text{{\rm are}}~  \text{{\rm compatible}~{\rm poles}}
   } }\frac{1}{ s_{I_1}s_{I_2}\cdots s_{I_{n-3}}}\,,}
   \ea
   ${\rm where}$  $
f={\rm flip}({\rho}[1,2,\cdots,n-1]|{\rho}'[1,2,\cdots,n-1])
$
${\rm (which}$  ${\rm comes}$  ${\rm from}$  ${\rm }$  \eqref{mab}${\rm)}$
${\rm will}$  ${\rm be}$  ${\rm described}$  ${\rm in}$  ${\rm a}$  ${\rm moment}$  .
   \end{theorem}

Here we first briefly show that the set of allowed poles on the RHS, which will be denoted as  $P(C_nC'_n)$ , are the intersection of $P(C_n),P(C'_n)$  .
Divide $P(C_n)$ into several subsets $P_m(C_n)$  with $m=2,3,\cdots,n-2$ according to the number of particles of a pole, then
any pole $s_I\in P_m(C)\cap P_m(C')$ must have $m-1$ lines in $C$ and $C'$ using the rule in \cite{Baadsgaard:2015voa}, so it has $2m-2$ lines in $CC'$ and   $s_I\in P_m(CC')$.  Reversely, any   $s_I\in P_m(CC')$, it must have $r$ lines in $C$ and $2m-r$ lines in $C'$. However, $r\leq m-1$ and  $2m-r\leq m-1$ or subcycle appears in $C$ or $C'$. So $r=2m-r=m-1$, {\it i.e.}  $s_I\in P_m(C)\cap P_m(C')$.So
 \ba\label{pccprime}
P_m(CC')=P_m(C)\cap P_m(C')\,,
\ea
thus we have proved the main part of Theorem \ref{identicalCcp}
 .

Now we turn to the  overall sign.
 Note that if we require the orientation of the linking edge is from  $C^1$ to $C^2$ in \eqref{recursion2}, we provides a canonical way to stretch all legs of a Feynman diagram, which gives us a ordering denoted as ${\rho}[1,2,\cdots,n-1]$. So does that of $C'_n$ denoted as  ${\rho}'[1,2,\cdots,n-1]$. Then
 $(-1)^{{\rm flip}({\rho}[1,2,\cdots,n-1]|{\rho}'[1,2,\cdots,n-1])}$  gives the sign in Theorem \ref{identicalCcp}, see the proof in Appendix \ref{appb}.

  For example, take
   \tikzset{
particle/.style={draw=black, postaction={decorate},
    decoration={markings,mark=at position .5 with {\arrow[draw=blue,scale=1.5]{>}}}}
 }
\def  \layersep {.6cm}
 $
C_6=
\raisebox{-.5cm}{

}
$.
Then the allowed poles are given by
\ba\label{exampafaf}
P(C_6C'_6)=
P(C_6)\cap P(C'_6)=\{s_{1,2},s_{3,4},s_{2,3,4},s_{1,2,3,4},s_{2,3,4,5}\}\,.
\ea
All $n-3=3$ compatible pole sets are
\ba\label{wahaha}
&&\Big\{
\{s_{1,2},s_{3,4},s_{1,2,3,4}\},
\{s_{3,4},s_{2,3,4},s_{1,2,3,4}\},
\{s_{3,4},s_{2,3,4},s_{2,3,4,5}\}
\Big\}
\nl
&=&
\Big\{
   \raisebox{-.55cm}{
%
}
\Big\}\,.
\ea
As all Feynman diagrams share the same sign in Theorem \ref{identicalCcp}, we take the first one above as a representative one.
The canonical way to draw it according to the recursion \eqref{recursion2} and the orientation of $C_6$ is
     \raisebox{-1cm}{
 %
  }
, which gives a ordering ${\rho}'[1,2,3,4,5]=(2,1,3,4,5)$. So
  \ba
  f={\rm flip}({\rho}[1,2,3,4,5]|{\rho}'[1,2,3,4,5])=
  {\rm flip}(5,3,4,1,2|2,1,3,4,5)=3\,.
  \ea
 Here we can either count the times whether $(5,3),(3,4),(4,1),(1,2)$
flip in $(2,1,3,4,5)$ or count  the times whether $(2,1),(2,3),(3,4),(4,5)$
flip in $(5,3,4,1,2)$.

Thus
\ba
  \int \dif \mu_6 \!\!
  \raisebox{-.5cm}{
%
}.
\ea

\begin{figure}[!htb]
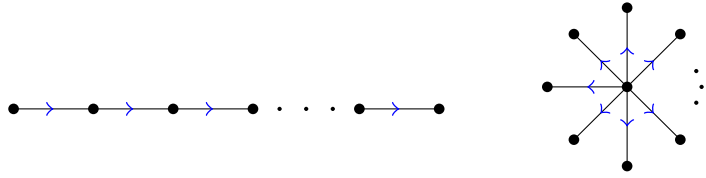

    \centering
    \subfloat{
   %
}
\caption{\label{S graph2}Natural orientation for $C_n^H,C_n^S$}
\end{figure}

When one Cayley function in Theorem \ref{identicalCcp} is a PT factor, it picks out all Feynman diagrams of the other Cayley function compatible with ordering. For example the CHY formula of a
Hamiton graph and a star graph is given by
\def\layersep{.7cm}
 \tikzset{
particle/.style={draw=black, postaction={decorate},
    decoration={markings,mark=at position .5 with {\arrow[draw=blue,scale=1.5]{>}}}}
 }
\ba
\int \dif \mu_n
{\rm PT}(\alpha,i,\beta,n)C_n^{S}(i)=(-1)^{|\a|}\sum_{\rho\in \a^{-1} \sqcup\! \sqcup \b }
\raisebox{-0.4cm}{
 %
    }
\ea
here we take the natural orientation of a
Hamiton graph and a star graph, see figure \ref{S graph2}.
$\a^{-1}$ is the reverse list of $\a$. $\a^{-1} {\sqcup\! \sqcup} \b $ are the permutations with the ordering of particles from  $\a^{-1}$  and $\b$ unchanged respectively. For example,
\ba
\int \dif \mu_6
{\rm PT}(3,2,4,1,5,6)C_6^{S}(4)
&=&
  \raisebox{-.55cm}{
%
}\,.
\nl
\ea
When two Cayler function in Theorem \ref{identicalCcp} are PT factors, it comes back to double amplitude  \eqref{mab}.

Here is another typical CHY formula of two star graphs with different center points,
 \ba\label{starstar}
 \int \dif \mu_n
C_n^{S}(i) C_n^{S}(j)
=-\sum
\raisebox{-1cm}{
 %
}
\ea

The CHY integral of any two arbitrary star graphs also has a closed formula, see
Appendix \ref{appc}.

As all the Feynman diagrams on the RHS of Theorem \ref{identicalCcp} share the same sign, it has a geometry description which is the intersection of the polytopes mapped from $C_n, C'_n$, see an example in figure \ref{intersectionpoly}.

\begin{figure}[!htb]
\centering
  %
}
  --++(0:2) coordinate (4)
  --++(-60:2) coordinate (5)
  --++(-120:2) coordinate (6)
  --+(180:2) ;

\draw [line width=.03cm,blue] ($(1)+(-135:.15)$)--($(2)+(180:.15)$)
--($(3)+(135:.15)$)--($(4)+(2,0)+(45:.15)$)
--($(6)+(2,0)+(-45:.15)$)--($(1)+(-135:.15)$);

 \end{tikzpicture}
 \caption{\label{intersectionpoly}
 Intersection of polytopes mapped from ${\rm PT}(1,2,3,4,5)$ and $C^S_5(2)$
 }
  gives the CHY intergral of them
\end{figure}

\subsection{A new basis of Cayley functions}

As shown in \eqref{idkk}, KK basis provides a basis for the space of all Cayley functions. However, we are also interested in a new basis with elements that have a special property. Given a Parke-Taylor factor, we would like the CHY formula of PT and an element to give either a {\it single} Feynamn diagram or {\it zero}.
In the study of Z integrals \cite{Mafra:2016mcc}, the authors have proposed an algorithm for constructing an alternative basis of rational functions of $n$ punctures, which we believe should be the same as our new basis. We have checked explicitly that up to $n=8$ they coincide and we leave it to a future work to show this for all multiplicities. Below we first present the basis for all $n$, and then study its applications in both CHY and disk integrals. Without loss of generality, we choose ${\rm PT}(1,2,\cdots,n)$ and align the particles in the labelled tree in this ordering.

The new basis are obtained recursively using the map $M$ defined as following:

\begin{enumerate}
\item $M$ maps an {\color {blue}ordered} particle label set to a connected subgraph set.
\item As starting point,  $M(\{i\})=\{
\raisebox{-.5cm}{
\begin{tikzpicture}[scale=.5]
\fill (0,0) circle (0.05);
\node at (0,-.5) {$i$};
\end{tikzpicture}}
\}$.
\item The map is defined recursively, via the function $\Lambda$,
\ba\label{MP}
M(\{i_1,i_2,\cdots,i_k\})
=\!\!\!\!\!\!\bigsqcup\limits_{\substack{{I_1\sqcup I_2 \sqcup\cdots \sqcup I_r= \{i_2,\cdots,i_k\}}
}}\!\!\!\!\!\! \Lambda_{i_1}(M(I_1)\otimes M(I_2)\otimes\cdots \otimes M(I_r)).\quad
\ea
Here  $i_1<i_2<\cdots<i_k$ and we always pick out the {\color{blue} left-most} particle $i_1$ as the starting point to drawing lines and divide the remaining sequence into all possible disjoint sets $I_1, I_2 ,\cdots , I_r$ with $r=1,\cdots,k-1$.
$\otimes$ means direct product and
$M(I_1)\otimes M(I_2)\otimes\cdots \otimes M(I_r)$ is a set of disajoint subgraphs with $r$ parts. What $\Lambda_{i_1}$ does for each   non-connected subgraph  is to draw a line from $i_1$ to    the {\color{blue}{right-most}} particle label of each connected part respectively. So
$\Lambda_{i_1}$ actually acts on each elements of $M(I_1),M(I_2),\cdots,M(I_r)$ respectively as shown below,
\ba
\Lambda_{i_1}(M(I_1)\otimes\cdots \otimes M(I_r))
 =
 \{
  \raisebox{-.3cm}{
\begin{tikzpicture}[scale=.8]

\draw (0,0) .. controls (.2,.1) .. (1,0);
\draw (0,0) .. controls (.3,.3) .. (2,0);
\draw (0,0) .. controls (.5,.7) .. (4,0);

\fill (2.7,0) circle (0.03);
\fill (3,0) circle (0.03);
\fill (3.3,0) circle (0.03);

\node at (0,-.2){$i_1$};
\node at (.8,-.2){$m_1$};
\node at (1.8,-.2){$m_2$};
\node at (3.8,-.2){$m_r$};

\end{tikzpicture}}
|
 m_1 \in  M(I_1),\cdots,
 m_r \in  M(I_r)
 \}\,.
 \nl
  \ea
Note that $m_1,\cdots,m_r$ are connected subgraphs and they are linked to $i_1$ from their right-most point.
\end{enumerate}
 There is always a trivial line $\{1,n-1\}$ in each element of $M(\{1,2,\cdots,n-1\})$
as 1 is always the minimum particle in its particle set and $n-1$ is always the right-most particle point of its subgraph.
 Sometimes we  draw a dashed line instead for later convenience.

For example
\ba\label{3ptbasis}
M(\{1,2\})=\Lambda_1(M(\{2\}))=\{
  \raisebox{-.3cm}{
\begin{tikzpicture}[scale=.8]

\draw (0,0) .. controls (.2,.1) .. (1,0);

\node at (0,-.3){$1$};
\node at (1,-.3){$2$};

\fill (0,0) circle (0.03);
\fill (1,0) circle (0.03);

\end{tikzpicture}}
 \}\,.
\ea

\ba\label{4ptbasis}
M(\{1,2,3\})
&=&
\Lambda_1(M(\{2,3\}))\bigsqcup \Lambda_1(M(\{2\})\otimes M(\{3\}))
\nl
&=&
\{
  \raisebox{-.3cm}{
\begin{tikzpicture}[scale=.8]

\draw (0,0) .. controls (.2,.1) .. (1,0);
\draw (-1,0) .. controls (-.5,.4) .. (1,0);

\node at (-1,-.3){$1$};
\node at (0,-.3){$2$};
\node at (1,-.3){$3$};

\fill (0,0) circle (0.03);
\fill (1,0) circle (0.03);
\fill (-1,0) circle (0.03);

\end{tikzpicture}}
,
  \raisebox{-.3cm}{
\begin{tikzpicture}[scale=.8]

\draw (-1,0) .. controls (-.8,.1) .. (0,0);
\draw (-1,0) .. controls (-.5,.4) .. (1,0);

\node at (-1,-.3){$1$};
\node at (0,-.3){$2$};
\node at (1,-.3){$3$};

\fill (0,0) circle (0.03);
\fill (1,0) circle (0.03);
\fill (-1,0) circle (0.03);

\end{tikzpicture}}
 \}\,.
\ea
Here the right-most particle label of $M(\{2,3\})$ is $3$, so we draw a line from 1 to 3.

\ba
M(\{1,2,3,4\})
&=&
\Lambda_1(M(\{2,3,4\}))\bigsqcup \Lambda_1(M(\{2,3\})\otimes M(\{4\}))\bigsqcup \Lambda_1(M(\{2\})\otimes M(\{3,4\}))
\nl
&&
\bigsqcup \Lambda_1(M(\{3\})\otimes M(\{2,4\}))
\bigsqcup \Lambda_1(M(\{2\})\otimes M(\{3\})\otimes M(\{4\}))\,,
\ea
where $\Lambda_1$ acts on the two graphs of $M(\{2,3,4\})$ respectively
\ba\label{p1m}
\Lambda_1(M(\{2,3,4\}))
=
\{
  \raisebox{-.3cm}{
\begin{tikzpicture}[scale=.8]

\draw (0,0) .. controls (.2,.1) .. (1,0);
\draw (-1,0) .. controls (-.5,.4) .. (1,0);
\draw (-2,0) .. controls (-1.2,.8) .. (1,0);
\node at (-2,-.3){$1$};
\fill (-2,0) circle (0.03);

\node at (-1,-.3){$2$};
\node at (0,-.3){$3$};
\node at (1,-.3){$4$};

\fill (0,0) circle (0.03);
\fill (1,0) circle (0.03);
\fill (-1,0) circle (0.03);

\end{tikzpicture}}
,
  \raisebox{-.3cm}{
\begin{tikzpicture}[scale=.8]

\draw (-1,0) .. controls (-.8,.1) .. (0,0);
\draw (-1,0) .. controls (-.5,.4) .. (1,0);
\draw (-2,0) .. controls (-1.2,.8) .. (1,0);
\node at (-2,-.3){$1$};
\fill (-2,0) circle (0.03);

\node at (-1,-.3){$2$};
\node at (0,-.3){$3$};
\node at (1,-.3){$4$};

\fill (0,0) circle (0.03);
\fill (1,0) circle (0.03);
\fill (-1,0) circle (0.03);

\end{tikzpicture}}
 \}\,,
\ea
and crossing lines come out because of the non consecutive sequence $\{2,4\}$,
\ba\label{p13}
\Lambda_1(M(\{3\}\otimes  M(\{2,4\}))=\{
  \raisebox{-.3cm}{
\begin{tikzpicture}[scale=.8]

\draw (0,0) .. controls (.3,.3) .. (2,0);
\draw (1,0) .. controls (1.3,.3) .. (3,0);
\draw (0,0) .. controls (.5,.8) .. (3,0);

\node at (0,-.2){$1$};
\node at (1,-.2){$2$};
\node at (2,-.2){$3$};
\node at (3,-.2){$4$};

\fill (0,0) circle (0.03);
\fill (1,0) circle (0.03);
\fill (2,0) circle (0.03);
\fill (3,0) circle (0.03);

\end{tikzpicture}}
 \}\,.
\ea
So
\ba \label{5ptbasis}
M(\{1,2,3,4\})
&=&
\{
 \raisebox{-.3cm}{
\begin{tikzpicture}[scale=.8]

\draw (0,0) .. controls (.2,.1) .. (1,0);
\draw (-1,0) .. controls (-.5,.4) .. (1,0);
\draw[>=stealth,  thick,dashed] (-2,0) to [bend left] (1,0) ;
\node at (-2,-.3){$1$};
\fill (-2,0) circle (0.03);

\node at (-1,-.3){$2$};
\node at (0,-.3){$3$};
\node at (1,-.3){$4$};

\fill (0,0) circle (0.03);
\fill (1,0) circle (0.03);
\fill (-1,0) circle (0.03);

\end{tikzpicture}}
,
  \raisebox{-.3cm}{
\begin{tikzpicture}[scale=.8]

\draw (-1,0) .. controls (-.8,.1) .. (0,0);
\draw (-1,0) .. controls (-.5,.4) .. (1,0);
\draw[>=stealth,  thick,dashed] (-2,0) to [bend left] (1,0) ;
\node at (-2,-.3){$1$};
\fill (-2,0) circle (0.03);

\node at (-1,-.3){$2$};
\node at (0,-.3){$3$};
\node at (1,-.3){$4$};

\fill (0,0) circle (0.03);
\fill (1,0) circle (0.03);
\fill (-1,0) circle (0.03);

\end{tikzpicture}}
,
\raisebox{-.3cm}{
\begin{tikzpicture}[scale=.8]

\draw (0,0) .. controls (.2,.1) .. (1,0);
\draw (-1,0) .. controls (.3,.3) .. (1,0);
\draw[>=stealth,  thick,dashed] (-1,0) to [bend left] (2,0) ;
\node at (2,-.3){$4$};
\fill (2,0) circle (0.03);

\node at (-1,-.3){$1$};
\node at (0,-.3){$2$};
\node at (1,-.3){$3$};

\fill (0,0) circle (0.03);
\fill (1,0) circle (0.03);
\fill (-1,0) circle (0.03);

\end{tikzpicture}}
,
\nl
&&\,\;
\raisebox{-.3cm}{
\begin{tikzpicture}[scale=.8]

\draw (0,0) .. controls (.4,.1) .. (1,0);
\draw (2,0) .. controls (2.2,.1) .. (3,0);
\draw[>=stealth,  thick,dashed] (0,0) to [bend left] (3,0) ;
\node at (0,-.2){$1$};
\node at (1,-.2){$2$};
\node at (2,-.2){$3$};
\node at (3,-.2){$4$};

\fill (0,0) circle (0.03);
\fill (1,0) circle (0.03);
\fill (2,0) circle (0.03);
\fill (3,0) circle (0.03);

\end{tikzpicture}}
,
  \raisebox{-.3cm}{
\begin{tikzpicture}[scale=.8]

\draw (0,0) .. controls (1,.3) .. (2,0);
\draw (1,0) .. controls (2,.3) .. (3,0);
\draw[>=stealth,  thick,dashed] (0,0) to [bend left] (3,0) ;
\node at (0,-.2){$1$};
\node at (1,-.2){$2$};
\node at (2,-.2){$3$};
\node at (3,-.2){$4$};

\fill (0,0) circle (0.03);
\fill (1,0) circle (0.03);
\fill (2,0) circle (0.03);
\fill (3,0) circle (0.03);

\end{tikzpicture}}
,
 \raisebox{-.3cm}{
\begin{tikzpicture}[scale=.8]

\draw (0,0) .. controls (-.2,.1) .. (-1,0);
\draw (1,0) .. controls (.5,.4) .. (-1,0);
\draw[>=stealth,  thick,dashed] (-1,0) to [bend left] (2,0) ;
\node at (2,-.3){$4$};
\fill (2,0) circle (0.03);

\node at (1,-.3){$3$};
\node at (0,-.3){$2$};
\node at (-1,-.3){$1$};

\fill (0,0) circle (0.03);
\fill (1,0) circle (0.03);
\fill (-1,0) circle (0.03);

\end{tikzpicture}}
\}\,.
\ea

There are
 \href{https://en.wikipedia.org/wiki/Stirling_number}
{Stirling number of the second kind }
of terms in the union in \eqref{MP} .
 So, using recursion, one can
 easily prove that
 $|M(\{i_1,\cdots,i_k\})|=(k-1)!$
 ,
 {\it i.e.},
 there are $(k-1)!$ connected subgraphs in $M(\{i_1,\cdots,i_k\})$ .
Thus there  $(n-2)!$ Cayley functions in  $M(\{1,2,\cdots,n-1\})$ and we believe they compose a new set of basis  denoted as $C^{\rm basis}$, which we have checked  up to 10pts and are the same as those in \cite{Mafra:2016mcc} up to 8pts. Using the recursion above and the transition rule \eqref{idkk}, a proof based on direct inspection should be straightforward.

\begin{figure}[!htb]
\centering
\subfloat[$K_1$]{
\includegraphics[width=0.17\textwidth]{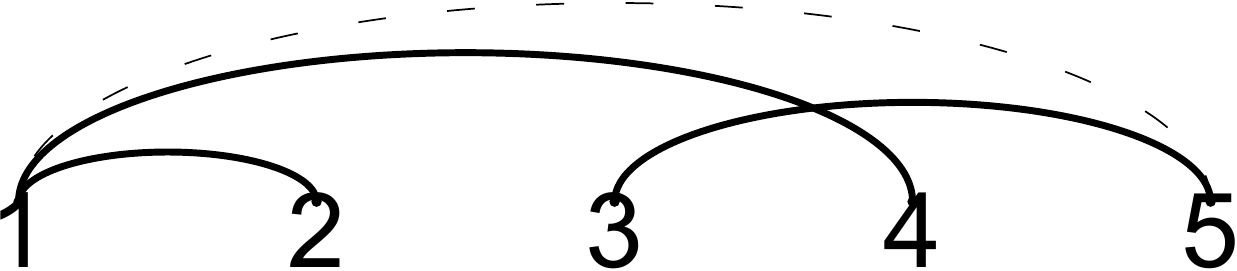}}
\subfloat[$K_2$]{
\includegraphics[width=0.17\textwidth]{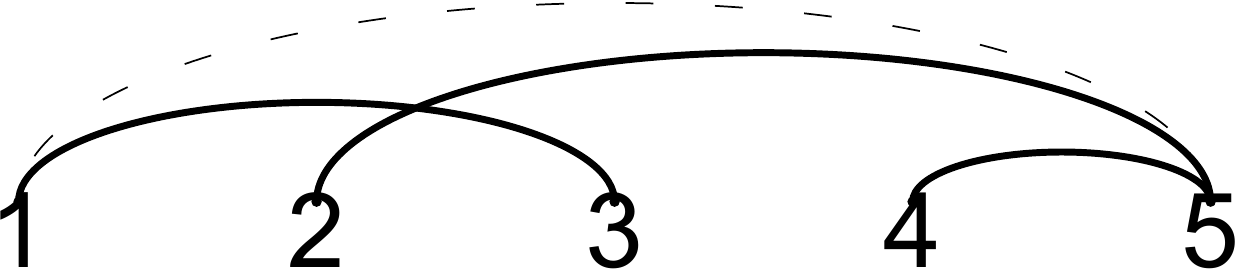}}
\subfloat[$K_3$]{
\includegraphics[width=0.17\textwidth]{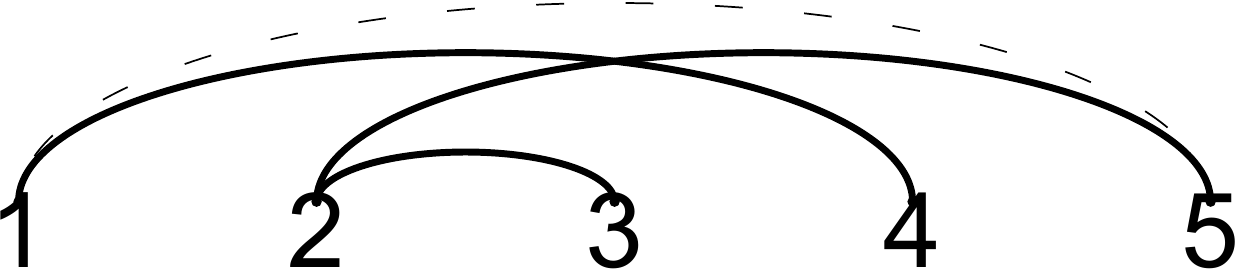}}
\subfloat[$K_4$]{
\includegraphics[width=0.17\textwidth]{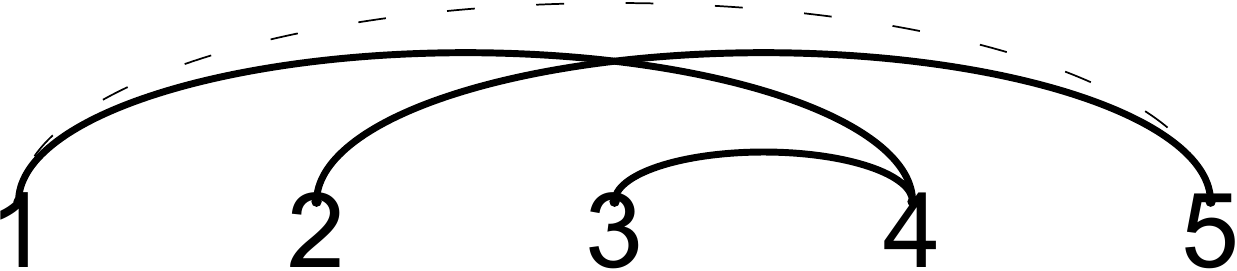}}
\subfloat[$K_5$]{
\includegraphics[width=0.17\textwidth]{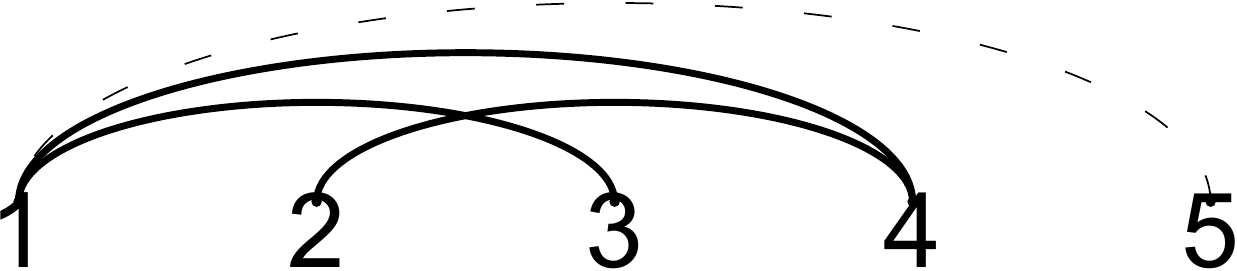}}

\subfloat[$K_6$]{
\includegraphics[width=0.17\textwidth]{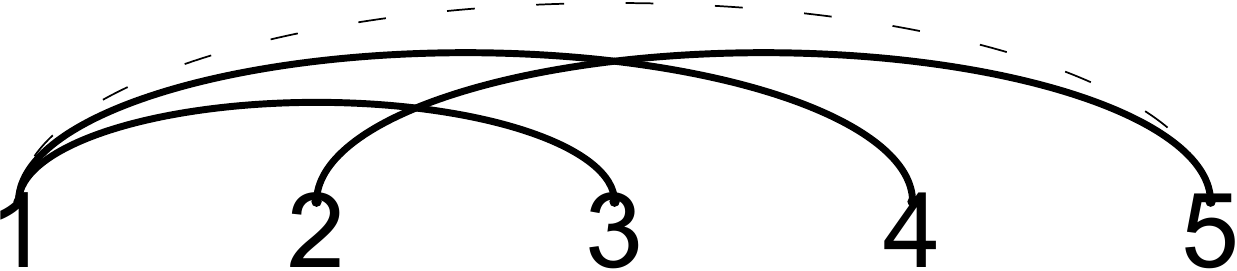}}
\subfloat[$K_7$]{
\includegraphics[width=0.17\textwidth]{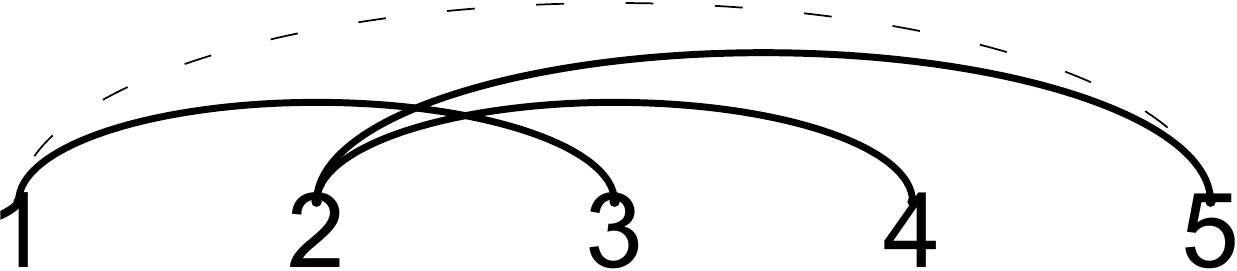}}
\subfloat[$K_8$]{
\includegraphics[width=0.17\textwidth]{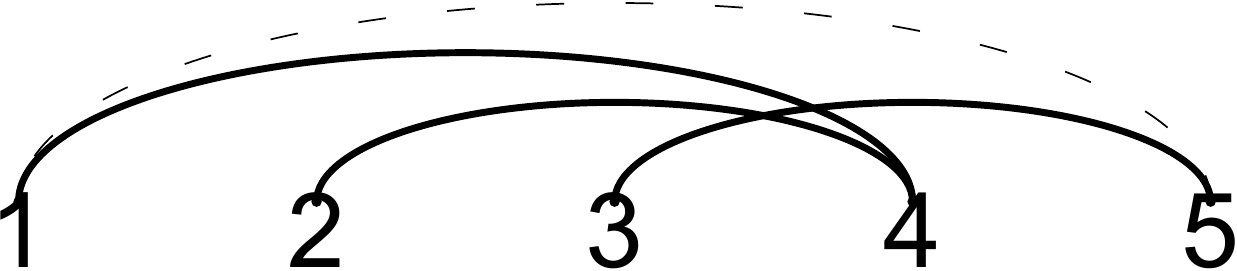}}
\subfloat[$K_9$]{
\includegraphics[width=0.17\textwidth]{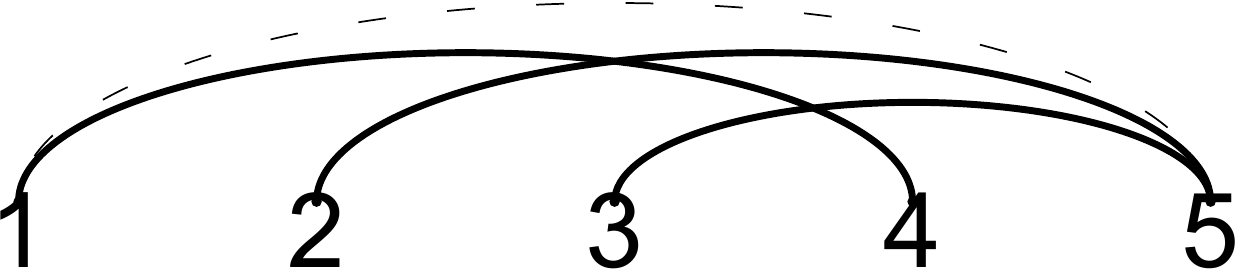}}
\subfloat[$K_{10}$]{
\includegraphics[width=0.17\textwidth]{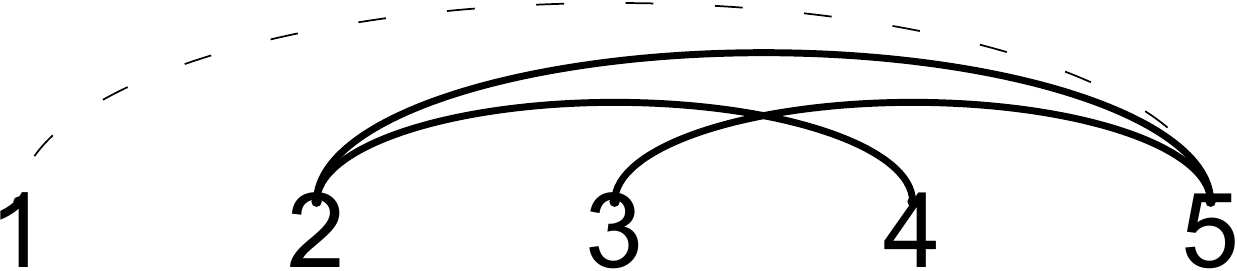}}
\caption{\label{kernel6} 10 $C^{\rm kernel}$ for 6pt}
\end{figure}

For example, the basis of 3,4,5pt have been shown in \eqref{3ptbasis},\eqref{4ptbasis},\eqref{5ptbasis}.
There are 24 basis in 6pt
, 10 of which  have crossing lines as shown in figure \ref{kernel6}.

As it turns out, there are ${\rm Cat}_{n-3}$ elements without crossing lines denoted as $C^{\rm single}$ and $(n-2)!-{\rm Cat}_{n-3}$ elements with crossing lines $C^{\rm kernel}$. Now we show, with ${\rm PT}(1,2,\cdots,n)$, how the CHY integral of any $C^{\rm single}$ gives a single Feynman diagram and why  the CHY integral of any $C^{\rm kernel}$ gives zero.

\subsubsection*{Elements for a single graph}
If we restrict the union in \eqref{MP} with a additional rule that $I_1,\cdots,I_r$ must be consecutive sequences and denote this new map as $M^{\rm single}$, then
 there are no crossing lines coming out ,see \eqref{p13}, and $M^{\rm single}(\{1,2,\cdots,n-1\})$ gives all $C^{\rm single}$. Equivalently, we can obtain all  $C^{\rm single}$ in one step: any $n-2$ pairs $\{i_a,j_a\}$ with $i_a< j_a$ which are interval mutually compatible ,{\it i.e.} either $[i_a,j_a]\subset [i_b,j_b] $ or $[i_b,j_b]\subset [i_a,j_a] $ or $[i_a,j_a]\cup [i_b,j_b]=\emptyset$, corresponds to a $C^{\rm single}$.
Obviously, there is always a line $\{1,n-1\}$ and there are ${\rm Cat}_{n-2}$ of $C^{\rm single}$
. Ignoring this trivial line, we can read out the single Feynman diagram directly from the left $n-3$ lines ,
\ba\label{singlept}
\boxed{
\int \dif \mu_n {\rm PT}(1,\cdots,n) C^{\rm single}
(\{i_1,j_1\},\cdots,\{i_{n-3},j_{n-3}\},\{1,n\!-\!1\})=\frac{1}{s_{i_1,\cdots,j_1}\cdots s_{i_{n-3},\cdots,j_{n-3}}}}\,.\nl
\ea
Here $s_{i,\cdots,j}$ is the abbreviation of cyclic pole $s_{i,i+1,\cdots,j}$.
For example
\ba\label{cpt}
\int\dif \mu_5 {\rm PT}(1,\cdots,5)\frac{1}{\sigma _{1,2} \sigma _{1,3}\sigma _{1,4}}
&=&
\frac{1}{s _{1,2} s _{1,2,3}}\,,
\nl
\int\dif \mu_6 {\rm PT}(1,\cdots,6)\frac{1}{\sigma _{1,3} \sigma _{2,3}\sigma _{4,5}\sigma _{1,5}}
&=&
\frac{1}{s _{1,2,3} s _{2,3}s _{4,5}}\,.
\ea

Now we give a brief proof of \eqref{singlept} using \eqref{pccprime}. Thanks to ${\rm PT}(1,\cdots,n)$, we only need to consider the  cyclic poles of $C^{\rm single}$. Each pair $\{i,j\}\in \{\{i_1,j_1\},\cdots,\{i_{n-3},j_{n-3}\}\}$ corresponds to a connected subgraph which is made up of all lines $\{i_r,j_r\}$ with $i\leq i_r<j_r \leq j$. Note that all points $i,i+1,\cdots,j$ locate in this connected subgraph as their are no crossing lines in $C^{\rm single}$, so this subgraph corresponds to a
Pole $s_{i,i+1,\cdots,j}$. So we obtain $n-3$ allowed poles as shown in \eqref{singlept}.
While for any pair $\{i',j'\}\notin \{\{i_1,j_1\},\cdots,\{i_{n-3},j_{n-3}\}\}$ with $i'<j'$, there are no connected line from $i'$ to $j'$ restrained in the region $[i',j']$ or non mutually compatible lines seen in Figure \ref{twosides} appear, let alone a connected subgraph contained $i',i'+1,\cdots,j'$ located in $[i',j']$. So $s_{i',i'+1,\cdots,j'}$ is forbidden and no more poles comes out. Obviously, $s_{i_1,\cdots,j_1},\cdots,s_{i_{n-3},\cdots,j_{n-3}}$ are compatible each other  and these give the cubic graph
 shown in \eqref{singlept}.

As known that the CHY integral of two ${\rm PT}(1,\cdots,n)$ gives ${\rm Cat}_{n-2}$ of planar Feynman diagrams. Now we translate each planar cubic graph
 to a $C^{\rm single}$, which is consistent to the following identity,
\ba
{\rm PT}(1,2,\cdots,n)= \sum^{{\rm Cat}_{n-2}} C^{\rm single}
(\{i_1,j_1\},\cdots,\{i_{n-3},j_{n-3}\},\{1,n-1\})\,,
\ea
with the gauge  fixing $\s_1\rightarrow 0,\s_{n-1}\rightarrow 1,\s_n\rightarrow \infty$. Here we sum over all ${\rm Cat}_{n-2}$ $C^{\rm single}$.
 In \cite{Arkani-Hamedsonghe}, we will see that this identity can be interpreted as a triangulation of the associahedron into ${\rm Cat}_{n-2}$ simplices.

Eq. \eqref{singlept} is a very clean identity, using which, reversely, we can  translate any cubic Feynman diagram to CHY integral directly.
For example, given a cubic Feynman diagram,
\raisebox{-1cm}{
\begin{tikzpicture}[scale=.5]
  \draw (0,0) node [left=0pt] {\tiny 1}
  --++(1,0) coordinate (a) -- +(0,1) node [above=0pt] {\tiny 2};
  \draw (a)--++(1,0) coordinate (b) --++(0,1)coordinate (c) --++(0,1)coordinate (d)--+(0,1)node [above=0pt]{\tiny 3};
  \draw (c)--+(.5,0) node [right=-3pt]{\tiny 5};
  \draw (d)--+(.5,0) node [right=-3pt]{\tiny 4};
  \draw (b)--++(1.1,0) coordinate (e)--+(1,0) node [right=0pt] {\tiny 8};
  \draw (e)--++(0,1) coordinate (f)--+(0,1) node [above=0pt] {\tiny 6};
   \draw (f)--+(.5,0) node [right=0pt]{\tiny 7};
\end{tikzpicture}}
 ~(here without loss of general, we let the particle labels are $1,2,\cdots,8$ as other cases are just relabelling), as the poles are $s_{1,2},s_{3,4},s_{3,4,5},s_{1,2,3,4,5},s_{6,7}$, the pairs in Cayley function we need are $\{1,2\},\{3,4\},\{3,5\},\{1,5\},\{6,7\}$. Thus the full CHY formula for this Feynman diagram is
 \ba
\raisebox{-1cm}{
\begin{tikzpicture}[scale=.5]
  \draw (0,0) node [left=0pt] {\tiny 1}
  --++(1,0) coordinate (a) -- +(0,1) node [above=0pt] {\tiny 2};
  \draw (a)--++(1,0) coordinate (b) --++(0,1)coordinate (c) --++(0,1)coordinate (d)--+(0,1)node [above=0pt]{\tiny 3};
  \draw (c)--+(.5,0) node [right=-3pt]{\tiny 5};
  \draw (d)--+(.5,0) node [right=-3pt]{\tiny 4};
  \draw (b)--++(1.1,0) coordinate (e)--+(1,0) node [right=0pt] {\tiny 8};
  \draw (e)--++(0,1) coordinate (f)--+(0,1) node [above=0pt] {\tiny 6};
   \draw (f)--+(.5,0) node [right=0pt]{\tiny 7};
\end{tikzpicture}}
\!\!\!=\!
\int \dif \mu_8 {\rm PT}(1,2,\cdots,8) C(\{1,2\},\{3,4\},\{3,5\},\{1,5\},\{6,7\},\{1,7\}).\qquad
 \ea

Last but not least, we briefly comment on a corollary of  \eqref{singlept}, namely it can be used to give a large class of CHY formulas for $\phi^p$ graphs . The idea is that one can blow up any $\phi^p$ graph to a cubic graph, which can be translated into a formula via \eqref{singlept}, and the result is given by further  multiplying with those additional inverse propagators. There are many ways of blowing up the $\phi^p$ graph, and any way of doing it gives such a formula. For example, we can write
 $
 \raisebox{-.6cm}{
\begin{tikzpicture}[scale=.4]
  \draw (-1,0) node [left =-2pt]{\tiny 2} --(1,0);
  \draw (0,-1) node [below =-2pt]{\tiny 1} --(0,1)node [above =-2pt]{\tiny 3};
\end{tikzpicture}}
=s_{1,2}
 \raisebox{-.6cm}{
\begin{tikzpicture}[scale=.4]
  \draw (-1,0) node [left =-2pt]{\tiny 2} --(1,0);
  \draw (-.2,-1) node [below =-2pt]{\tiny 1}--(-.2,0);
   \draw (.2,0)--(.2,1)node [above =-2pt]{\tiny 3};
\end{tikzpicture}}
$,
$
 \raisebox{-.6cm}{
\begin{tikzpicture}[scale=.4]
  \draw (0,0)--(-72:1) node [below =-2pt]{\tiny 1} ;
 \draw (0,0)--(-144:1) node [left =-2pt]{\tiny 2} ;
  \draw (0,0)--(144:1) node [left =-2pt]{\tiny 3} ;
   \draw (0,0)--(72:1) node [above =-2pt]{\tiny 4} ;
   \draw (0,0)--(1.2,0);
\end{tikzpicture}}
=
s_{1,2}s_{1,2,3}
\raisebox{-.4cm}{
\begin{tikzpicture}[scale=.4]
  \draw (0,0)node [left=-2pt] {\tiny 1}
  --++(1,0) coordinate (a) --+(0,1) node [above =-2pt] {\tiny 2};
  \draw (a)  --++(.4,0) coordinate (b) --+(0,1) node [above =-2pt] {\tiny 3};
  \draw (b)  --++(.4,0) coordinate (c) --+(0,1) node [above =-2pt] {\tiny 4};
  \draw (c)--+(1,0);
\end{tikzpicture}}
$ . To illustrate the method, we consider a 10-pt $\phi^4$ Feynman diagram,
 \raisebox{-1cm}{
\begin{tikzpicture}[scale=.6]
  \draw (0,0) coordinate (a)--++(1,0) coordinate (b)--+(1,0) node [right=-2pt]{\tiny 9};
  \draw (b)--+(0,-1) node [below=-2pt] {\tiny 10};
  \draw (b)--++(0,1) coordinate (c)--+(0,.5) node [above=-2pt] {\tiny 5};
  \draw (c)--+(-.5,0) node [left=-2pt] {\tiny 4};
\draw (c)--++(.5,0)coordinate (d)--+(.5,0)  node [right=-2pt] {\tiny 7};
\draw (d)--+(0,.5) node [above=-2pt] {\tiny 6};
\draw (d)--+(0,-.5) node [below=-2pt] {\tiny 8};
\draw (a)--+(-.5,0) node [left=-2pt] {\tiny 2};
\draw (a)--+(0,-.5) node [below=-2pt] {\tiny 1};
\draw (a)--+(0,.5) node [above=-2pt] {\tiny 3};
\end{tikzpicture} }
; one way to rewrite it is
$s_{1,2}s_{4,5}s_{6,7}s_{9,10}
\raisebox{-1cm}{
\begin{tikzpicture}[scale=.6]
  \draw (0,0) coordinate (a)--++(1,0) coordinate (b);
  \draw ($(b)+(0,-.2)$)--+(1,0) node [right=-2pt]{\tiny 9};
  \draw (b)--+(0,-1) node [below=-2pt] {\tiny 10};
  \draw (b)--++(0,1) coordinate (c)--+(0,.5) node [above=-2pt] {\tiny 5};
  \draw ($(c)+(0,.2)$)--+(-.5,0) node [left=-2pt] {\tiny 4};
\draw (c)--++(.5,0)coordinate (d)--+(.5,0)  node [right=-2pt] {\tiny 7};
\draw (d)--+(0,.5) node [above=-2pt] {\tiny 6};
\draw ($(d)+(-.2,0)$)--+(0,-.5) node [below=-2pt] {\tiny 8};
\draw (a)--+(-.5,0) node [left=-2pt] {\tiny 2};
\draw ($(a)+(-.2,0)$)--+(0,-.5) node [below=-2pt] {\tiny 1};
\draw (a)--+(0,.5) node [above=-2pt] {\tiny 3};
\end{tikzpicture} }$
, and by \eqref{singlept} we obtain its CHY formula as
 \ba
\raisebox{-1cm}{
\begin{tikzpicture}[scale=.6]
  \draw (0,0) coordinate (a)--++(1,0) coordinate (b)--+(1,0) node [right=-2pt]{\tiny 9};
  \draw (b)--+(0,-1) node [below=-2pt] {\tiny 10};
  \draw (b)--++(0,1) coordinate (c)--+(0,.5) node [above=-2pt] {\tiny 5};
  \draw (c)--+(-.5,0) node [left=-2pt] {\tiny 4};
\draw (c)--++(.5,0)coordinate (d)--+(.5,0)  node [right=-2pt] {\tiny 7};
\draw (d)--+(0,.5) node [above=-2pt] {\tiny 6};
\draw (d)--+(0,-.5) node [below=-2pt] {\tiny 8};
\draw (a)--+(-.5,0) node [left=-2pt] {\tiny 2};
\draw (a)--+(0,-.5) node [below=-2pt] {\tiny 1};
\draw (a)--+(0,.5) node [above=-2pt] {\tiny 3};
\end{tikzpicture} }
&=&
\int \dif \mu_{10} {\rm PT}(1,2,\cdots,10)\, s_{1,2}s_{4,5}s_{6,7}s_{9,10}
\nl&&
\times C(\{1,2\},\{1,3\},\{4,5\},\{4,8\},\{6,7\},\{6,8\},\{1,8\},\{1,9\}).
\ea
In this way, we find a large class of simple CHY integrands for any $\phi^p$ graph (and  \cite{Baadsgaard:2015ifa} corresponds to a symmetrized version ; see also \cite{Cachazo:2014xea,Baadsgaard:2016fel} for other methods ).

\subsubsection*{Elements in the kernel }
\begin{figure}[!htb]
\centering
\begin{tikzpicture}[scale=1.]
\fill (0,0) circle (0.03);
\fill (1,0) circle (0.03);
\fill (1.4,0) circle (0.03);
\fill (2.5,0) circle (0.03);
\draw (0,0) arc (180:0:.7);
\draw (1,0) arc (180:0:.75);
\node at (0,-.3) {$i$};
\node at (1,-.3) {$j$};
\node at (1.4,-.3) {$k$};
\node at (2.5,-.3) {$l$};
\end{tikzpicture}
\caption{\label{overlap} Crossing lines}
\end{figure}
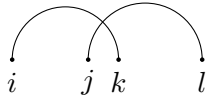

Now we move to the second kind, $C^{\rm kernel}$ and prove they produce zero in its CHY formula with a PT factor briefly.
For any $C^{\rm kernel}$, we pick out two lines $\{i,k\}$ and $\{j,l\}$ which are crossing each other, as shown in figure \ref{overlap}.

Because of the construction way of $M$, there are no connected line from $j$ to $i$ without passing $l$ or
connected line from $k$ to $l$ without passing $i$.
An immediate observation is that any $s_{A\cup \{k\}}$ or $s_{A\cup \{j\}}$  with $k,j\notin A$ is a non-planar pole and can't appear in the CHY integral. The only possible way for $j,k$ appearing in a pole is  $s_{I\cup \{j,k\}}$  with $i,l\in I$. Before using the lines $\{i,k\}$ and $\{j,l\}$, the other $n-4$ lines can only provide $n-5$ compatible poles at most, denoted as $\{\cdots, s_{I_1},s_{I_2},\cdots,s_{I_r}\}$ with $I_1\subset I_2 \cdots \subset I_r$.  Now we consider to use the two lines $\{i,k\}$ and $\{j,l\}$. However, there is at most one more compatible pole coming out denoted as
$s_{I_t\cup\{j,k\}}$ and the $n-4$ compatible pole set becomes $\{\cdots, s_{I_1},\cdots,s_{I_t},s_{I_t\cup\{j,k\}},s_{I_{t+1}\cup\{j,k\}},\cdots,s_{I_r\cup\{j,k\}}\}$. So there are no Feynman diagrams coming out.

\subsubsection*{More about the new basis}

Owing to the clear property of $C^{\rm single }$ and $C^{\rm kernel }$, we can't wait to expand any Cayley functions or even general CHY half integrand (without subcycle about $\s$) to these basis.  For example,
\ba
C(\{1,2\},\{2,3\},\{1,4\})&=&C(\{1,2\},\{1,3\},\{1,4\})
+C(\{1,3\},\{2,3\},\{1,4\})\,,\nl
C(\{1,3\},\{2,3\},\{2,4\},\{1,5\})&=&C(\{1,3\},\{2,3\},
\{1,4\},\{1,5\})
+C(\{2,3\},\{1,4\},\{2,4\},\{1,5\})\,,
\nl&&
-C(\{1,3\},\{1,4\},\{2,4\},\{1,5\})\,,
\nl
C(\{1,3\},\{1,2\},\{2,4\},\{2,5\})&=&
C(\{1,2\},\{1,3\},\{1,4\},\{1,5\})
+C(\{1,3\},\{1,4\},\{2,4\},\{1,5\})\nl
&&+C(\{1,3\},\{2,4\},\{2,5\},\{1,5\})\,.
\ea
Then calculating their CHY integral  with the canonical PT factor becomes as easy as consulting a dictionary, see below
\ba\label{cpt}
\int\dif \mu_5 {\rm PT}(1,\cdots,5)
C(\{1,2\},\{2,3\},\{1,4\})
&=&
\frac{1}{s _{1,2} s _{1,2,3}}+\frac{1}{s _{2,3} s _{1,2,3}}\,,\nl
\int\dif \mu_6 {\rm PT}(1,\cdots,6)
C(\{1,3\},\{2,3\},\{2,4\},\{1,5\})
&=&
\frac{1}{s _{1,2,3} s _{2,3}s _{1,2,3,4}}+\frac{1}{s _{2,3} s _{1,2,3,4} s _{2,3,4}}\,,\nl
\int\dif \mu_6 {\rm PT}(1,\cdots,6)
C(\{1,3\},\{1,2\},\{2,4\},\{2,5\})
&=&
\frac{1}{s _{1,2} s _{1,2,3}s _{1,2,3,4}}\,.
\ea
More application will be seen in next section.

\begin{figure}[!htb]
 \centering
\begin{tikzpicture}
\fill (0,0) circle (0.03) node [below=0pt] {$i$};
\fill (1,0) circle (0.03) node [below=0pt] {$k_1$};
\fill (1.5,0) circle (0.03) node [below=0pt] {$k_2$};
\fill (0.5,0) circle (0.03) node [below=0pt] {$j$};
\fill (3,0) circle (0.03) node [below=0pt] {$l$};

\draw (0,0) arc (180:0:.75);
\draw[line width=.05cm] (1,0) arc (180:0:.25);
\draw[line width=.05cm] (0.5,0) arc (180:0:.25);
\draw (.5,0) arc (180:0:1.25);
\end{tikzpicture}
\caption{\label{complicated}an example for  more complicated cases}
\end{figure}
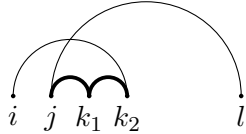

 Above we have shown a constructive way   to get $C^{\rm basis}$.  How to identify whether an arbitrary Cayley function  is a $C^{\rm basis}$ ?
Motivated by the rule that a line is always drawn from a left-most point to a right-most point of a subgraph,
such construction shown in figure \ref{twosides}
 \begin{figure}[!htb]
 \centering
  \subfloat{
 \begin{tikzpicture}
\fill (0,0) circle (0.03);
\fill (1,0) circle (0.03);
\fill (2.5,0) circle (0.03);
\draw (0,0) arc (180:0:.5);
\draw (1,0) arc (180:0:.75);

\node at (0,-.3) {$i$};
\node at (1,-.3) {$j$};
\node at (2.5,-.3) {$l$};

\node at (1.3,-1) { Containing $\frac{1}{\s_{i,j}\s_{j,l}}$ with $i<j<l$
 };
\end{tikzpicture}
}
 \subfloat{
\begin{tikzpicture}
\fill (0,0) circle (0.03);
\fill (1,0) circle (0.03);
\fill (1.5,0) circle (0.03);
\fill (3,0) circle (0.03);
\draw (0,0) arc (180:0:.75);
\draw (1,0) arc (180:0:.25);
\draw (1,0) arc (180:0:1);

\node at (0,-.3) {$i$};
\node at (1,-.3) {$j$};
\node at (1.5,-.3) {$k$};
\node at (3,-.3) {$l$};

\node at (1.5, -1) {Containing $\frac{1}{\s_{i,k}\s_{j,k}\s_{j,l}}$  with $i<j<k<l$
};
\end{tikzpicture}
}
\caption{\label{twosides}}
\end{figure}
can't appear in $C^{\rm basis}$. Reversely,   as long as they don't have these two constructions, which actually excludes many Cayley functions, see an example for more complicated cases in figure \ref{complicated}, and makes sure that the left-most particle of  any  connected subgraph , like here $i$, has to be linked to the right-most particle, like here $l$,
 there is always a way to construct them by the map $M$ and so they belong to the basis.

Though the CHY integral of canonical PT factor and $C^{\rm kernel}$  is zero, it will contribute in string integral, as  we discuss now.

\section{Cayley functions and disk integrals}\label{sec4}

In this section, we study the natural appearance of Cayley functions in certain disk integrals of open superstring theory. The basic objects we are interested in are a class of disk integrals with Cayley functions as (half) integrands, which we collectively call $Z$ integrals~\cite{Broedel:2013tta,Carrasco:2016ygv,Mafra:2016mcc,Carrasco:2016ldy}
\be\label{Zgen}
Z(12 \cdots,n| \{i,j\}):=(\alpha')^{n{-}3}\int_{(12\cdots n)}\dif^{n{-}3} z\,\prod_{i<j}^{n{-}1} |z_{i j}|^{\alpha' s_{i j}}~\frac 1 {z_{i_1 j_1}} \cdots \frac 1 {z_{i_{n{-}2}, j_{n{-}2}}}\,,
\ee
where we have chosen to fix the ${\rm PSL}(2,\mathbb R)$ redundancy by setting  {\it e.g.} $(z_1, z_{n{-}1}, z_n)=(0,1,\infty)$ and the domain for integrals over $\dif^{n{-}3} z$, denoted as $(12\cdots n)$, means $0<z_2<\cdots< z_{n{-}2}<1$. In addition to the Koba-Nielsen factor, we insert the SL(2)-fixed $C(\{i,j\})$ in the integrand, which can be rewritten in a SL(2) covariant form as before. In the special case that $C=$PT$(\beta)$, it reduces to the more familiar $Z$ integrals which depend on another ordering $\beta$:
\be
Z(12\cdots n| \beta)=(\alpha')^{n{-}3}\int_{(12\cdots n)}\dif^{n{-}3} z\,\prod_{i<j}^{n{-}1} |z_{i j}|^{\alpha' s_{i j}}~{\rm PT}(\beta)\,,
\ee
These $Z$ integrals have played important roles not only for gluon amplitudes in open superstring theory, but also for higher-order corrections to NLSM and other theories~\cite{Carrasco:2016ygv,Carrasco:2016ldy}.
To see this, let's recall the main results of \cite{Mafra:2011nv}: it has been shown that any $n$-pt tree amplitude in type I superstring theory is a linear combination of $(n{-}3)!$ partial amplitudes in super-Yang-Mills theory (SYM), with ordering $(1, \pi(2), \cdots, \pi(n{-}2), n{-}1, n)$
\be
{\cal M}_n^{\rm type~I} (1,2,\cdots,n)=\sum_{\pi \in S_{n{-}3}} F(12\cdots n| \pi) M_n^{\rm SYM} (1, \pi(2), \cdots, \pi(n{-}2), n{-}1,n)\,,
\ee
where all the $\alpha'$-dependence is encoded in the $(n{-}3)!$ disk integrals $F$'s defined as
\be
F(12\cdots n | \pi):=(\alpha')^{n{-}3} \int_{(12\cdots n)}\dif^{n{-}3} z\,\prod_{i<j}^{n{-}1} |z_{i j}|^{\alpha' s_{i j}}~\prod_{b=2}^{n{-}2}\sum_{a=1}^{b{-}1} \frac{s_{\pi(a), \pi(b)}}{z_{\pi(a), \pi(b)}}\,,
\ee
with $\pi(1)=1$ ($\pi(n{-}1)=n{-}1$ though that is not used here). The RHS is nothing but a sum of $(n{-}3)!$ $Z$ integrals, dressed by products of $n{-}3$ poles:
\be
F(12\cdots n | \pi)=
\sum_{\substack{\pi^{-1} (i_a)<a\\a=2,\cdots, n{-}2}} \prod_{a=2}^{n{-}2} s_{i_a, j_a}\,Z(12\cdots, n|\{1,n{-}1\}, \{i_2, \pi(2)\}, \cdots, \{i_{n{-}2}, \pi(n{-}2)\})\,,
\ee
where after fixing $i_1=1, j_1=n{-}1$, we have Cayley functions with $j_a=\pi(a)$ and each $i_a$  precedes $j_a$ in the ordering $\pi$, for $a=2, \cdots, n{-}2$ (there are $(n{-}3)!$ of them). For example, $F(1234|2)=s_{12} Z(1234|\{1,2\},\{1,3\})$ and for $n=5$ (we suppress the ordering $(12345)$ and the overall edge $\{1,4\}$ in $Z$ integrals):
\begin{align}
F(12345|23)&=s_{12}~\left(s_{13} Z(\{1,2\},\{1,3\})+ s_{23} Z(\{1,2\},\{2,3\}\right)\,, \\\nonumber
F(12345|32)&=s_{13}~\left(s_{12} Z(\{1,2\}, 13)+ s_{23} Z(\{2,3\},\{1,3\})\right)\,.
\end{align}
Thus we have seen that the complete $\alpha'$-dependence of tree amplitudes in type I theory is encoded in these $Z$ integrals, \eqref{Zgen}. The $\alpha'$ expansion of generic $Z$ integrals can be computed, but it suffices to do so for those where the $C$ functions form a basis. A convenient choice is to focus on a $(n{-}2)!$ basis given by $Z(12\cdots|\beta)$ where PT$(\beta)$'s form a KK basis ~\cite{Broedel:2013tta}, and it is well known that such Z integrals give double-partial amplitudes in the $\alpha'\to 0$ limit:
\be
Z(\alpha|\beta)=m(\alpha|\beta) + {\cal O}(\alpha'^2)\,,
\ee
where the first correction starts at ${\cal O}(\alpha'^2)$ since ${\cal O}(\alpha')$ term vanishes identically
, which
follows
from supersymmetry of open string amplitudes.
 In the following, we will focus on $Z$ integrals with Cayley functions in the $(n{-}2)!$ new basis, and as we will see shortly, they play a special role in the $\alpha'$ expansion of disk integrals. In fact, such $Z$ integrals have been studied in \cite{Broedel:2013tta,Mafra:2016mcc} , where these integrals are called pole-channel basis.
Our discussion here will focus on a graphic way of reading off nice properties of this Z-integral basis from the structures of Cayley functions.

Note that the new basis consists of ${\rm Cat}_{n{-}2}$ $C^{\rm single}$'s and $(n{-}2)!- {\rm Cat}_{n{-}2}$ $C^{\rm kernel}$'s, thus at ${\cal O}(1)$ in the $\alpha'$ expansion, we have either a single cubic tree graph or zero:
\be\label{Zzero}
Z(12\cdots n | \{i,j\})= \begin{cases} \displaystyle \frac 1 {s_{i_1 \cdots j_1}} \cdots \cdots \frac 1 {s_{i_{n{-}3} \cdots j_{n{-}3}}}  + {\cal O}(\alpha'^2)\,, &\quad {\rm for}~~C^{\rm single}(\{i,j\})\,,\\
0 + {\cal O}(\alpha'^2)\,, &\quad {\rm for}~~C^{\rm kernel}(\{i,j\})\,,\\
\end{cases}
\ee
where note that we have suppressed the trivial edge $(1,n{-}1)$.
For $n=5$ we have 5 elements with single graph, {\it e.g.} $Z(12345|\{1,2\},\{1,3\})=\frac 1 {s_{12}} \frac 1 {s_{123}} + {\cal O}(\alpha'^2)$,  and the kernel one gives $Z(12345|\{1,3\}, \{2,4\})={\cal O}(\alpha'^2)$. A natural question is, can we say something about higher order corrections, especially in the case of $C^{\rm kernel}$ ? We propose that one can obtain {\it pole structures} of the leading non-vanishing $\alpha'$ order directly from corresponding Cayley tree graphs. 

{\bf Proposal}: For any $C(\{i,j\})$ in the new basis, the pole structure for the first non-vanishing order in the $\alpha'$-expansion of $Z(12\cdots n | \{i,j\})$ is determined by its {\it maximal subgraph without crossing}, $M$. Let's assume that $M$ has $m$ edges which, without loss of generality, are denoted as $\{i_1, j_1\}, \cdots, \{i_m, j_m\}$ (out of all the $n{-}3$ edges $\{i_1, j_1\}, \cdots, \{ i_{n{-}3}, j_{n{-}3}\}$), then the Z integral has the leading non-vanishing order at ${\cal O}(\alpha'^{n{-}3{-}m})$:
\be\label{Zleading}
Z(12\cdots n | \{i,j\})=\frac{(\alpha')^{n{-}3{-}m}\,c_m}{s_{i_1, \cdots, j_1} \cdots s_{i_m, \cdots, j_m} } + {\cal O}(\alpha'^{n{-}2{-}m})\,.
\ee
where $c_m$ is a multiple zeta value of transcendental  weight $n{-}3{-}m$. For $C^{\rm single}$,\eqref{Zleading} reduces to the ${\cal O}(1)$ cubic tree of \eqref{Zzero}, since by definition it has all $m=n{-}3$ non-crossing edges. The other extreme is the case that there is no non-crossing subgraph, $m=0$, and we predict that the first non-vanishing order is at ${\cal O}((\alpha')^{n{-}3})$, which is given by a multiple zeta value with weight $n{-}3$.

Note that \eqref{Zleading} is also consistent with the absence of ${\cal O}(\alpha')$: for $C^{\rm kernel}$ we have $m<n{-}3$ but we can at most have $m=n{-}5$ which corresponds to only two edges crossed. Thus in the general case $0<m\leq n{-}5$, we have at leading order ${\cal O}((\alpha')^{n{-}3{-}m})$, product of $m$ compatible propagators (a subset of a cubic tree).
We believe that the proposal can be proved using the Berends-Giele recursion for Z integrals given in \cite{Mafra:2016mcc} (which in turn was based on methods of $\alpha'$ expansion in \cite{Broedel:2013tta,Mafra:2011nw}).

Let's illustrate the result with more examples. For $n=6$, there are 10 $C^{\rm kernel}$'s shown in figure \ref{kernel6} . We see that $K_1, K_2, K_3, K_4$ all have an edge ($m=1$ subgraph) that does not cross others, while others have no non-crossing subgraph ($m=0$), thus
\ba
&\{Z(K_1), Z(K_2), Z(K_3), Z(K_4)\} \sim \alpha'^2 \zeta_2 \{\frac 1 {s_{12}}, \frac 1 {s_{45}}, \frac 1 {s_{23}}, \frac 1 {s_{34}}\} + {\cal O}(\alpha'^3)\,,\nl
&\{Z(K_5), Z(K_6), Z(K_7), Z(K_8), Z(K_9), Z(K_{10})\} \sim \alpha'^3 \zeta_3 + {\cal O}(\alpha'^4)\,,
\ea
where we have suppressed the overall ordering $(12\cdots 6)$, and ignored overall constants.

\begin{figure}[!htb]
\centering
\subfloat[$\sim \alpha'^2 \zeta_2 \frac {1}{s_{1,2}s_{1,2,3}}+\cdots$]{
\includegraphics[width=0.23\textwidth]{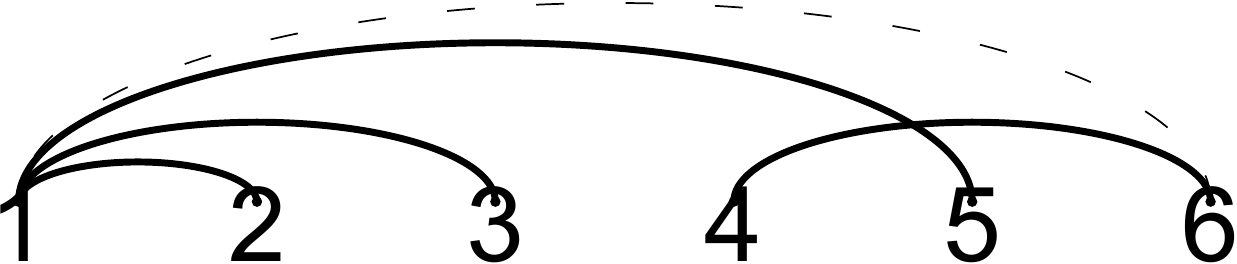}}
\subfloat[$\sim \alpha'^2 \zeta_2 \frac 1 {s_{1,2}s_{6,7}}+\cdots$]{
\includegraphics[width=0.23\textwidth]{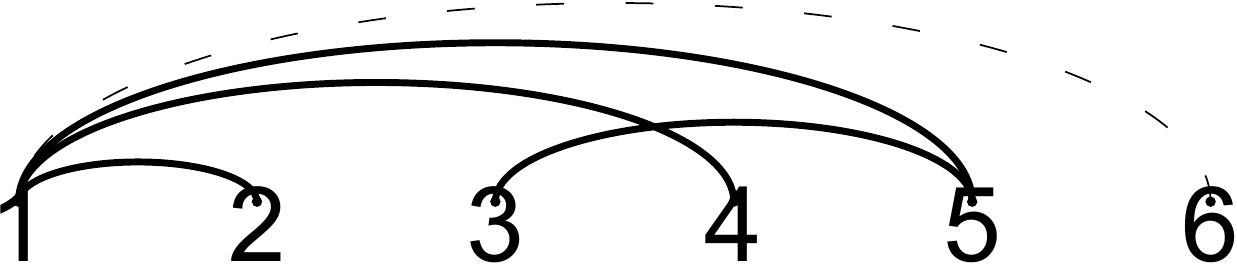}}
\subfloat[$\sim \alpha'^3 \zeta_3 \frac 1 {s_{1,2}}+\cdots$]{
\includegraphics[width=0.23\textwidth]{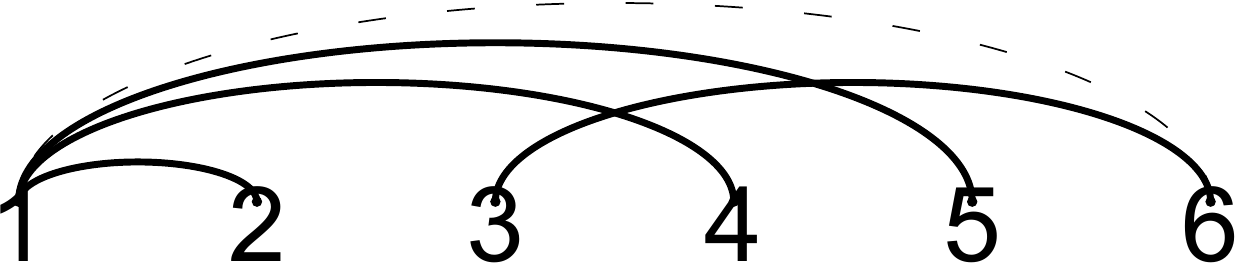}}
\subfloat[$\sim \alpha'^4 \zeta_4+\cdots $]{
\includegraphics[width=0.23\textwidth]{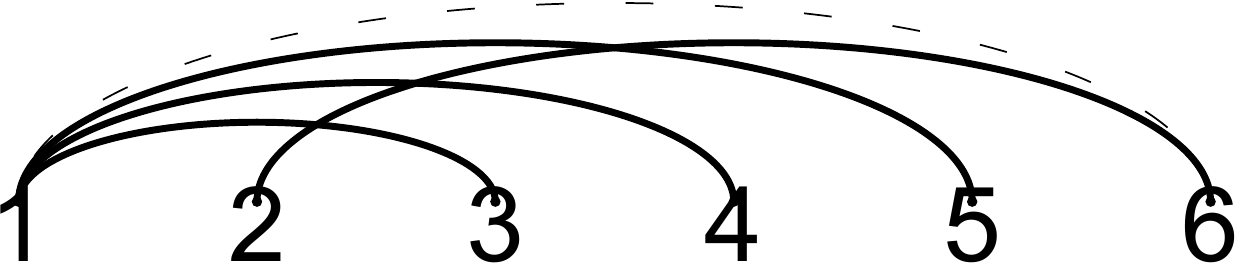}}
\caption{\label{kernel7} Some disk integrals for 7pt}
\end{figure}

For $n=7$, there are  all 78 $C^{\rm kernel}$'s and we find that all of them fall into three categories according to their leading non-vanishing order:
(a): $m=2$: $\alpha'^2$ order with two compatible poles,
(b): $m=1$: $\alpha'^3$ order with one pole, and (c): $m=0$: $\alpha'^4$ order without any pole. Here are examples for these three cases seen in figure \ref{kernel7}.

\begin{figure}[!htb]
\centering
\subfloat[$\sim  \frac {\alpha'^2 \zeta_2}{s_{1,2}s_{4,5}s_{4,5,6}}+\cdots$]{
\includegraphics[width=0.23\textwidth]{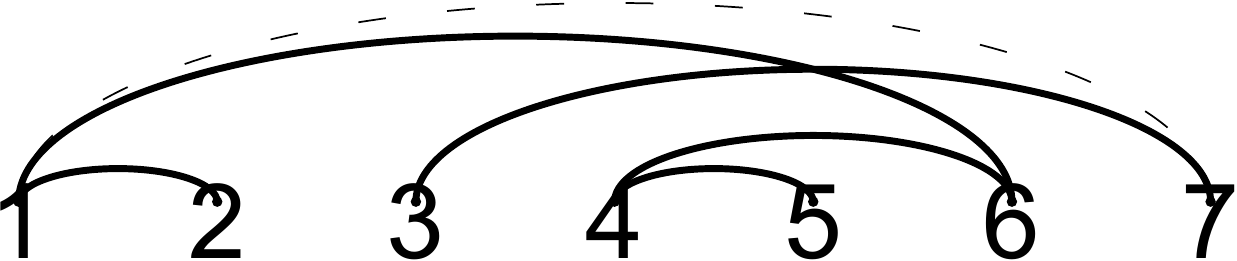}}
\subfloat[$\sim \alpha'^3 \zeta_3 \frac 1 {s_{1,2}s_{1,2,3}}+\cdots$]{
\includegraphics[width=0.23\textwidth]{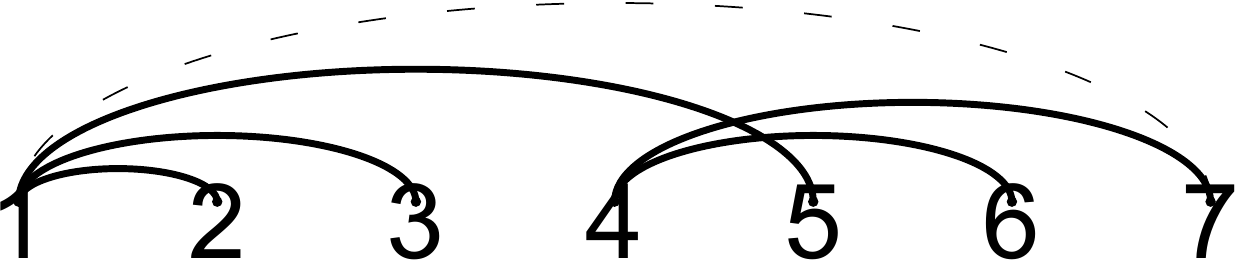}}
\subfloat[$\sim \alpha'^4 \zeta_4 \frac 1 {s_{1,2}}+\cdots$]{
\includegraphics[width=0.23\textwidth]{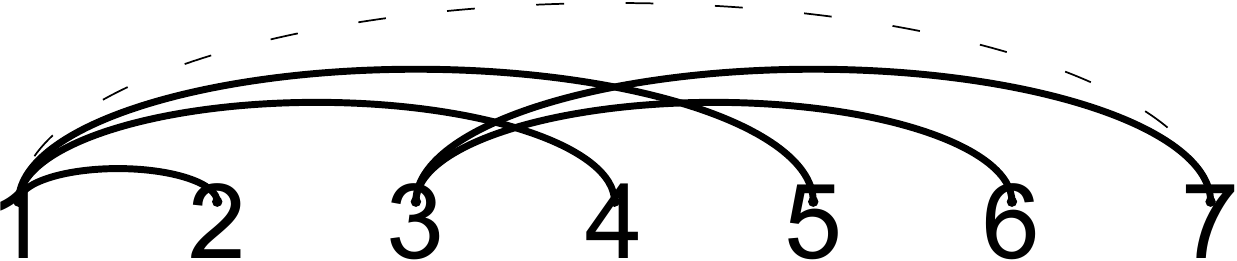}}
\subfloat[$\sim \alpha'^5 \zeta_5 +\cdots$]{
\includegraphics[width=0.23\textwidth]{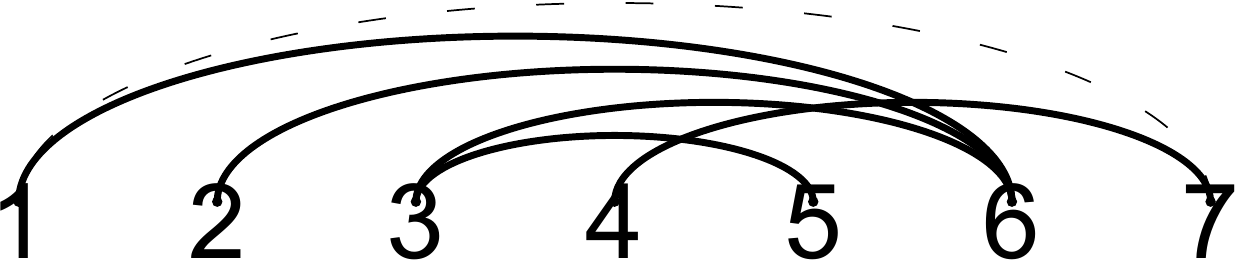}}
\caption{\label{kernel8} Some disk integrals for 8pt}
\end{figure}

Finally, let's present the following $n=8$ examples seen in figure \ref{kernel8}.

\section{Discussions and Outlook}\label{sec5}

In this note, we have introduced Cayley functions as a new class of half integrands in CHY formulas; they naturally generalize the Parke-Taylor factor (PT), which arises from a line or Hamiltonian tree, to general cases of labelled trees. We have discussed important aspects and applications of Cayley functions. First of all, we have presented a diagrammatic way to directly read off the sum of cubic Feynman diagrams, as given by the CHY formula with $C^2$. Combinatorically, a collection of such cubic trees correspond to a polytope, thus providing a one-to-one map between Cayley functions and certain polytopes; we classified such polytopes as ranging from the associahedron (Hamiltonian tree graph) to permutohedron (star tree graph).The CHY formula with $C C'$ produces Feynman diagrams that correspond to the intersection of the two polytopes.

Furthermore, we have studied the linear space of all half integrands without forming subcycles. The dimension of the space is $(n{-}2)!$ since any such half integrands can be reduced to the KK basis of PT factors, and we have found a nice formula for the reduction. We have introduced a new basis where each element has the property that under CHY formula with a given PT, it gives either a single diagram or zero. Finally, we have briefly discussed how these Cayley functions and especially the new basis can be used in disk integrals of superstring theory.  In the following, we will briefly mention more aspects of Cayley functions that have not been covered above, especially open questions along several directions.

\subsection*{Beyond Cayley functions: from $G(2,n)$ to ${\cal M}_{0,n}$}

One of the most important properties of a Cayley function is that it maps to a sum of cubic Feynman diagrams (with coefficients $+1$).
Of course they are just special cases of half-integrands that have this property, and we suspect that they are the simplest ones. As a first step towards going beyond Cayley functions, we find a larger class of such half-integrands, which are in one-to-one correspondence with MHV non-planar on-shell diagrams in $N=4$ SYM~\cite{Arkani-Hamed:2014bca}, and ${\bf C}$'s are just special cases of these functions.

%
%

Any MHV on-shell-diagram gives a rational function, ${\bf B}(\lambda_1, \cdots, \lambda_n)$ defined on G$(2,n)$ with weight -2: ${\bf B} \to \prod_{i=1}^n x_i^{-2} {\bf B}$ for $\lambda^\alpha_i \to x_i \lambda^\alpha_i$ with $a=1,2, \cdots, n$. Such a function is related to our functions on ${\cal M}_{0,n}$ via $(\lambda_i^{\alpha=1}, \lambda_i^{\alpha=2})=t_i (1,\sigma_i)$ (thus $\langle \lambda_i \lambda_j \rangle=t_i t_j \sigma_{i,j}$). The simplest case is again a Parke-Taylor factor (the demominator of original "Parke-Taylor" formula) which is the same as our {\bf PT} up to an overall prefactor:
\be\label{PTG2n}
\frac 1 {\langle 1\,2\rangle \cdots \langle n\, 1\rangle}=\frac 1 {\prod_{i=1}^n t_i^2}~{\bf PT}(1,2,\cdots,n)\,.
\ee
In this way, any function with weight -2 on G$(2,n)$ can be converted to that on ${\cal M}_{0,n}$.

As shown in~\cite{Arkani-Hamed:2014bca}, a generic MHV on-shell-diagram is characterized by $n{-}2$ triplets of labels $(i_a, j_a, k_a)$ for $1\leq a\leq n{-}2$ (we assume that all labels $1,2,\cdots, n$ are covered), and its rational function is always a positive sum of \eqref{PTG2n} with different orderings
\be\label{expB}
{\bf B}(\{i,j,k\})=\sum_{\substack{\pi \in S_n/Z_n\\i<j<k~{\rm in}~\pi}} \frac 1 {\langle \pi(1), \pi(2) \rangle \cdots \langle \pi(n), \pi(1) \rangle}\,.
\ee
Here $i<j<k$ is a cyclic ordering in $\pi$.
Generally, a ${\bf B}$ function takes a form more complicated than ${\bf C}$'s~\cite{Arkani-Hamed:2014bca}, but it is straightforward to see when it can reduce to a ${\bf C}$ (with the prefactor as in\eqref{PTG2n}): if all the $n{-}2$ triplets share a label, {\it e.g.} $k_1=\cdots=k_{n{-}2}=n$ for all $a$, ${\bf B}$ reduces to ${\bf C}$ with the same $\{i,j\}$:
\be
{\bf B}(\{i_1,j_1,n\}, \cdots, \{i_{n{-}2}, j_{n{-}2}, n\})=\prod_{a=1}^n t_a^{-2} {\bf C} (\{i,j\})\,,
\ee
and we see that \eqref{expB} reduces to \eqref{id} if we fix $n$ to be at the end of all orderings.

One can show that ${\bf B}$ functions also have the property that ${\bf C}$'s have: any CHY formula with ${\bf B}^2$ gives a sum of Feynman diagrams, which can be encoded in a polytope as that for a Cayley function. We will leave the generalization of theorem \ref{identicalCc} and full classifications of these more general polytopes to a future work.


\subsection*{Open questions for Cayley functions}

There are other open questions regarding Cayley functions in CHY formulas. The most obvious question is to understand better the origin of the map from Cayley functions to polytopes, and what is the significance of these polytopes in mathematics, see \cite{GraphAssociahedra,ConciniCProcesi,AlexanderPostnikov,AlexandePostnikovVictorReiner}
for some previous work.
One possible direction is to consider the class of graph associahedra based on Dynkin diagrams, which are known to tile the compactied moduli space of punctured Riemann spheres \cite{MDavisTJanuszkiewiczRScott}. It would be fascinating to explore whether this class of graphs has special properties in the context of CHY formulae or disk integrals.
Besides, it would be highly desirable to generalize our study of Cayley functions and polytopes of Feynman diagrams to loop level, along the line of $\phi^3$ loop amplitudes from CHY-like constructions~\cite{Geyer:2015bja, He:2015yua,Feng:2016nrf,Baadsgaard:2015hia,He:2017spx}.

We would also like to understand better the meaning of the new basis. For example, it is well known that one can expand those half integrands appearing in CHY formulas of other theories (such as the reduced Pfaffian {\it etc.}) in the KK basis; a natural question is when we expand them in our new basis, what is the interpretation of the coefficients? Moreover, we know that in twistor-string formula for ${\cal N}=4$ SYM~\cite{Witten:2003nn,Roiban:2004yf}, Parke-Taylor factors are mapped to color-ordered amplitudes. Similarly a Cayley function is mapped to a certain sum of such color-ordered amplitudes, which in turn form a basis different than the usual KK basis. It may be interesting to study their properties as well.

Another direction concerns higher-order $\alpha'$ corrections in Z integrals and other integrals in superstring theory. We have only studied leading non-zero order in the $\alpha'$-expansion of Z integrals, and it would be intriguing to extract sub-leading pole structures from the graph.
For Z integrals with PT's, such sub-leading terms can be obtained systematically using the method in \cite{Broedel:2013tta}, which can be turned into results for Z integrals in the new basis. What is remained to be done is a more direct (and preferably diagrammatic) way of extracting higher-order terms from Cayley functions.
 Moreover, it is possible that the combinatorical polytope structures generalize to disk integrals (see \cite{Mizera:2017cqs} for related work which studies certain combinatoric structures in closed-string integrals).

\section*{Acknowledgements}
S.H. would like to thank Nima Arkani-Hamed, Yuntao Bai and Gongwang Yan for stimulating discussions and collaborations on related projects. We also thank Chi-Sing Lam,  Sebastian Mizera and  Chi Zhang  for useful discussions and  especially Oliver Schlotterer for very helpful comments on the draft.
We also thank Freddy Cachazo, Nick Early and a referee for comments on v2 of the paper.
S.H.'s research is supported in part by the
Thousand Young Talents program and the Key Research Program of Frontier Sciences of CAS.

\appendix

\section{ A sketch of proof for \eqref{recursiongraph} using factorization}\label{appa}

All Feynman diagrams must have $n-3$ compatible poles, so they must be contained on the RHS of \eqref{recursiongraph}. The only problem is that maybe some terms on the RHS of \eqref{recursiongraph} don't appear on the RHS of Theorem \ref{identicalCc}. So we assign each of them of a undetermined coefficient and use to determine them.
First we see a simple case,
  \def \layersep {.7cm}
 \ba\label{4ptafaf}
 \raisebox{-.55cm}{
\begin{tikzpicture}[shorten >=0pt,draw=black,
        node distance = \layersep,
        neuron/.style = {circle, minimum size=3pt, inner sep=0pt,  fill=black } ]

     \node[neuron] (1) {};
     \node[ neuron,right of = 1] (2)  {};
   \node[ neuron,right of = 2] (3)  {};

        \draw (1) node[below=4pt]{$1$} --(2)node[below=4pt]{$2$}
        --(3)node[below=4pt]{$3$};
    \end{tikzpicture}}
    =
  x_1  \raisebox{-.55cm}{
\begin{tikzpicture}[shorten >=0pt,draw=black,
        node distance = \layersep,
        neuron/.style = {circle, minimum size=3pt, inner sep=0pt,  fill=black } ]

     \node[neuron] (1) {};
     \node[ neuron,right of = 1] (2)  {};
   \node[ neuron,right of = 2] (3)  {};

        \draw[dashed] (1) node[below=4pt]{$1$} --(2)node[below=4pt]{$2$};

       \draw (2)   --(3)node[below=4pt]{$3$};
    \end{tikzpicture}}
    +
   x_2 \raisebox{-.55cm}{
\begin{tikzpicture}[shorten >=0pt,draw=black,
        node distance = \layersep,
        neuron/.style = {circle, minimum size=3pt, inner sep=0pt,  fill=black } ]

     \node[neuron] (1) {};
     \node[ neuron,right of = 1] (2)  {};
   \node[ neuron,right of = 2] (3)  {};

        \draw (1) node[below=4pt]{$1$} --(2)node[below=4pt]{$2$};
           \draw[dashed] (2)  --(3)node[below=4pt]{$3$};
            \end{tikzpicture}}\,,
 \ea
According to the analyzing in \eqref{pc} ,  \raisebox{-.55cm}{
\begin{tikzpicture}[shorten >=0pt,draw=black,
        node distance = \layersep,
        neuron/.style = {circle, minimum size=3pt, inner sep=0pt,  fill=black } ]

     \node[neuron] (1) {};
     \node[ neuron,right of = 1] (2)  {};
   \node[ neuron,right of = 2] (3)  {};

        \draw (1) node[below=4pt]{$1$} --(2)node[below=4pt]{$2$}
        --(3)node[below=4pt]{$3$};
    \end{tikzpicture}}
    have poles
     \raisebox{-.55cm}{
\begin{tikzpicture}[shorten >=0pt,draw=black,
        node distance = \layersep,
        neuron/.style = {circle, minimum size=3pt, inner sep=0pt,  fill=black } ]

     \node[neuron] (1) {};
     \node[ neuron,right of = 1] (2)  {};

        \draw (1) node[below=4pt]{$1$} --(2)node[below=4pt]{$2$};
    \end{tikzpicture}}
    ,
     \raisebox{-.55cm}{
\begin{tikzpicture}[shorten >=0pt,draw=black,
        node distance = \layersep,
        neuron/.style = {circle, minimum size=3pt, inner sep=0pt,  fill=black } ]

     \node[neuron] (1) {};
     \node[ neuron,right of = 1] (2)  {};

        \draw (1) node[below=4pt]{$2$} --(2)node[below=4pt]{$3$};
    \end{tikzpicture}},
    which means it must contains both
      \raisebox{-1cm}{
 \begin{tikzpicture}[shorten >=0pt,draw=black,scale=.25,
        node distance = .2cm,
        neuron3/.style = {circle, minimum size=.1pt, inner sep=0pt,  fill=black } ]
     \node[neuron3] {}
     child {node[neuron3] {}node[below=0pt]{$1$}}
        child {node[neuron3] {}
        child {node[neuron3] {} node[below=0pt]{$2$}}
           child {node[neuron3] {}node[below=0pt]{$3$}}}
           ;
     \draw (0,0)--(.5,1) node[right=0pt]{$n$};
    \end{tikzpicture}  }
    and
 \raisebox{-1cm}{
 \begin{tikzpicture}[shorten >=0pt,draw=black,scale=.25,
        node distance = .2cm,
        neuron3/.style = {circle, minimum size=.1pt, inner sep=0pt,  fill=black } ]
     \node[neuron3] {}
        child {node[neuron3] {}
        child {node[neuron3] {} node[below=0pt]{$1$}}
           child {node[neuron3] {}node[below=0pt]{$2$}}}
           child {node[neuron3] {}node[below=0pt]{$3$}};
     \draw (0,0)--(.5,1) node[right=0pt]{$n$};
    \end{tikzpicture}  }
    corresponding to the two terms
    on the RHS of  \eqref{4ptafaf}, {\it i.e.} $x_1\neq 0,x_2\neq 0$. The next thing is to determine their relative sign appearing in Theorem \ref{identicalCc}.
    While they must be the same as we can't allow the value of one expression after being taken the residue of its pole is 1 while the other is $-1$, so $x_1=x_2=1$.

    Now we move the 5pt cases,
       \def \layersep {.7cm}
 \ba\label{5ptafaff}
 \raisebox{-.55cm}{
\begin{tikzpicture}[shorten >=0pt,draw=black,
        node distance = \layersep,
        neuron/.style = {circle, minimum size=3pt, inner sep=0pt,  fill=black } ]

     \node[neuron] (1) {};
     \node[ neuron,right of = 1] (2)  {};
   \node[ neuron,right of = 2] (3)  {};
      \node[ neuron,right of = 3] (4)  {};
        \draw (1) node[below=4pt]{$1$} --(2)node[below=4pt]{$2$}
        --(3)node[below=4pt]{$3$}--(4)node[below=4pt]{$4$};
    \end{tikzpicture}}
    =
x_1    \raisebox{-.55cm}{
\begin{tikzpicture}[shorten >=0pt,draw=black,
        node distance = \layersep,
        neuron/.style = {circle, minimum size=3pt, inner sep=0pt,  fill=black } ]

     \node[neuron] (1) {};
     \node[ neuron,right of = 1] (2)  {};
   \node[ neuron,right of = 2] (3)  {};
      \node[ neuron,right of = 3] (4)  {};
        \draw[dashed] (1) node[below=4pt]{$1$} --(2)node[below=4pt]{$2$};

       \draw (2)   --(3)node[below=4pt]{$3$}--(4)node[below=4pt]{$4$};
    \end{tikzpicture}}
    +
 x_2   \raisebox{-.55cm}{
\begin{tikzpicture}[shorten >=0pt,draw=black,
        node distance = \layersep,
        neuron/.style = {circle, minimum size=3pt, inner sep=0pt,  fill=black } ]

     \node[neuron] (1) {};
     \node[ neuron,right of = 1] (2)  {};
   \node[ neuron,right of = 2] (3)  {};
      \node[ neuron,right of = 3] (4)  {};
        \draw (1) node[below=4pt]{$1$} --(2)node[below=4pt]{$2$};
           \draw[dashed] (2)  --(3)node[below=4pt]{$3$};
            \draw (3) --(4)node[below=4pt]{$4$};
    \end{tikzpicture}}
    +
  x_3  \raisebox{-.55cm}{
\begin{tikzpicture}[shorten >=0pt,draw=black,
        node distance = \layersep,
        neuron/.style = {circle, minimum size=3pt, inner sep=0pt,  fill=black } ]

     \node[neuron] (1) {};
     \node[ neuron,right of = 1] (2)  {};
   \node[ neuron,right of = 2] (3)  {};
      \node[ neuron,right of = 3] (4)  {};
        \draw (1) node[below=4pt]{$1$} --(2)node[below=4pt]{$2$}
        --(3)node[below=4pt]{$3$};
       \draw[dashed] (3)   --(4)node[below=4pt]{$4$};
    \end{tikzpicture}}\,,
 \ea
 According to pole analyzing, it must contains the Feynman diagrams in the first and the last term on the RHS of above equation, {\it i.e.} $x_1\neq 0,x_3\neq 0$. While if we take the factorization $s_{3,4}\rightarrow 0$, only the first two terms survives,
    \def \layersep {.7cm}
 \ba\label{4ptafafff}
 \raisebox{-.55cm}{
\begin{tikzpicture}[shorten >=0pt,draw=black,
        node distance = \layersep,
        neuron/.style = {circle, minimum size=3pt, inner sep=0pt,  fill=black } ]

     \node[neuron] (1) {};
     \node[ neuron,right of = 1] (2)  {};
   \node[ neuron,right of = 2] (3)  {};

        \draw (1) node[below=4pt]{$1$} --(2)node[below=4pt]{$2$}
        --(3)node[below=4pt]{$I$};
    \end{tikzpicture}}
    =
   x_1 \raisebox{-.55cm}{
\begin{tikzpicture}[shorten >=0pt,draw=black,
        node distance = \layersep,
        neuron/.style = {circle, minimum size=3pt, inner sep=0pt,  fill=black } ]

     \node[neuron] (1) {};
     \node[ neuron,right of = 1] (2)  {};
   \node[ neuron,right of = 2] (3)  {};

        \draw[dashed] (1) node[below=4pt]{$1$} --(2)node[below=4pt]{$2$};

       \draw (2)   --(3)node[below=4pt]{$I$};
    \end{tikzpicture}}
    +
   x_2 \raisebox{-.55cm}{
\begin{tikzpicture}[shorten >=0pt,draw=black,
        node distance = \layersep,
        neuron/.style = {circle, minimum size=3pt, inner sep=0pt,  fill=black } ]

     \node[neuron] (1) {};
     \node[ neuron,right of = 1] (2)  {};
   \node[ neuron,right of = 2] (3)  {};

        \draw (1) node[below=4pt]{$1$} --(2)node[below=4pt]{$2$};
           \draw[dashed] (2)  --(3)node[below=4pt]{$I$};
            \end{tikzpicture}}\,,
 \ea
where $I$ is the internal particle. In factorization limits, it just reduces to the case of 4pt, \eqref{4ptafaf}, which means $x_1=x_2$. Similarly, $x_2=x_3=x_1=1$.
The Feynman diagrams in each term of the RHS of \eqref{5ptafaff} share the same sign , so all 5 Feynman diagrams of  \raisebox{-.55cm}{
\begin{tikzpicture}[shorten >=0pt,draw=black,
        node distance = \layersep,
        neuron/.style = {circle, minimum size=3pt, inner sep=0pt,  fill=black } ]

     \node[neuron] (1) {};
     \node[ neuron,right of = 1] (2)  {};
   \node[ neuron,right of = 2] (3)  {};
      \node[ neuron,right of = 3] (4)  {};
        \draw (1) node[below=4pt]{$1$} --(2)node[below=4pt]{$2$}
        --(3)node[below=4pt]{$3$}--(4)node[below=4pt]{$4$};
    \end{tikzpicture}} share the same sign in Theorem \ref{identicalCc}.

    Generally, for a arbitrary Cayley function, according to the pole analyzing, some terms on the RHS of \eqref{identicalCc22} must appear in Theorem \ref{identicalCc} with nonzero coefficient. After we take all kinds of factorization, other terms   on the RHS of \eqref{identicalCc22} appear to join them with the same sign using the results of lower points recursively. So
 any $n-3$ compatible poles corresponds to a Feynman diagram in Theorem \ref{identicalCc} with coefficient $+1$.

\section{Proof of the sign in Theorem \ref{identicalCcp}}\label{appb}
As for the subtle all overall sign in Theorem \ref{identicalCcp}, in principle, we can expand $C_n,C'_n$ into PT's using the identity \eqref{id}, then their CHY formulas becomes a summation of double partial amplitudes , see \eqref{mab}
    \ba\label{twob}
   \int \dif \mu_n C_n C'_n=
   \sum_{\alpha,\beta}m(\alpha|\beta)\,.
   \ea
   However there is huge cancellation between these double partial amplitudes and a clever way is to  find a dominant one to determine the sign in Theorem  \ref{identicalCcp}.
  Our idea is to use the factorization , see figure \ref{ccpgraph}, recursively until we find the dominating PT from the expansion of $C_n,C'_n$ repectively.

   As we will find all Feynman diagrams on the RHS of Theorem \ref{identicalCcp} share the same sign, {\it i.e.} we can pick out any one denoted as the represetative Feynman diagram, {\it i.e.} any $n-3$ compatible poles both belonging to $P(C_n)$ and  $P(C'_n)$ to represent all cases.
  There are always two ``biggest'' poles $s_I,s_{\bar I}$ between these $n-3$ compatible poles (here we  means all particles of other poles without using the particle $n$ sit on these two ``biggest'' poles ) whose corresponding subgraphs together make up the labelled tree of $C_n$ (so does $C'_n$) up to an oriented edge
\def\layersep{1cm}
 \tikzset{
particle/.style={draw=black, postaction={decorate},
    decoration={markings,mark=at position .5 with {\arrow[draw=blue,scale=1.5]{>}}}}
 }
 \raisebox{.1cm}{
 \tikz{
  \draw[particle,dashed] (0,0)
--(1,0);}}, see figure \ref{ccpgraph}
.
\begin{figure}[!htb]
\centering
\subfloat{
\begin{tikzpicture}[shorten >=0pt,draw=black,
        node distance = \layersep,
        neuron2/.style = {circle, minimum size=41pt, inner sep=0pt,  fill=black!20 } ]

  \node[neuron2] (1) {$I$};
    \node[neuron2] at (2.4,0) (2) {~~~~~~${\bar I}$~~~~~~};

     \coordinate (A) at (1.2, 0);
           \coordinate [ below of = A] (n) ;

\draw (1)--(2);
\draw (A)--(n)
node[below=0pt]{$n$};
\node at ($(n)+4*(n)-4*(A)$){~};
\node at ($(n)+1*(n)-1*(A)$){
$s_I,s_{\bar I}$ are two biggest poles in this target
};
\node at ($(n)+1.5*(n)-1.5*(A)$){
 Feynman diagram without using $n$};
\end{tikzpicture}
}\subfloat{
\begin{tikzpicture}
\node {
\begin{tikzpicture}[shorten >=0pt,draw=black,
        node distance = \layersep,
        neuron/.style = {circle, minimum size=3pt, inner sep=0pt,  fill=black } ]

     \node[neuron] (1) {};
     \node[ neuron,right of = 1] (2)  {};
   \node[ neuron,above of = 2] (3)  {};
     \node[ neuron,right of = 2] (4)  {};
    \node[ neuron,right of = 4] (8)  {};
     \node[ neuron,right of = 8] (9)  {};
    \node[ neuron,above of = 4] (5)  {};
    \node[ neuron,above of = 5] (6)  {};
     \node[ neuron,right of = 5] (7)  {};
     \node at (.5,2.75) {$s_I \rightarrow 0$};
     \node at (.5,2.3) {$s_{\bar I} \rightarrow 0$};

    \draw[particle] (1)--(2);
       \draw[particle] (2) --(3);
    \draw[particle] (5)--(4);
        \draw[particle] (6)--(5);
        \draw[particle] (4)--(8);
         \draw[particle] (9)--(8);
     \draw[particle] (5)--(7);
\draw[particle,dashed] (2)node[below=0pt,green!50!red,xshift=-.05cm]{$i$}--(4)
node[below=0pt,green!50!red]{${\bar i}$};

\draw[blue!30!yellow,dashed] (0.5,0.5) ellipse (.8 and 1.2);
\node[blue!30!yellow] at (.5,1.1) {$I$};

\draw[blue!30!yellow,dashed] (3,1) ellipse (1.6 and 2.1);
\node[blue!30!yellow] at (3.4,2.1) {${\bar I}$};

\node at (.5,-.4) {$C^1$};
\node at (3,-.4) {$C^2$};

\node at ($.5*(1)+.5*(9)-1.5*(5)+1.5*(4)$) {pick out ${\rm PT}(I,{\bar I},n)$ from the expansion of $C_n$};

\end{tikzpicture}
}
node [below=70pt]{
\begin{tikzpicture}[shorten >=0pt,draw=black,
        node distance = \layersep,
        neuron/.style = {circle, minimum size=3pt, inner sep=0pt,  fill=black } ]

     \node[neuron] (1) {};
     \node[ neuron,right of = 1] (2)  {};
   \node[ neuron,above of = 1] (3)  {};
     \node[ neuron,right of = 2] (4)  {};
    \node[ neuron,right of = 4] (8)  {};
     \node[ neuron,right of = 8] (9)  {};
    \node[ neuron,above of = 4] (5)  {};
    \node[ neuron,above of = 5] (6)  {};
     \node[ neuron,right of = 6] (7)  {};
     \node at (.5,2.75) {$s_I \rightarrow 0$};
     \node at (.5,2.3) {$s_{\bar I} \rightarrow 0$};

    \draw[particle] (1)--(2);
       \draw[particle] (1) --(3);
    \draw[particle] (5)--(4);
        \draw[particle] (5)--(6);
        \draw[particle] (8)--(4);
         \draw[particle] (9)--(8);
     \draw[particle] (7)--(6);
\draw[particle,dashed] (4)
node[below=0pt,green!50!red]{${\bar i}'$}
--(2)node[below=0pt,green!50!red,xshift=-.05cm]{$i'$};

\draw[blue!30!yellow,dashed] (0.5,0.5) ellipse (.8 and 1.2);
\node[blue!30!yellow] at (.5,1.1) {$I$};

\draw[blue!30!yellow,dashed] (3,1) ellipse (1.6 and 2.1);
\node[blue!30!yellow] at (3.4,2.1) {${\bar I}$};

\node at (.5,-.4) {$C'^1$};
\node at (3,-.4) {$C'^2$};

\node at ($.5*(1)+.5*(9)-1.5*(5)+1.5*(4)$) {pick out ${\rm PT}({\bar I},I,n)$ from the expansion of $C'_n$};
\end{tikzpicture}
};
\end{tikzpicture}
}
\caption{\label{ccpgraph}}
  \end{figure}
Thus the represetative Feynman diagram corresponds to a particular  factorization $s_I\rightarrow 0,s_{\bar I}\rightarrow 0$ under which only certain Feynman diagrams survive.  Among the PT's from the expansion of $C_n$, only those which can be divided into two subgraphs with the particles $I$ and ${\bar I}$ respectively could contribute under this factorization. This decides the contributing PT's are either ${\rm PT}(I,{\bar I},n)$ or ${\rm PT}({\bar I},I,n)$,
   which  is further decided by the orientation of the linked edge
    \raisebox{.1cm}{
 \tikz{
  \draw[particle,dashed] (0,0)
--(1,0);}} .
    This was the time we saw the importance of the orientation of $C_n$ in the CHY integral of two distinguished Cayley functions.
    The subgraphs $C^1,C^2$  themselves are labelled trees,  so we can do this factorization recursively and the range of surviving PT's becomes more and more narrow until a single one comes out.  One can see the procedure to find the dominating PT is just to draw the representative Feynman diagram in the canonical way described in main tex . While we can also define a map ${\rho}$
based on a representative Feynman diagram and a oriented labelled tree of $C_n$ (or $C'_n$ ) to find the ordering of the dominating PT more abstractly,
 \begin{enumerate}
   \item ${\rho}$ maps an unordered sequence to an ordered sequence.
   \item As starting point, $({\rho}[i])=(i)$.
   \item The map is defined recursively,
   \ba
   ({\rho}[I])=({\rho}[I_1],{\rho}[I_2]) \,,
   \ea
      where
         $I$ is a particle set of a  labelled tree and
    $I_1\sqcup I_2=I$  are particle sets of
    subgraphs linked by an edge with the orientation from $i_1$ to $i_2$ which correspond to the
    two biggest poles of those from
 of the  $n-3$ compatible poles made up by the particles $I$.
 \end{enumerate}
Then at last, we obtain the dominating ${\rm PT}({\rho}[1,2,\cdots,n-1],n)$ of $C_n$ and similarly that of $C'_n$ denoted as ${\rm PT}({\rho}'[1,2,\cdots,n-1],n)$. Note that  $m\big({\rho}[1,2,\cdots,n-1],n\big|{\rho}'[1,2,\cdots,n-1],n\big)$ provides more than the representative Feynman diagram in general , however only this double partial amplitude provides this representative Feynman diagram among all $m(\a|\b)$ on the RHS of \eqref{twob}. So the representative Feynman diagram on the RHS of Theorem \ref{identicalCcp} must share the same as that of $m\big({\rho}[1,2,\cdots,n-1],n|{\rho}'[1,2,\cdots,n-1],n\big)$ , see \eqref{mab}, {\it i.e.} \ba
f={\rm flip}({\rho}[1,2,\cdots,n-1]|{\rho}'[1,2,\cdots,n-1])\,.
\ea

Let's repeat the procedure about the example (above \eqref{exampafaf})in main text.
Take the first Feynman diagram in \eqref{wahaha}
as a representative one. The two biggest poles are
$s_{1,2,3,4}$ and ``$s_{5}$''. The corresponding subgraphs of $C_6$ are   \raisebox{-.5cm}{
\begin{tikzpicture}[shorten >=0pt,draw=black,
        node distance = .7cm,
        neuron/.style = {circle, minimum size=3pt, inner sep=0pt,  fill=black } ]
     \node[neuron] (1) {};
       \node[ neuron,right of = 1] (2)  {};
       \node[ neuron,above of = 2] (3)  {};
   \node[ neuron,right of = 2] (4)  {};
      \node[ neuron,right of = 4] (5)  {};
        \draw[particle] (1) node[below=4pt]{$1$}  --(2)node[below=4pt]{$2$};
      \draw[dashed,particle](3)node[right=0pt]{$5$}--(2);
        \draw[particle] (4)node[below=4pt]{$3$}--(2) ;
          \draw[particle] (4) --(5)node[below=4pt]{$4$};
    \end{tikzpicture}
  }. Note that the orientation of the link edge is from 5 to 2, so
  \ba
  ({\rho}[1,2,3,4,5])=({\rho}[5],{\rho}[1,2,3,4])
  =(5,{\rho}[1,2,3,4])\,.
  \ea
Now we look at new Feynman diagrams made from the factorization. While in this case, one is a trivial point and we only need to take  the other one into consideration,
   \raisebox{-.55cm}{
\begin{tikzpicture}[node distance=.4cm]
 \coordinate (1);
 \coordinate [ right of = 1] (a1) ;
  \coordinate [ right of = a1] (a2) ;
   \coordinate [ right of = a2] (a3) ;
       \coordinate [ above of = a1] (2) ;
      \coordinate [ above of = a2] (a4) ;
       \coordinate [  above right of = a4] (6) ;
           \coordinate [  above left of = a4] (3) ;
          \draw (1) node[below=0pt]{\tiny  1} --(a2);
      \draw (a1)  --(2)node[left=0pt]{{\tiny 2}};
         \draw (a2) --(a4)--(3)node[above=0pt]{{\tiny 3}};
           \draw (a4)--(6)node[above=0pt]{{\tiny 4}};
            \draw[line width=.1cm] (a2)--(a3);
    \end{tikzpicture}}, which corresponds to the subgraph
 \raisebox{-.5cm}{
\begin{tikzpicture}[shorten >=0pt,draw=black,
        node distance = .7cm,
        neuron/.style = {circle, minimum size=3pt, inner sep=0pt,  fill=black } ]
     \node[neuron] (1) {};
       \node[ neuron,right of = 1] (2)  {};
         \node[ neuron,right of = 2] (4)  {};
      \node[ neuron,right of = 4] (5)  {};
        \draw[particle] (1) node[below=4pt]{$1$}  --(2)node[below=4pt]{$2$};
             \draw[particle] (4)node[below=4pt]{$3$}--(2) ;
          \draw[particle] (4) --(5)node[below=4pt]{$4$};
    \end{tikzpicture}
  }. The two biggest  poles of    \raisebox{-.55cm}{
\begin{tikzpicture}[node distance=.4cm]
 \coordinate (1);
 \coordinate [ right of = 1] (a1) ;
  \coordinate [ right of = a1] (a2) ;
   \coordinate [ right of = a2] (a3) ;
       \coordinate [ above of = a1] (2) ;
      \coordinate [ above of = a2] (a4) ;
       \coordinate [  above right of = a4] (6) ;
           \coordinate [  above left of = a4] (3) ;
          \draw (1) node[below=0pt]{\tiny  1} --(a2);
      \draw (a1)  --(2)node[left=0pt]{{\tiny 2}};
         \draw (a2) --(a4)--(3)node[above=0pt]{{\tiny 3}};
           \draw (a4)--(6)node[above=0pt]{{\tiny 4}};
            \draw[line width=.1cm] (a2)--(a3);
    \end{tikzpicture}}
    are $s_{1,2},s_{3,4}$
, corresponding to the factorization
 \raisebox{-.5cm}{
\begin{tikzpicture}[shorten >=0pt,draw=black,
        node distance = .7cm,
        neuron/.style = {circle, minimum size=3pt, inner sep=0pt,  fill=black } ]
     \node[neuron] (1) {};
       \node[ neuron,right of = 1] (2)  {};
         \node[ neuron,right of = 2] (4)  {};
      \node[ neuron,right of = 4] (5)  {};
        \draw[particle] (1) node[below=4pt]{$1$}  --(2)node[below=4pt]{$2$};
             \draw[dashed,particle] (4)node[below=4pt]{$3$}--(2) ;
          \draw[particle] (4) --(5)node[below=4pt]{$4$};
    \end{tikzpicture}
  }. Because of
   \raisebox{-.5cm}{
\begin{tikzpicture}[shorten >=0pt,draw=black,
        node distance = .7cm,
        neuron/.style = {circle, minimum size=3pt, inner sep=0pt,  fill=black } ]
     \node[neuron] (1) {};
       \node[ neuron,right of = 1] (2)  {};
                \draw[dashed,particle] (2)node[below=4pt]{$3$}--
                (1) node[below=4pt]{$2$}  ;
                \end{tikzpicture}
  }, we have
    \ba
  ({\rho}[1,2,3,4,5])=(5,{\rho}[1,2,3,4])
  =(5,{\rho}[3,4],{\rho}[1,2])\,.
  \ea
  Finally, because of
   \raisebox{-.5cm}{
\begin{tikzpicture}[shorten >=0pt,draw=black,
        node distance = .7cm,
        neuron/.style = {circle, minimum size=3pt, inner sep=0pt,  fill=black } ]
     \node[neuron] (1) {};
       \node[ neuron,right of = 1] (2)  {};
                \draw[dashed,particle]
                (1) node[below=4pt]{$1$}
                 --  (2)node[below=4pt]{$2$}  ;
                \end{tikzpicture}
  } and
     \raisebox{-.5cm}{
\begin{tikzpicture}[shorten >=0pt,draw=black,
        node distance = .7cm,
        neuron/.style = {circle, minimum size=3pt, inner sep=0pt,  fill=black } ]
     \node[neuron] (1) {};
       \node[ neuron,right of = 1] (2)  {};
                \draw[dashed,particle] (1) node[below=4pt]{$3$}
                 --  (2)node[below=4pt]{$4$}  ;
                \end{tikzpicture}
  }, we have
      \ba
  ({\rho}[1,2,3,4,5])=(5,{\rho}[3,4],{\rho}[1,2])
  =(5,{\rho}[3],{\rho}[4],{\rho}[1],{\rho}[2])
  = (5,3,4,1,2)\,.
  \ea
Similarly,
      \ba
  ({\rho}'[1,2,3,4,5])=({\rho}'[1,2,3,4],5)
  =({\rho}'[2,1],{\rho}'[3,4],5)
  = (2,1,3,4,5)\,.
  \ea

If we choose the second Feynman diagram in \eqref{wahaha} as the representative Feynman diagram, ${\rho}[1,2,3,4,5]$ and  ${\rho}'[1,2,3,4,5]$  will usually be different,
      \ba
  ({\rho}[1,2,3,4,5])&=&(5,{\rho}[1,2,3,4])
  =(5,1,{\rho}[2,3,4])
  = (5,1,{\rho}[3,4],2)
  = (5,1,3,4,2)\,,
  \nl
    ({\rho}'[1,2,3,4,5])&=&({\rho}'[1,2,3,4],5)
  =({\rho}'[2,3,4],1,5)
  = (2,{\rho}'[3,4],1,5)  = (2,3,4,1,5)\,,
  \nl
  \ea
  while their flip times share the same  odd-even property
  \ba
  f={\rm flip}({\rho}[1,2,3,4,5]|{\rho}'[1,2,3,4,5])=
  {\rm flip}(5,1,3,4,2|2,3,4,1,5)=3\,.
  \ea
One can take the last Feynman diagram in \eqref{wahaha} as the representative Feynman diagram and see the odd-even property of flip times  doesn't change,either,
  \ba
  f=
  {\rm flip}(1,5,3,4,2|2,3,4,5,1)=3\,.
  \ea

\section{CHY formula of two arbitrary star graphs}\label{appc}

In main text, we mainly consider such  Cayley functions with $n$ sent to infinity and they are characterised by $n-2$ pairs.  They may not be characterised by $n-2$ pairs again if we send another puncture of their covariant form to infinity. For general Cayley functions which have an arbitrary puncture that is special to sent to infinity,
  the CHY integral of themselves squared is well defined  as it is just a relabelling.  While those of two different Cayley functions may meet an illness as ${\rm SL}(2,{\mathbb C})$ redundancy only allow to send one puncture to infinity and might not satisfy both requirement of two different Cayley functions.  This time,
  it seems we couldn't use the technical described  in Theorem \ref{identicalCcp} to do their CHY integral while it is not.  Many properties are inherited, such as $P({\bm C}{\bm C}')=P({\bm C})\cap P({\bm C}')$
 and the Feynman diagrams obtained by the CHY integral of two distinct Cayley functions is still the intersection of those obtained by the CHY integral of
 of Cayley function squared, except that these Feynman diagrams may not share an overall sign again and we have to determine them one by one.

 For example,
to do the CHY integral of  two star graphs
with different punctures which is expected to be sent to infinity , we have to use their  ${\rm SL}(2,{\mathbb C})$ covariant \eqref{cccova}, denoted as
${\bm C}_n^{S}(i;n), {\bm C}_n^{S}(j;n')$ respectively
\ba
{\bm C}_n^{S}(i;n)&=&
\frac {\sigma_{i,n}^{n{-}3}}{\sigma_{i,1}\cdots \sigma_{i,i-1} \sigma_{i,i+1}\cdots \sigma_{i,n{-}1} \sigma_{1,n} \cdots \sigma_{n{-}1, n}}\,,\nl
{\bm C}_n^{S}(j;n')&=&
\frac {\sigma_{j,n'}^{n{-}2}}{\sigma_{j,1}\cdots \sigma_{j,j-1} \sigma_{j,j+1}\cdots \sigma_{j,n} \sigma_{1,n'} \cdots \sigma_{n'-1,n'} \cdots \sigma_{n'+1,n'} \sigma_{n, n'}}\,.
\ea
Then do the original CHY integral \eqref{sm}. Owing the symmetry of $j,n'$ in ${\bm C}_n^{S}(j;n')$, one can expect the CHY formulas of ${\bm C}_n^{S}(i;n),{\bm C}_n^{S}(i;n')$ should be analogue to \eqref{starstar}. So here we only consider the case with $i,j,n,n'$ four different particles and it turns out that
 \ba\label{starstar}
 \int \dif {\bm \mu}_n
{\bm C}_n^{S}(i;n) {\bm C}_n^{S}(j;n')
&=&\sum
\raisebox{-1cm}{
 \begin{tikzpicture}[scale=.7]
    \draw (0,0)--(4.5,0);
    \draw (1,0)--(1,1);
    \draw (3,0)--(3,1);
       \draw (3.5,0)--(3.5,1);
    \draw (1.5,0)--(1.5,1);
    \fill (2,.7) circle (.02);
    \fill (2.3,.7) circle (.02);
    \fill (2.6,.7) circle (.02);
    \node at (4.7,0) {$n$};
    \node at (-.2,0) {$i$};
      \node at (3.5,1.2) {$n'$};
            \node at (1,1.2) {$j$};

    \draw [decorate,decoration={brace,mirror,amplitude=10pt}]
    (3.1,1.1)--(1.4,1.1)  node [midway,yshift=0.25in] {$(n{-}4)!$ permutations};
    \end{tikzpicture}
    }
    \nl&&+
 (-1)^n  \sum
\raisebox{-1cm}{
 \begin{tikzpicture}[scale=.7]
    \draw (0,0)--(4.5,0);
    \draw (1,0)--(1,1);
    \draw (3,0)--(3,1);
       \draw (3.5,0)--(3.5,1);
    \draw (1.5,0)--(1.5,1);
    \fill (2,.7) circle (.02);
    \fill (2.3,.7) circle (.02);
    \fill (2.6,.7) circle (.02);
    \node at (4.7,0) {$n$};
    \node at (-.2,0) {$i$};
      \node at (3.5,1.2) {$j$};
            \node at (1,1.2) {$n'$};

    \draw [decorate,decoration={brace,mirror,amplitude=10pt}]
     (3.1,1.1)--(1.4,1.1)  node [midway,yshift=0.25in] {$(n{-}4)!$ permutations};
    \end{tikzpicture}
    }\,.\nl
 \ea
 Here we see the results are the intersection of Feynman diagrams of the CHY formulas of $\big({\bm C}_n^{S}(i;n)\big)^2$ and those of $ \big({\bm C}_n^{S}(j;n')\big)^2$. While we also see there may be relative sign between Feynman diagrams.

For example,
 \ba
  \int \dif {\bm \mu}_n
{\bm C}_5^{S}(3;5) {\bm C}_6^{S}(4;1)
&=&
  \raisebox{-.55cm}{
\begin{tikzpicture}[node distance=.4cm]
 \coordinate (1);
 \coordinate [ right of = 1] (a1) ;
  \coordinate [ right of = a1] (a2) ;
   \coordinate [ right of = a2] (a3) ;
    \coordinate [ right of = a3] (5) ;
    \coordinate [ above of = a1] (2) ;
      \coordinate [ above of = a2] (3) ;
         \coordinate [ above of = a3] (4) ;
      \draw (1) node[below=0pt]{{\tiny 3}}  --(5)node[below=0pt]{{\tiny 5}};
      \draw (a1)  --(2)node[above=0pt]{{\tiny 4}};
         \draw (a2) --(3)node[above=0pt]{{\tiny 2}};
          \draw (a3) --(4)node[above=0pt]{{\tiny 1}};
\end{tikzpicture}}
-
  \raisebox{-.55cm}{
\begin{tikzpicture}[node distance=.4cm]
 \coordinate (1);
 \coordinate [ right of = 1] (a1) ;
  \coordinate [ right of = a1] (a2) ;
   \coordinate [ right of = a2] (a3) ;
    \coordinate [ right of = a3] (5) ;
    \coordinate [ above of = a1] (2) ;
      \coordinate [ above of = a2] (3) ;
         \coordinate [ above of = a3] (4) ;
      \draw (1) node[below=0pt]{{\tiny 3}}  --(5)node[below=0pt]{{\tiny 5}};
      \draw (a1)  --(2)node[above=0pt]{{\tiny 1}};
         \draw (a2) --(3)node[above=0pt]{{\tiny 2}};
          \draw (a3) --(4)node[above=0pt]{{\tiny 4}};
\end{tikzpicture}}\,,
\nl
 \int \dif {\bm \mu}_n
{\bm C}_6^{S}(2;6) {\bm C}_6^{S}(3;1)
&=&
  \raisebox{-.55cm}{
\begin{tikzpicture}[node distance=.4cm]
 \coordinate (1);
 \coordinate [ right of = 1] (a1) ;
  \coordinate [ right of = a1] (a2) ;
   \coordinate [ right of = a2] (a3) ;
    \coordinate [ right of = a3] (a4) ;
     \coordinate [ right of = a4] (6) ;
    \coordinate [ above of = a1] (2) ;
      \coordinate [ above of = a2] (3) ;
         \coordinate [ above of = a3] (4) ;
    \coordinate [ above of = a4] (5) ;

     \draw (1) node[below=0pt]{{\tiny 2}}  --(6)node[below=0pt]{{\tiny 6}};
      \draw (a1)  --(2)node[above=0pt]{{\tiny 3}};
         \draw (a2) --(3)node[above=0pt]{{\tiny 4}};
          \draw (a3) --(4)node[above=0pt]{{\tiny 5}};
          \draw (a4) --(5)node[above=0pt]{{\tiny 1}};
\end{tikzpicture}}
+
  \raisebox{-.55cm}{
\begin{tikzpicture}[node distance=.4cm]
 \coordinate (1);
 \coordinate [ right of = 1] (a1) ;
  \coordinate [ right of = a1] (a2) ;
   \coordinate [ right of = a2] (a3) ;
    \coordinate [ right of = a3] (a4) ;
     \coordinate [ right of = a4] (6) ;
    \coordinate [ above of = a1] (2) ;
      \coordinate [ above of = a2] (3) ;
         \coordinate [ above of = a3] (4) ;
    \coordinate [ above of = a4] (5) ;

     \draw (1) node[below=0pt]{{\tiny 2}}  --(6)node[below=0pt]{{\tiny 6}};
      \draw (a1)  --(2)node[above=0pt]{{\tiny 3}};
         \draw (a2) --(3)node[above=0pt]{{\tiny 5}};
          \draw (a3) --(4)node[above=0pt]{{\tiny 4}};
          \draw (a4) --(5)node[above=0pt]{{\tiny 1}};
\end{tikzpicture}}
\nl
&&+
  \raisebox{-.55cm}{
\begin{tikzpicture}[node distance=.4cm]
 \coordinate (1);
 \coordinate [ right of = 1] (a1) ;
  \coordinate [ right of = a1] (a2) ;
   \coordinate [ right of = a2] (a3) ;
    \coordinate [ right of = a3] (a4) ;
     \coordinate [ right of = a4] (6) ;
    \coordinate [ above of = a1] (2) ;
      \coordinate [ above of = a2] (3) ;
         \coordinate [ above of = a3] (4) ;
    \coordinate [ above of = a4] (5) ;

     \draw (1) node[below=0pt]{{\tiny 2}}  --(6)node[below=0pt]{{\tiny 6}};
      \draw (a1)  --(2)node[above=0pt]{{\tiny 1}};
         \draw (a2) --(3)node[above=0pt]{{\tiny 4}};
          \draw (a3) --(4)node[above=0pt]{{\tiny 5}};
          \draw (a4) --(5)node[above=0pt]{{\tiny 3}};
\end{tikzpicture}}
+
  \raisebox{-.55cm}{
\begin{tikzpicture}[node distance=.4cm]
 \coordinate (1);
 \coordinate [ right of = 1] (a1) ;
  \coordinate [ right of = a1] (a2) ;
   \coordinate [ right of = a2] (a3) ;
    \coordinate [ right of = a3] (a4) ;
     \coordinate [ right of = a4] (6) ;
    \coordinate [ above of = a1] (2) ;
      \coordinate [ above of = a2] (3) ;
         \coordinate [ above of = a3] (4) ;
    \coordinate [ above of = a4] (5) ;

     \draw (1) node[below=0pt]{{\tiny 2}}  --(6)node[below=0pt]{{\tiny 6}};
      \draw (a1)  --(2)node[above=0pt]{{\tiny 1}};
         \draw (a2) --(3)node[above=0pt]{{\tiny 5}};
          \draw (a3) --(4)node[above=0pt]{{\tiny 4}};
          \draw (a4) --(5)node[above=0pt]{{\tiny 3}};
\end{tikzpicture}}\,.
\ea

\bibliographystyle{utphys}

\bibliography{mybibliography}

\end{document}